\documentclass[a4paper,11pt]{article}

\pdfoutput=1

\usepackage{jheppub}
\usepackage{booktabs} 
\usepackage[T1]{fontenc}
\usepackage{xcolor}
\usepackage{textcomp}
\usepackage{amsmath}
\usepackage{amssymb}
\usepackage{graphicx}
\usepackage{array}
\usepackage{multirow}
\usepackage{epstopdf}
\usepackage{slashed}
\usepackage{caption}
\usepackage[utf8]{inputenc}
\usepackage{mathtools}
\usepackage{subfigure}

\newcommand{\ra}[1]{\renewcommand{\arraystretch}{#1}}
\newcommand{\be}{\begin{equation}}
\newcommand{\ee}{\end{equation}}

\newcommand{\average}[1]{\ensuremath{\left\langle #1 \right\rangle}}
\newcommand{\parentheses}[1]{\ensuremath{\left( #1 \right)}}
\newcommand{\commutator}[1]{\ensuremath{\left[ #1 \right]}}
\newcommand{\anticommutator}[1]{\ensuremath{\left\lbrace #1 \right\rbrace}}

\newcommand{\bra}[1]{\langle #1 |}
\newcommand{\ket}[1]{| #1 \rangle}

\def\sba  {s_{\beta-\alpha}}
\def\cba  {c_{\beta-\alpha}}

\newcommand{\ULd}{U_L^d}\newcommand{\ULdD}{U_L^{d\dagger}}
\newcommand{\URd}{U_R^d}
\newcommand{\ULuD}{U_L^{u\dagger}}
\newcommand{\URu}{U_R^u}
\newcommand{\Xqx}[2]{X^{#1}_{#2}}
\newcommand{\XdL}{\Xqx{d}{\rm L}}\newcommand{\XdR}{\Xqx{d}{\rm R}}
\newcommand{\XuL}{\Xqx{u}{\rm L}}\newcommand{\XuR}{\Xqx{u}{\rm R}}
\newcommand{\CKMmat}{V}
\newcommand{\CKMLmat}{V_L^{\phantom{\dagger}}}\newcommand{\CKMLmatdag}{V_L^\dagger}
\newcommand{\CKMRmat}{V_R^{\phantom{\dagger}}}\newcommand{\CKMRmatdag}{V_R^\dagger}
\newcommand{\CKM}[1]{\CKMmat_{#1}}
\newcommand{\CKMc}[1]{\CKMmat_{#1}^\ast}
\newcommand{\BR}{\text{BR}}
\newcommand{\refEQ}[1]{Eq.~\eqref{#1}}

\newcommand{\Heff}{\mathcal H_{\rm eff}}
\newcommand{\HEFF}[1]{\mathcal H_{\rm eff}^{#1}}
\newcommand{\MBs}{M_{B_s}}
\newcommand{\DMBs}{\Delta\MBs}

\preprint{\\ADP-19-14/T1094,\, CFTP/19-023,\, SISSA 19/2019/FISI}

\DeclareUnicodeCharacter{00A0}{ }

\title{\boldmath Higgs Quark Flavor Violation: Simplified Models and Status of General Two-Higgs-Doublet Model}

\def\coepp{ARC Center of Excellence for Particle Physics at Terascale, University of Adelaide, Adelaide, SA 5005, Australia}
\def\sissa{SISSA/INFN, Via Bonomea 265, I-34136 Trieste, Italy}
\def\lisbon{Centro de Física Teórica de Partículas (CFTP), Instituto Superior Técnico (IST), U. de Lisboa (UL), Av. Rovisco Pais 1, P-1049-001 Lisboa, Portugal}

\author[a]{\textbf{Juan Herrero-Garcia,}\vspace*{0mm}}
\author[c]{\textbf{Miguel Nebot,}\vspace*{0mm}}
\author[b]{\textbf{Filip Rajec,}\vspace*{0mm}}
\author[b]{\textbf{Martin White}\vspace*{0mm}}
\author[b]{\textbf{and Anthony G. Williams\,}\vspace*{0mm}}

\affiliation[a]{\sissa}
\affiliation[b]{\coepp}
\affiliation[c]{\lisbon}

\emailAdd{jherrero@sissa.it}
\emailAdd{miguel.r.nebot.gomez@tecnico.ulisboa.pt}
\emailAdd{filip.rajec@adelaide.edu.au}
\emailAdd{anthony.williams@adelaide.edu.au}
\emailAdd{martin.white@adelaide.edu.au}

\abstract{
We study quark flavor violating interactions mediated by the Higgs boson $h$. We consider observables involving a third generation quark, of both the up and the down quark sectors, like $h\rightarrow bs$ and $t\rightarrow ch$. Using an effective field theory approach we systematically list all the possible tree-level ultraviolet completions, which comprise models with vector-like quarks and/or extra scalars. We provide upper bounds on the flavor violating transitions allowed by current limits stemming from low energy processes, such as meson mixing and $b \to s \gamma$. We find that scenarios with vector-like quarks always have very suppressed flavor-violating transitions, while a general two Higgs doublet model may have a sizeable rate. To study the latter case in detail, we perform a full numerical simulation taking into account all relevant theoretical and phenomenological constraints. Our results show that ${\rm BR}(t\rightarrow ch)$ [${\rm BR}(h\rightarrow bs)$] are still allowed at the subpercent [percent] level, which are being [may be] explored at the LHC [future colliders]. Finally, we have found that the mild mass-splitting discrepancy with respect to the SM in the $B_s$ meson system can be accommodated in the Two-Higgs-Doublet Model. If confirmed, it yields the prediction ${\rm BR}(h\to bs)\simeq 10^{-4}$, if the new contribution to the mass-splitting is dominated by tree-level Higgs boson exchange.}

\makeatother
\begin{document} 
\maketitle

\section{Introduction} \label{intro}

In the Standard Model (SM), neutral flavor-changing transitions are absent at tree level. They arise at the one loop level with various (additional) sources of suppression like, for example, small elements of the Cabibbo-Kobayashi-Maskawa (CKM) mixing matrix or the Glashow-Iliopoulos-Maiani (GIM) mechanism. It is then clear that they constitute a privileged arena in the search for physics beyond the SM. The discovery in 2012 \cite{Aad:2012tfa,Chatrchyan:2012xdj} of a Higgs-like scalar, $h$ in the following, opened the possibility of exploring a new domain in neutral flavor-changing transitions and a strong experimental effort has followed, targeting processes like $t\to ch,uh$ or $h\to \bar t^\ast c\,(u)\to W^-\bar bc\,(u)$, and potentially also $h \rightarrow b s,\,bd$. We generically denote these processes as Higgs Quark Flavor Violation (HQFV).

Since $m_t\simeq v/\sqrt 2$, with $v$ the electroweak symmetry breaking vacuum expectation value of the SM Higgs doublet, the Higgs-top Yukawa coupling is close to 1: if new physics is present, one may expect that such large couplings also manifest in observable transitions of the top quark to up or charm quarks mediated by $h$. Different studies of top flavor-changing neutral decays can be found in Refs.~\cite{AguilarSaavedra:2004wm, Yang:2004af, Larios:2006pb, Ferreira:2008cj, Craig:2012vj, Harnik:2012pb, Degrande:2014tta, Abbas:2015cua, Altunkaynak:2015twa, Cirigliano:2016nyn}. 
Current experimental bounds on the branching ratios of those processes are at the $10^{-3}$ level (see for example Refs.~\cite{Patrignani:2016xqp,Aaboud:2017mfd,Sirunyan:2017uae,Aaboud:2018pob}):
\be  \label{eq:topFlChDecays:Exp1}
\BR(t\to hq)<7.9\cdot 10^{-3},\quad \BR(t\to hc)<2.2\cdot 10^{-3},\qquad \BR(t\rightarrow hu)<2.4\cdot 10^{-3},
\ee
at 95\% CL. 
Limits on flavour changing couplings of the top quark to the $Z$ boson are also quite stringent (see Refs.~\cite{Abazov:2011qf,Patrignani:2016xqp,Chatrchyan:2013nwa,Sirunyan:2017kkr,Aad:2015uza,Aaboud:2018nyl}): 
\be\label{eq:topFlChDecays:Exp}
\BR(t\to Zc)<2.4\cdot 10^{-4},\qquad \BR(t\rightarrow Zu)<1.7\cdot 10^{-4},
\ee
at 95\% CL. Similar constraints apply to $\BR(t\to q\gamma,qg)$ (see for example Refs.~\cite{Aad:2015gea,Khachatryan:2015att}). Concerning flavor-changing couplings of $h$ to other quarks, the LHC experiments have little direct sensitivity \cite{Blankenburg:2012ex}, while the ILC could in principle reach subpercent sensitivity for the branching ratios of $h\to bs,bd$ \cite{Barducci:2017ioq}; in Ref.~\cite{Crivellin:2017upt}, it was found that $\BR(h\to bs)$ can be as large as $10^{-1}$ in Two-Higgs-Doublet Models (2HDM). Indirect constraints can also be obtained from transitions (i.e. mixing) in the different neutral meson systems, $K^0$--$\bar K^0$ (ds), $D^0$--$\bar D^0$ (cu), $B_d^0$--$\bar B_d^0$ (bd) and $B_s^0$--$\bar B_s^0$ (bs), and from rare decays like $b\to s\gamma$.

In this paper we concentrate on transitions involving the third and second quark generations. The goal is to answer what are the largest possible values in the case of UV completions, beyond the effective approach. The paper is organised as follows. In section \ref{EFT} we discuss quark flavor violation in the SM and beyond using an EFT approach. We list all possible UV completions and show how the general two Higgs doublet model -- type III -- is the most promising scenario for large HQFV. In section \ref{sec:2HDM} we concentrate on the relevant aspects of the latter. Flavor related constraints are addressed in section \ref{sec:ConstraintsFlavor}. A numerical analysis is then presented in section \ref{sec:numerical_analysis}. Additional details are covered in the appendices.


\section{Quark Flavor Violation in the SM and Beyond} \label{EFT}

In this section we discuss different aspects of quark flavor violation in the SM and beyond. We define the effective Yukawa couplings of the Higgs boson to up and down quarks as
\be \label{Yuk_eff}
\mathcal{L}^{\rm eff}_{\rm Yuk} \equiv -\bar q_u\, y_{\rm u}\, q_u \,h - \bar q_d \,y_{\rm d}\, q_d\, h\,+ {\rm H.c.},
\ee
with summation over omitted generation indices understood: $q_u=(u,c,t)$ and $q_d=(d,s,b)$ are vectors in generation space (the quark fields are in their mass bases) and $y_{\rm u}$ and $y_{\rm d}$ are $3\times 3$ complex Yukawa coupling matrices.

\subsection{Effective Field Theory for Higgs Quark Flavor Violation}

\begin{figure}[htb]
\begin{center}
\subfigure[]{\includegraphics[height=0.14\textheight]{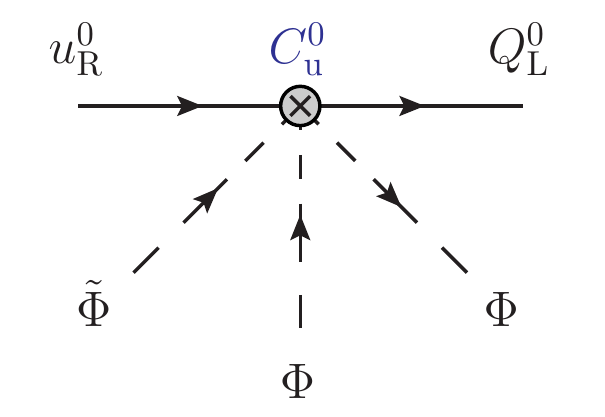}}\qquad
\subfigure[]{\includegraphics[height=0.14\textheight]{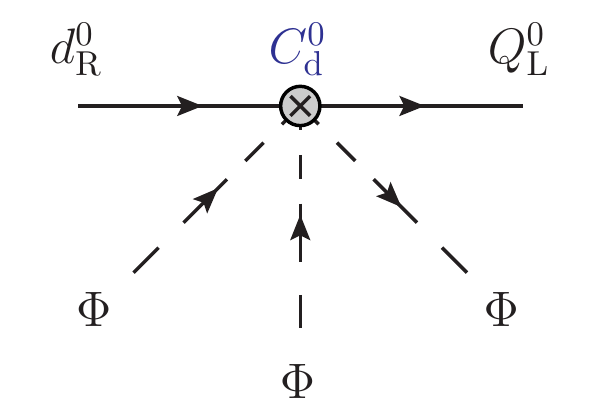}}
\caption{\emph{Yukawa operators} in Eqs.~\eqref{Yuk_op_quarks_up} and \eqref{Yuk_op_quarks_down}.\label{fig:YukawaOperators}}
\end{center}
\end{figure}

In the SM the quark kinetic terms at the renormalizable level read
\be   \label{kin_quarks}
\mathcal{L}_{\rm kin}= \bar Q^0 i \slashed{D} Q^0+\bar u_{\rm R}^0i  \slashed{D} u_{\rm R}^0+\bar d_{\rm R}^0i  \slashed{D} d_{\rm R}^0 + {\rm H.c.},\,
\ee
where, under $SU(2)_{\rm L}$, $Q^0=(u_{\rm L}^0,\,d_{\rm L}^0)$ are the quark doublets, and $u_{\rm R}^0$, ($d_{\rm R}^0$) the up-type (down-type) quark singlets. ``0'' superscripts correspond to fields in a weak basis while the mass eigenstate basis is unlabelled. $D$ denotes the covariant derivative for the different SM transformations. The SM Yukawa Lagrangian for the up and down-type quarks is
\be   \label{Yuk_quarks}
\mathcal{L}_{\rm Yuk}= - \bar Q^0\,Y_{\rm u}\,u_{\rm R}^0\tilde{\Phi}-\bar Q^0\,Y_{\rm d}\,d_{\rm R}^0\Phi+ {\rm H.c.}\,,
\ee
where $\Phi=(\Phi^+,\,\Phi_0)^T$ is the SM Higgs doublet. Electroweak symmetry is spontaneously broken by $\langle \Phi \rangle=\frac{v}{\sqrt 2}\left(\begin{smallmatrix}0\\ 1\end{smallmatrix}\right)$, with $v\simeq 246$ GeV, and thus $\mathcal{L}_{\rm Yuk}$ includes mass terms
\be \label{eq:mass_quarks}
\mathcal{L}_{\rm m_q}=-\bar u_{\rm L}^0\,\frac{v}{\sqrt 2}Y_{\rm u}\,u_{\rm R}^0-\bar d_{\rm L}^0\,\frac{v}{\sqrt 2}Y_{\rm d}\,d_{\rm R}^0+ {\rm H.c.}.
\ee
The effective Higgs interactions of Eq.~\eqref{Yuk_eff}, already written in the quark mass basis, have the simple form $-\frac{m_q}{v}\bar qqh$ for each quark $q$, with mass $m_q$. That is, at tree level, Higgs couplings to quarks do not violate flavor in the SM. This is an accidental symmetry of the SM, like gauge coupling universality, lepton flavor/number or baryon number, and will be violated at the loop level or via effective operators. Indeed, at one loop, in the SM $\BR(h \to bs)\sim 10^{-7}$ while $\BR(t \to ch)\sim 10^{-15}$ \cite{Benitez-Guzman:2015ana} (the smallness of $t\to ch$ is due to the extra GIM suppression for virtual down quarks). Beyond Eq.~\eqref{Yuk_quarks}, the lowest dimension quark flavor-changing operators involving the Higgs field appear at dimension 6. We refer to them in the following as \emph{Yukawa operators}. Denoting the scale of new physics by $\Lambda$, the effective Lagrangians for up and down quarks read respectively
\be \label{Yuk_op_quarks_up}
\mathcal{L}^{\rm eff}_{\rm up}=\frac{-1}{\Lambda^2} \bar{Q}^0\,C_{\rm u}^0\,u_{\rm R}^0\tilde{\Phi}\,(\Phi^\dagger \Phi)+{\rm H.c.}\,,
\ee
and 
\be \label{Yuk_op_quarks_down}
\mathcal{L}^{\rm eff}_{\rm down}=\frac{-1}{\Lambda^2} \bar{Q}^0\,C_{\rm d}^0\,d_{\rm R}^0\Phi\,(\Phi^\dagger \Phi)+{\rm H.c.}\,.
\ee
They are represented in Figure \ref{fig:YukawaOperators}.
After electroweak symmetry breaking (EWSB), the diagonalization of the complete quark mass matrices is
\be  \label{mass_quarks_u}
D_{\rm u} = \ULuD\frac{v}{\sqrt{2}}\Big(Y_{\rm u}+C_{\rm u}^0\frac{v^2}{2\Lambda^2}\Big)\URu\,,
\ee
\be   \label{mass_quarks_d}
D_{\rm d} = \ULdD\frac{v}{\sqrt{2}}\Big(Y_{\rm d}+C_{\rm d}^0\frac{v^2}{2\Lambda^2}\Big)\URd\,.
\ee
Now the effective Yukawas of the Higgs in Eq.~\eqref{Yuk_eff} read
\be    \label{Y_quarks}
y_{\rm u}= \frac{D_u}{v}+C_{\rm u} \frac{v^2}{\sqrt{2}\Lambda^2}\,, \qquad y_{\rm d}= \frac{D_d}{v}+C_{\rm d} \frac{v^2}{\sqrt{2}\Lambda^2}\,,
\ee
where $C_{\rm u}=\ULuD C_{\rm u}^0\URu$ and $C_{\rm d}=\ULdD C_{\rm d}^0\URd$: the Higgs Yukawa interactions are no longer diagonal at tree level, generating HQFV. Without loss of generality, we can use the mass basis for the up-type quarks. The quark charged current interactions read $\mathcal{L}_{W}=\frac{g}{\sqrt 2}\bar{u}_{\rm L}\gamma^{\mu}\CKMmat d_{\rm L}W^+_\mu+\text{H.c.}$, where the CKM matrix is $\CKMmat=\ULd$.

\subsection{Simplified Models} \label{UV}

\begin{figure}
\begin{center}
\subfigure[Topology A]{\includegraphics[height=0.17\textheight]{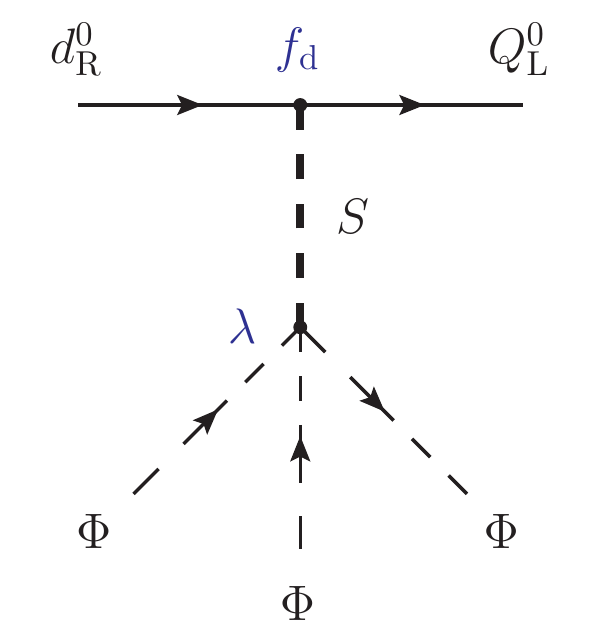}}\qquad
\subfigure[Topology B]{\includegraphics[height=0.17\textheight]{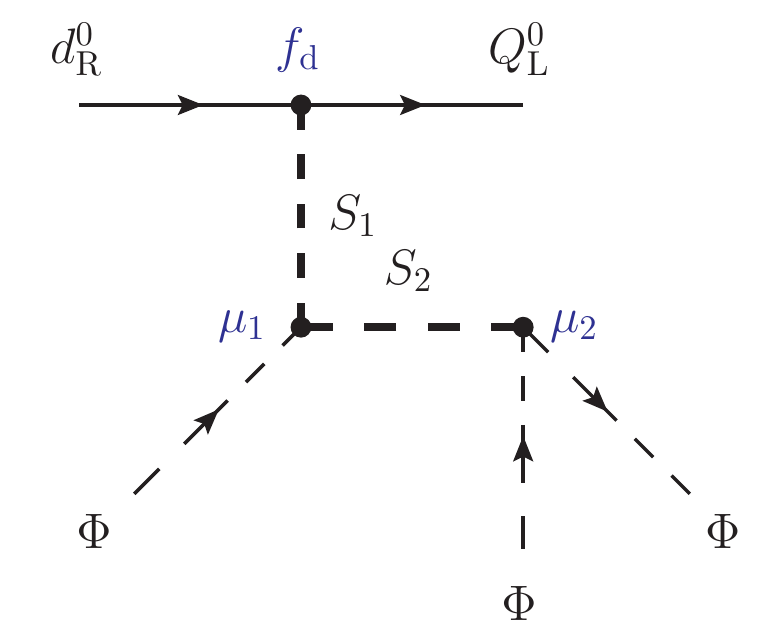}}\qquad
\subfigure[Topology C]{\includegraphics[height=0.17\textheight]{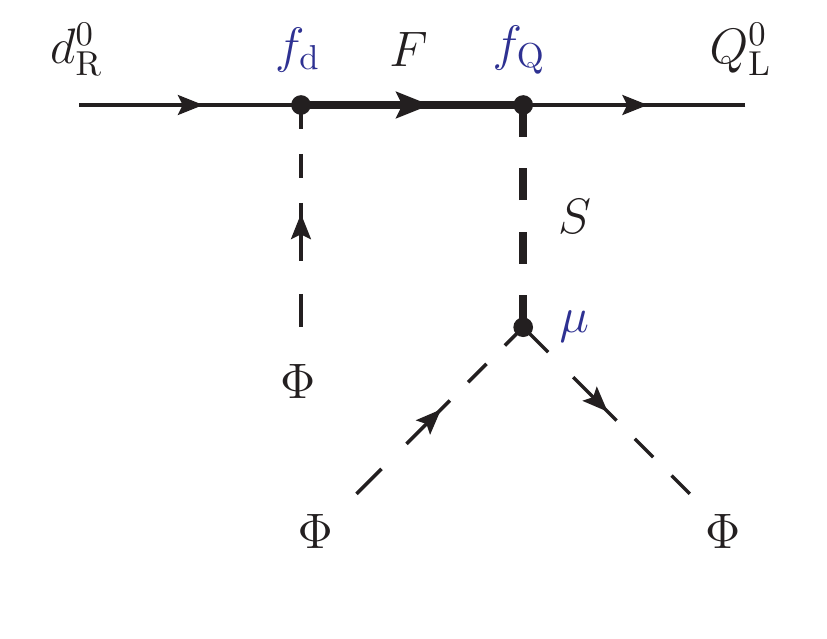}}\qquad
\subfigure[Topology D]{\includegraphics[width=0.48\textwidth]{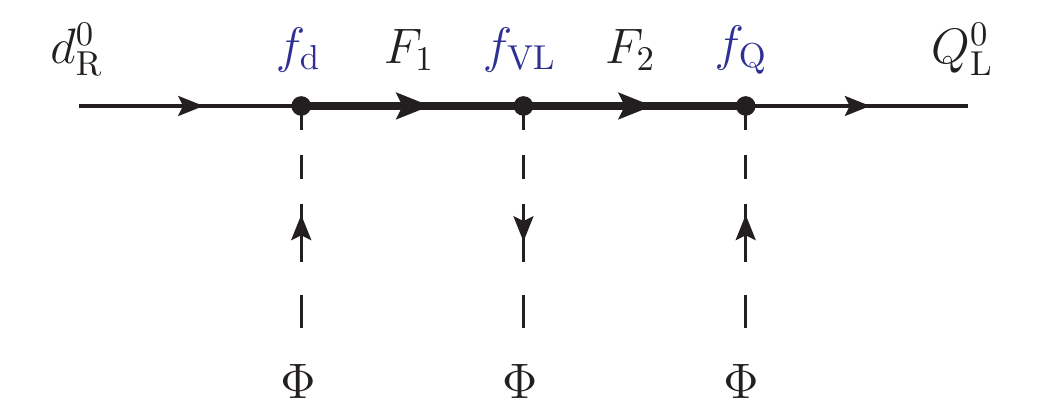}}
\caption{Tree-level topologies of the \emph{Yukawa operator} for down-type quarks, see Table \ref{tab:topologiesH_down}. Similar diagrams exist for up-type quarks. \label{fig:topologiesH_down}}
\end{center}
\end{figure}

In this section we discuss tree level simplified models by ``opening'' the Yukawa operators for up and down quarks given in Eqs.~\eqref{Yuk_op_quarks_up} and \eqref{Yuk_op_quarks_down}, respectively. For down quarks the operators are represented in Figure \ref{fig:topologiesH_down}. Tables \ref{tab:topologiesH_up} (for up quarks) and \ref{tab:topologiesH_down} (for down quarks) list all the possible UV completions. We follow the same approach used for the case of Higgs lepton flavor violation in Ref.~\cite{Herrero-Garcia:2016uab}. We have considered 2 extra particles at most: the new particles considered in each model are given in the second column of Tables \ref{tab:topologiesH_up} and \ref{tab:topologiesH_down}, where S stands for scalar and F for fermion. The ${\rm (SU(2)_L, Y)}$ quantum numbers are given in the third column. In the last column, the form of the contributions to $C_{\rm u,d}$ is provided in terms of the masses of the new particles ($m_F$ or $m_{F_j}$ for fermions, $m_S$ or $m_{S_j}$ for scalars) and of generic new physics couplings: scalar quartic couplings $\lambda$, dimensionful trilinear scalar couplings $\mu$, $\mu_j$, and Yukawa-type couplings $f_{\rm q}$, where ${\rm q=u,d,Q}$ refers to the SM field involved (in that order of preference if two SM fields are involved) or ${\rm q=VLQ}$ for an interaction term involving two new vector-like quarks. Note that the CKM matrix $\CKMmat$ enters in the expression of $C_{\rm d}$ (down quarks).

\subsubsection{The Yukawa Operators} \label{Yukawa}

The tree level UV completions of the up and down \emph{Yukawa operators} that only involve  scalars are identical to the ones discussed for Higgs lepton flavor violation in Ref.~\cite{Herrero-Garcia:2016uab} (see Table 3 therein), denoted by topologies A and B. Topology A corresponds to the 2HDM, which has been extensively studied in all its variants (see for example Refs.~\cite{Ivanov:2017dad,Atwood:1996vj,Mahmoudi:2009zx,Crivellin:2013wna,Celis:2013rcs,Dumont:2014wha,Botella:2015hoa, Arhrib:2015maa,Crivellin:2017upt}). In sections \ref{sec:2HDM} to \ref{sec:ConstraintsFlavor} different aspects of the general 2HDM are discussed, and a full numerical analysis of HQFV is presented in section \ref{sec:numerical_analysis}. Topology B adds new scalars to topology A, while the flavor structure is still dominated by the couplings of the scalar doublet to the quark bilinears. Topologies $C^i$ ($i=1,..,4$) and $D^j$ ($j=1,..,3$)  correspond to models where vector-like quarks (VLQ) are also present (VLQ are, of course, color triplets). Models like these have been studied in the literature, see for instance Refs.~\cite{delAguila:1982fs,Branco:1986my,Lavoura:1992qd,delAguila:1998tp,delAguila:2000aa,delAguila:2000rc,Barenboim:2001fd,Botella:2012ju,Fajfer:2013wca,Aguilar-Saavedra:2013qpa,Ellis:2014dza,Ishiwata:2015cga,Bobeth:2016llm}.

\begin{table} 
\centering
\begin{tabular}{ | c | c |  c |c |}
\hline
Topology & Particles & Representations & $C_{\rm u}/\Lambda^2$\\ \hline \hline 
$A_{\rm u}$ & $S$ &$(2,1/2)_S$& $\frac{f_{\rm u} \lambda}{m_S^2}$ \\ \hline \hline 

$B_{\rm u}$ &$S_1 \oplus S_2$ &$(2,1/2)_S$\,$\oplus$\,$(1,0)_S, (3,0)_S, (3,-1)_S$& $\frac{f_{\rm u} \mu_1 \mu_2}{m_{S_1}^2m_{S_2}^2}$\\ \hline \hline 

$C_{\rm u}^1$ &$S \oplus F$ &$(2,1/6)_F$\,$\oplus$\,$(1,0)_S, (3,0)_S$& \cr
$C_{\rm u}^2$ &$S \oplus F$ &$(2,7/6)_F$\,$\oplus$\,$(3,-1)_S$&  \cr
$C_{\rm u}^3$ &$S \oplus F$ &$(1,2/3)_F$\,$\oplus$\,$(1,0)_S$, $(3,2/3)_F$\,$\oplus$\, $(3,0)_S$& $\frac{f_{\rm Q}f_{\rm u}  \mu}{m_F m_{S}^2}$ \cr
$C_{\rm u}^4$ &$S \oplus F$ &$(3,-1/3)_F$\,$\oplus$\,$(3,-1)_S$& \\ \hline  \hline 

$D_{\rm u}^1$ &$F_1 \oplus F_2$ &$(2,7/6)_F$\,$\oplus$\,$(1,2/3)_F, (3,2/3)_F$& \cr
$D_{\rm u}^2$ &$F_1 \oplus F_2$ &$(2,1/6)_F$\,$\oplus$\,$(1,2/3)_F, (3,2/3)_F$&$\frac{ f_{\rm Q}  f_{\rm VLQ} f_{\rm u}}{m_{F_1} m_{F_2}}$ \cr
$D_{\rm u}^3$ &$F_1 \oplus F_2$ &$(2,1/6)_F$\,$\oplus$\,$(1,-1/3)_F, (3,-1/3)_F$& \\
\hline
\end{tabular}
\caption{Tree-level topologies of the \emph{Yukawa operator} for up-type quarks, see Eq.~\eqref{Yuk_op_quarks_up}. S stands for scalar, F for fermion, with the representation under ${\rm (SU(2)_L, Y)}$. All vector-like fermions are color triplets, while the scalars are color singlets.} \label{tab:topologiesH_up}
\end{table}

\begin{table} 
\centering
\begin{tabular}{ | c | c |  c |c |}
\hline
Topology & Particles & Representations & $C_{\rm d}/\Lambda^2$\\ \hline \hline 
$A_{\rm d}$ &$S$ &$(2,-1/2)_S$& $V^\dagger \frac{f_{\rm d}  \lambda}{m_S^2}$ \\ \hline 
 \hline 
$B_{\rm d}$ &$S_1  \oplus S_2$ &$(2,-1/2)_S$\,$\oplus$\,$(1,0)_S, (3,0)_S, (3,1)_S$& $V^\dagger \frac{f_{\rm d} \mu_1 \mu_2}{m_{S_1}^2m_{S_2}^2}$\\ \hline 
 \hline 
$C_{\rm d}^1$ &$S \oplus F$ &$(2,1/6)_F$\,$\oplus$\,$(1,0)_S, (3,0)_S$& \cr
$C_{\rm d}^2$ &$S \oplus F$ &$(2,-5/6)_F$\,$\oplus$\,$(3,1)_S$& \cr
$C_{\rm d}^3$ &$S \oplus F$ &$(1,-1/3)_F$\,$\oplus$\,$(1,0)_S$, $(3,-1/3)_F$\,$\oplus$\, $(3,0)_S$& $V^\dagger \frac{f_{\rm Q} f_{\rm d} \mu}{m_F m_{S}^2}$\cr
$C_{\rm d}^4$ &$S \oplus F$ &$(3,2/3)_F$\,$\oplus$\,$(3,1)_S$& \\ \hline  \hline 

$D_{\rm d}^1$ &$F_1 \oplus F_2$ &$(2,1/6)_F$\,$\oplus$\,$(1,2/3)_F, (3,2/3)_F$& \cr
$D_{\rm d}^2$ &$F_1 \oplus F_2$ &$(2,1/6)_F$\,$\oplus$\,$(1,-1/3)_F, (3,-1/3)_F$&$V^\dagger \frac{f_{\rm Q}  f_{\rm VLQ} f_{\rm d} }{m_{F_1} m_{F_2}}$ \cr
$D_{\rm d}^3$ &$F_1 \oplus F_2$ &$(2,-5/6)_F$\,$\oplus$\,$(1,-1/3)_F, (3,-1/3)_F$& \cr
\hline
\end{tabular}
\caption{Similar to Table~\ref{tab:topologiesH_up} for the \emph{Yukawa operator} of down-type quarks, see Eq.~\eqref{Yuk_op_quarks_down}.\label{tab:topologiesH_down}}
\end{table}

\subsubsection{The Derivative Operators} \label{derivative}

\begin{figure}
\begin{center}
\subfigure[Topology $E^1_d$]{\includegraphics[height=0.125\textheight]{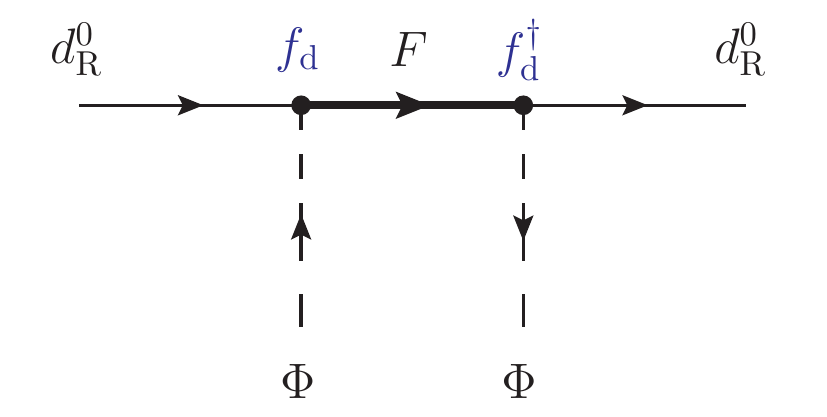}}\qquad
\subfigure[Topology $E^{3,4}_u$]{\includegraphics[height=0.125\textheight]{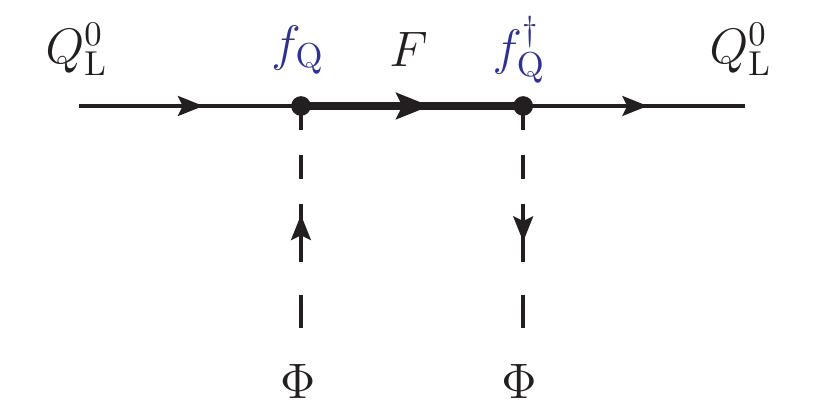}}
\caption{Examples of tree-level topologies (E) of the \emph{Derivative operators}, see Table \ref{tab:topologiesE}.\label{fig:topologiesE_down}}
\end{center}
\end{figure}

Besides the \emph{Yukawa operators}, there are other dimension 6 operators which generate HQFV. They involve covariant derivatives, and therefore we denote them as \emph{Derivative operators}. These are plotted in Figure \ref{fig:topologiesE_down} and are listed in Table \ref{tab:topologiesE}. They are related by the equations of motion (EOM) to the \emph{Yukawa operators} previously considered. This implies that, for instance, for up-type quarks their contribution to HQFV will be proportional to the quark masses. It is illustrative to consider them specifically. This is because some simple UV models directly generate them, and as we will show they are very constrained by limits from flavor-changing processes involving the Z boson. Moreover, some of the particles that generated the previous \emph{Yukawa operators} also generate these ones.

We show further details regarding the generation of flavor-changing neutral currents from these operators in Appendix~\ref{app:VLQ_details}. The key point is that the flavor-changing neutral currents appear because of a mismatch between the quantum
numbers on which the covariant derivative acts for the case of the Derivative operators and for the renormalizable kinetic terms of Eq.~\eqref{kin_quarks}~\cite{Herrero-Garcia:2016uab}. In Tab.~\ref{tab:topologiesE} we also list all the possible UV completions of the \emph{Derivative operators} (third column), as well as the new Z-mediated quark flavor violating, charged-current (CC) and HQFV interactions. The Higgs interactions in the last column are given in terms of the effective Yukawa couplings $y_q$ provided in Eq.~\eqref{Yuk_eff}. Notice that $Z$ and $W$-boson interactions are independent of the quark mass involved. The chirality of the quarks involved can be understood from the operators.

\begin{table} 
\centering 
\begin{tabular}{ | c |c | c |  c |c |c  |c |}
\hline
Operator&Topology & Particles & $Z u_\alpha u_\beta$ & $Z d_\alpha d_\beta$  & $W d_\alpha  u_\beta$  & $h q_\alpha q_\beta$ \\ \hline \hline 
$ (\bar u_{\rm R} \Phi^\dagger)i\slashed{D} (u_{\rm R}\, \Phi)$&$E_{\rm u}^1$ &$(2,7/6)_F$&-1&&& $\frac{|f_{\rm u}|^2 m_{\rm u} v}{m_F^2}$ \\ \hline
$ (\bar u_{\rm R} \Phi^T) i\slashed{D} (u_{\rm R}\,\Phi^*)$&$E_{\rm u}^2$  &$(2,1/6)_F$&+1&&& $\frac{|f_{\rm u}|^2 m_{\rm u} v}{m_F^2}$ \\ \hline

$ (\bar{Q} \tilde{\Phi})  i\slashed{D} (\tilde{\Phi}^\dagger Q)  $&$E_{\rm u}^{3}$ &$(1,2/3)_F$ &-1&&-1&$\frac{|f_{\rm Q}|^2 m_{\rm u} v}{m_F^2}$  \\ \hline

$(\bar{Q} \vec{\tau}\tilde{\Phi})i\slashed{D}  (\tilde{\Phi}^\dagger\vec{\tau}Q)   $&$E_{\rm u}^{4}$ &$(3,2/3)_F$&-1&-2&+1& $\frac{|f_{\rm Q}|^2 m_{\rm u} v}{m_F^2}$ \\\hline\hline

$ (\bar d_{\rm R} \Phi^\dagger) i\slashed{D} (d_{\rm R}\, \Phi)$&$E_{\rm d}^1$ &$(2,1/6)_F$&&-1&& $\frac{|f_{\rm d}|^2 m_{\rm d} v}{m_F^2}$ \\ \hline
$ (\bar d_{\rm R} \Phi^T) i\slashed{D} (d_{\rm R}\,\Phi^*)$&$E_{\rm d}^2$  &$(2,-5/6)_F$&&+1&& $\frac{|f_{\rm d}|^2 m_{\rm d} v}{m_F^2}$ \\ \hline

$  (\bar{Q} \Phi ) i\slashed{D} (\Phi^\dagger\,Q) $&$E_{\rm d}^{3}$  &$(1,-1/3)_F$&&+1&-1& $\frac{|V^\dagger f_{\rm Q}|^2 m_{\rm d} v}{m_F^2}$\\ \hline

$ (\bar{Q} \vec{\tau}\Phi) i\slashed{D} (\Phi^\dagger\,\vec{\tau}Q) $&$E_{\rm d}^{4}$ &$(3,-1/3)_F$&+2&+1&+1 &$\frac{|V^\dagger f_{\rm Q}|^2 m_{\rm d} v}{m_F^2}$\\ \hline
\end{tabular}
\caption{Tree-level topologies of the \emph{Derivative operators}. The Higgs interactions in the last column correspond to the effective Yukawa couplings $y_q$, provided in Eq.~\eqref{Yuk_eff}, with $q=u$ ($q=d$) for the first (last) four rows. $Z$ couplings are in units of $y_q v/m_q\,\times e/(2c_{W}s_{W})$, while $W$ ones are in units of $V\,y_q v/m_q\,\times e/(2\sqrt{2}s_{W})$.} \label{tab:topologiesE}
\end{table}

We have seen using the \emph{Derivative operators} that in the models with VLQ, both ZQFV and HQFV are related. We can derive the relationship among both explicitly, see also Ref.~\cite{Lavoura:1992qd,delAguila:2000aa,Herrero-Garcia:2016uab}. In models with VLQ, charged current couplings read
\be
\mathcal{L}_{W}=\frac{g}{\sqrt 2}\left(\bar{u}_{\rm L}\gamma^{\mu}\CKMLmat d_{\rm L}+\bar{u}_{\rm R}\gamma^{\mu}\CKMRmat d_{\rm R}\right)W_{\mu}^++\text{H.c.},
\ee
with $\CKMLmat$ the ``enlarged'' CKM matrix, and $\CKMRmat$ its right-handed counterpart (which arises when VLQ which are not $SU(2)_L$ singlets are considered). $\CKMLmat$ is now a $n_u\times n_d$ matrix for $n_u$ up quarks and $n_d$ down quarks, and it is not unitary.\footnote{It can be embedded, however, in a larger $n_L\times n_L$ unitary matrix, with $n_L$ the number of left-handed quark fields; for example, for a model with only an up-type singlet VLQ, $\CKMLmat$ is a $4\times 3$ submatrix of a $4\times 4$ unitary matrix.} The neutral current couplings read
\be\label{eq:ZVLQ}
\mathcal{L}_{Z}=\frac{g}{2c_{W}}\left(\bar{d}_{\rm L}\gamma^{\mu}\XdL d_{\rm L}+\bar{d}_{\rm R}\gamma^{\mu}\XdR d_{\rm R}-\bar{u}_{\rm L}\gamma^{\mu}\XuL u_{\rm L}-\bar{u}_{\rm R}\gamma^{\mu}\XuR u_{\rm R}-2s_{W}^{2}J_{\rm em}^{\mu}\right)Z_{\mu},
\ee
where
\begin{equation}
\XuL=\CKMLmat\CKMLmatdag,\quad \XuR=\CKMRmat\CKMRmatdag,\quad \XdL=\CKMLmatdag\CKMLmat,\quad \XdR=\CKMRmatdag\CKMRmat.
\end{equation}
The $Z$ flavor-changing interactions in $\mathcal L_{Z}$ are given by the non-unitarity of the mixing matrices. Similarly, the Yukawa couplings to $h$ read
\begin{multline}\label{eq:hVLQ}
\mathcal{L}_{h}=-\frac{h}{v}\bar{u}_{\rm L}\left(\XuL D_u+D_u\XuR-2\XuL D_u\XuR\right)u_{\rm R}\\
-\frac{h}{v}\bar{d}_{\rm L}\left(\XdL D_d+D_d\XdR-2\XdL D_d\XdR\right)d_{\rm R}+\text{H.c.}.
\end{multline}
Consider for example the $h\bar t_L c_R$ coupling; in the notation of \refEQ{Yuk_eff},
\begin{equation} \label{hVLQ}
v(y_{\rm u})_{tc}=(\XuL)_{tc}m_c+m_t(\XuR)_{tc}-2(\XuL D_u\XuR)_{tc}\,.
\end{equation}
The first term corresponds to the \emph{Derivative operator} $E_{\rm u}^{3\, (4)}$, where a VLQ singlet (triplet) is exchanged, while the second term corresponds to $E_{\rm u}^{1,2}$, where a VLQ doublet is exchanged. These contributions pick up a quark mass, as would be the case if one uses EOM to transform the operators. The last contribution corresponds to topology D of the up-quark  \emph{Yukawa operator}, where two types of VLQ are exchanged (a doublet plus a singlet or triplet). Therefore we can estimate the contribution as
\be 
v(y_{\rm u})_{tc}\approx \left| \frac{f_{Q}v}{m_{F_{1,3}}}\right|_{tc}^{2}m_{c}+\left|\frac{f_{u}v}{m_{F_{2}}}\right|_{tc}^{2}m_{t}- 2  \left(\frac{f_Q v}{m_{F_{1,3}}}f_{\rm VLQ}v\frac{f_u v}{m_{F_{2}}}\right)_{tc}\,.
\ee
In this example, the dominant term for top HQFV is the second or the last one. For bottom HQFV clearly the last term dominates unless $f_{\rm VLQ}v<m_{b}$.
Correspondingly, the deviations from $3\times 3$ unitarity of the CKM mixing matrix due to the presence of VLQ are
\begin{equation}
\CKMLmatdag\CKMLmat\approx \mathbf{1}- \frac{ |f_{\rm Q}|^2}{m_F^2} \,\frac{e\,  v^2}{\sqrt{2}s_{W}}\,,
\end{equation}
where now $f_{\rm Q}$ and $m_F$ are matrices in flavor space.

 \subsection{Estimates for Models with Vector-like Quarks}

The phenomenology of VLQ models has been scrutinised in the literature~\cite{delAguila:1982fs,Branco:1986my,Lavoura:1992qd,delAguila:1998tp,delAguila:2000aa,delAguila:2000rc,Barenboim:2001fd,Botella:2012ju,Fajfer:2013wca,Aguilar-Saavedra:2013qpa,Ellis:2014dza,Ishiwata:2015cga,Bobeth:2016llm,Harnik:2012pb}. For example, Ref.~\cite{Ishiwata:2015cga} addresses in some detail constraints arising from meson mixing.

For models with just VLQ, since ZQFV and deviations from $3\times 3$ unitarity of CC interactions are related to HQFV, one can estimate some simple upper bounds on $h\to bs$ and $t\to hc$. In the up sector, from Eq.~\eqref{eq:ZVLQ}, the leading contribution to $t\to Zc$ (which occurs at tree level), ignoring QCD corrections, is
\be\label{eq:tZc}
\Gamma(t\to Zc)=\frac{m_t^3}{32\pi v^2}\left(\left|(\XuL)_{ct}\right|^2+\left|(\XuR)_{ct}\right|^2\right) \mathcal{F}(m_Z, m_t)\,,
\ee
where $\mathcal{F}(m_X, m_t) =(1-m_X^2/m_t^2)^2 (1+2 m_X^2/m_t^2)$.
Since the top total decay width $\Gamma_t = \Gamma (t\to Wb)$ remains essentially unchanged, using $\mathcal{F}(m_W, m_t) \approx \mathcal{F}(m_Z, m_t)$, the experimental bound on $\BR(t\to Zc)$ in Eq.~\eqref{eq:topFlChDecays:Exp} gives
\be
\left(\left|(\XuL)_{ct}\right|^2+\left|(\XuR)_{ct}\right|^2\right)<\BR(t\to Zc)_{\rm exp}\,.
\ee
Concerning $t\to hc$ decays, with $m_c\ll m_h,m_t$, from the first line of Eq.~\eqref{eq:hVLQ} we have
\be
-\frac{h}{v}\bar{c}_{\rm L}\left[(\XuL)_{ct}m_t-2(\XuL)_{cq} m_q(\XuR)_{qt}\right]t_{\rm R}.
\ee
If the mixing with heavy VLQ is suppressed compared to the top exchange, the dominant contribution in the second term is $q=t$ and $(\XuR)_{tt}\simeq 1$, and thus we are left with an interaction term $-\frac{m_t}{v}h(\XuL)_{ct}\bar{c}_{\rm L}t_{\rm R}$. From the hermitian conjugate term in \eqref{eq:hVLQ}, the interaction term with flipped chiralities is $-\frac{m_t}{v}h(\XuR)_{ct}\bar{c}_{\rm R}t_{\rm L}$. Then, the leading prediction for $t\to hc$ (again, tree level, $m_c\to 0$ and no QCD corrections) is
\be\label{eq:thc}
\Gamma(t\to hc)=\frac{m_t^3}{32\pi v^2}\left(\left|(\XuL)_{ct}\right|^2+\left|(\XuR)_{ct}\right|^2\right)\mathcal{H}(m_h, m_t)\,,
\ee
where $\mathcal{H}(m_h, m_t) =(1-m_h^2/m_t^2)^2$. Combining Eqs.~\eqref{eq:tZc} and \eqref{eq:thc}, the experimental bound on $\BR(t\to Zc)$ translates into a bound
\be\label{eq:thc:val}
\BR(t\to hc)<\BR(t\to Zc)_{\rm exp}\, \frac{\mathcal{H}(m_h, m_t)}{\mathcal{F}(m_Z, m_t)}\simeq 7\times 10^{-5}\,,
\ee
which is two orders of magnitude smaller than the current sensitivity, Eq.~\eqref{eq:topFlChDecays:Exp1}.

For $b\to s$ transitions, although a similar reasoning would lead to straightforward bounds on the allowed values of $\BR(h\to bs)$ in the context of VLQ extensions of the SM, experimental input on $b-s$ transitions from $Z\to bs$ is much poorer than low energy constraints from $B_s$ mixing, $B_s\to\mu^+\mu^-$ or $b\to s\gamma$ transitions. For the latter, diagrams with chirality flips in the VLQ lines dominate the processes, in an analogous way to those discussed in Ref.~\cite{Herrero-Garcia:2016uab} for the lepton sector, further suppressing HQFV. Typical bounds on $Z_{bs}$ couplings from detailed studies in the literature are below the $10^{-4}$ level; one can thus estimate a rough upper bound on $\BR(h\to bs)$ in the context of VLQ extensions 
\be\label{eq:hbs:val}
\BR(h\to bs)<|Z_{bs}|^2\frac{3}{8\pi}\frac{m_h}{\Gamma_h}\simeq 10^{-5}\,.
\ee
Equations \eqref{eq:thc:val} and \eqref{eq:hbs:val} illustrate that HQFV in extensions with just VLQ, with branching ratios forced to be below the $10^{-5}$ level by ZQFV, are much less promising than scenarios with HQFV arising from a richer scalar sector. Moreover, the VLQ generating the \emph{Yukawa operators}, always generate the \emph{Derivative operators}, and therefore are subject to strong constraints. Therefore, in the following we focus on the simplest scalar scenario generating the \emph{Yukawa operators}: topology A, the Two-Higgs-Doublet Model.

\section{The General (Type III) Two-Higgs-Doublet Model} \label{sec:2HDM}
In this section we introduce the general 2HDM, also known as Type III 2HDM. Reviews addressing different 2HDMs can be found in Refs.~\cite{Gunion:2002zf,Davidson:2005cw, Djouadi:2005gj, Branco:2011iw, Ivanov:2017dad}. In section \ref{sec:pot} we discuss the scalar potential and in section \ref{YukawaCouplings} the Yukawa couplings. Aspects relevant for Higgs flavor-changing processes are studied in section \ref{HiggsFlavorChanging}. 

\subsection{The Scalar Potential}\label{sec:pot}

In a generic basis both Higgs scalar doublets $\Phi_{1}$ and $\Phi_{2}$ take VEVs denoted by $v_1$ and $v_2$, respectively. One can rotate to the Higgs basis \cite{Georgi:1978ri,Donoghue:1978cj,Botella:1994cs} where only one linear combination of $\Phi_{1}$ and $\Phi_{2}$, denoted by $H_1$, has a non-vanishing VEV, equal to $v=\sqrt{v^2_1+v^2_2} \simeq 246$~GeV, via the transformation
\begin{equation}
\label{eq:rotation_HB}
\parentheses{\begin{array}{c} H_1 \\ H_2 \end{array}} = \parentheses{\begin{array}{cc} c_\beta & s_\beta \\ -s_\beta & c_\beta \end{array}} \parentheses{\begin{array}{c} \Phi_1 \\ \Phi_2 \end{array}} \,,
\end{equation}
where the angle $\beta$ defines the mixing between the two doublets, with $\tan\beta\equiv v_2/v_1$, and the short-hand notations $s_x \equiv \sin x$ and $c_x \equiv \cos x$. We will also use $t_x \equiv \tan x$. In the Higgs basis, the doublets take the form
\begin{equation}
\label{eq:doublets_HB}
H_1 = \parentheses{\begin{array}{c} G^+ \\ \dfrac{1}{\sqrt{2}}\parentheses{v + \varphi_1 + i G^0} \end{array}} \,, \qquad H_2 = \parentheses{\begin{array}{c} H^+ \\ \dfrac{1}{\sqrt{2}}\parentheses{\varphi_2 + i A} \end{array}} \,,
\end{equation}
where $\varphi_1$ and  $\varphi_2$ are CP-even neutral Higgs fields, $A$ is a CP-odd neutral Higgs field, $H^+$ is a charged Higgs field, and $G^+$ and $G^0$ are the would-be Goldstone bosons, which provide the longitudinal polarizations of the $W^+$ and the $Z$ gauge bosons.
The most general scalar potential is given in the Higgs basis by \footnote{The transformations of parameters between different scalar bases can be found in Appendix A of Ref.~\cite{Davidson:2005cw}.}
\begin{align}
\label{eq:scalar_potential}
V &= M_{11}^2 H_1^\dagger H_1 + M_{22}^2 H_2^\dagger H_2 -  \parentheses{M_{12}^2 H_2^\dagger H_1 + {\rm H.c.}} + \dfrac{1}{2}\Lambda_1 \parentheses{H_1^\dagger H_1}^2 \nonumber\\
&+ \dfrac{1}{2}\Lambda_2 \parentheses{H_2^\dagger H_2}^2 + \Lambda_3 \parentheses{H_1^\dagger H_1} \parentheses{H_2^\dagger H_2} + \Lambda_4 \parentheses{H_1^\dagger H_2} \parentheses{H_2^\dagger H_1} \nonumber\\
&+ \anticommutator{\dfrac{1}{2}\Lambda_5 \parentheses{H_1^\dagger H_2}^2 + \commutator{\Lambda_6 \parentheses{H_1^\dagger H_1} + \Lambda_7 \parentheses{H_2^\dagger H_2}}H_1^\dagger H_2 + {\rm H.c.}}\,,
\end{align}
where $\Lambda_i$ ($i = 1,2,\ldots,7$) are the quartic couplings and $M^2_{ij}$  are bare mass-squared parameters. In general, $\Lambda_{5}$, $\Lambda_{6}$, $\Lambda_{7}$ and $M_{12}$ can be complex but, by redefining $H_1$ and $H_2$, one can, for example, choose $\Lambda_5$ to be real~\cite{Davidson:2005cw}. We assume, for simplicity, a CP-conserving scalar sector: all the parameters in \refEQ{eq:scalar_potential} are real. 

The minimisation conditions
\begin{equation}
M_{11}^2 = - \dfrac{1}{2}\Lambda_1 v^2 \,, \qquad M_{12}^2 = \dfrac{1}{2}\Lambda_6 v^2\,, \label{eq:mu1mu3}
\end{equation}
can be used to eliminate $M_{11}^2$ and $M_{12}^2$ as independent parameters. Inserting $\average{H_1} = (0, v/\sqrt{2})^T$ into Eq.~\eqref{eq:scalar_potential}, we obtain the squared mass of the charged scalar,
\begin{equation}
m_{H^\pm}^2 = M_{22}^2 +\dfrac{1}{2} v^2 \Lambda_3\,,
\end{equation}
and the mass matrix of the CP-even neutral scalars
\begin{equation}
\mathcal{M}_h^2= \left( \begin{array}{cc}
\Lambda_1 v^2 & \Lambda_6 v^2 \\
\Lambda_6 v^2 &\, m_{A}^2+\Lambda_5 v^2
\end{array} \right)\,,
\label{eq:cp_even_mass_matrix}
\end{equation}
where the mass of the CP-odd scalar is
\begin{equation}
m^2_{A} = m^2_{H^\pm} - \dfrac{1}{2} v^2 \parentheses{\Lambda_5 - \Lambda_4}\,.
\end{equation}
Thus, in the Higgs basis, the mass eigenstates $h$ and $H$ are a mixture of the CP-even states $\varphi_{1}$ and $\varphi_{2}$
\begin{equation}
\label{eq:CP-even_mixing}
\parentheses{\begin{array}{c} h \\ H \end{array}} = \parentheses{
\begin{array}{cc}
s_{\beta -\alpha} & c_{\beta -\alpha} \\
 c_{\beta -\alpha} & -s_{\beta -\alpha} \\
\end{array}
} \parentheses{\begin{array}{c} \varphi_1 \\ \varphi_2\end{array}},
\end{equation}
with masses
\begin{equation}\label{eq:lambda6}
m^2_{H,h} = \dfrac{1}{2} \left\{m^2_{A} + v^2 \parentheses{\Lambda_1 + \Lambda_5} \pm \sqrt{\commutator{m^2_{A}+ v^2 \parentheses{\Lambda_5 - \Lambda_1}}^2 + 4v^4 \Lambda_6^2 }\right\}\,.
\end{equation}
The mixing in \refEQ{eq:CP-even_mixing} is
\begin{equation}\label{eq:sinbminua}
s_{2(\beta -\alpha)}=  -\dfrac{2\Lambda_6 v^2}{m^2_{H}-m^2_{h}}.
\end{equation}
It will turn out useful to obtain $\Lambda_6$ by combining Eqs.~\eqref{eq:lambda6} and \eqref{eq:sinbminua}
\begin{equation}\label{eq:lamba6def}
\Lambda_6 = \frac{t_{2(\beta-\alpha)}}{2v^2} \biggl[ m_A^2 + v^2 (\Lambda_5 - \Lambda_1)  \biggr].
\end{equation}
Eq.~\eqref{eq:sinbminua} and Eq.~\eqref{eq:lamba6def} determine the sign of $\Lambda_6$ in terms of $\beta-\alpha$.
In the general 2HDM $t_\beta$ is not a physical parameter (see Ref.~\cite{Haber:2006ue} for a complete discussion regarding the significance of $t_\beta$). 
On the contrary, $\sba$ is a physical quantity; it needs to be sufficiently close to one (i.e., in the alignment or in the decoupling limit) so $h$ is an adequately SM-like Higgs boson, in agreement with current observations. In this limit $t_{2(\beta-\alpha)}$, and thus $\Lambda_6$, approach zero (for $\Lambda_6=0$ $h$ is exactly SM-like, with $m^2_h=\Lambda_1 v^2$, see \refEQ{eq:lambda6}).

\subsection{The Yukawa Lagrangian}\label{YukawaCouplings}

In order to have HQFV, both scalar doublets must couple to the quarks. The most general Yukawa Lagrangian in the generic scalar basis $\{\Phi_1,\Phi_2\}$ reads
\begin{equation}
-\mathcal{L}_{Q}=
\bar{Q}^0\, (Y^{\dagger}_{u1} \tilde \Phi_1 + Y^{\dagger}_{u2} \tilde \Phi_2)u_{\rm R}^0  + \bar{Q}^0\, (Y^{\dagger}_{d1} \Phi_1 + Y^{\dagger}_{d2} \Phi_2)d_{\rm R}^0  +{\rm H.c.}\,. \label{eq:yuk2d}
\end{equation}
$Y_{d1}$, $Y_{d2}$, $Y_{u1}$ and $Y_{u2}$ are completely general $3\times 3$ complex Yukawa matrices (generation indices are, again, understood and omitted). The lepton sector is assumed to be SM-like. The quark mass matrices are given by
\begin{equation} \label{mass_ch}
M_U=\frac{v}{\sqrt{2}}\parentheses{c_\beta Y^{\dagger}_{u1}+s_\beta Y^{\dagger}_{u2}},\,\qquad M_D=\frac{v}{\sqrt{2}}\parentheses{c_\beta Y^{\dagger}_{d1}+s_\beta Y^{\dagger}_{d2}}\,.
\end{equation}
We can rotate $\{\Phi_1,\Phi_2\}$ into the Higgs basis,
\begin{equation}
-\mathcal{L}_{Q}\, = \bar{Q}^0 \left[\dfrac{\sqrt{2} M_U}{v} \tilde H_1 + \xi^U \tilde H_2 \right] u_{\rm R}^0  +\bar{Q}^0 \left[\dfrac{\sqrt{2} M_D}{v} H_1 + \xi^D H_2 \right] d_{\rm R}^0 + \mathrm{H.c.} \,,
\label{eq:yuk2c}
\end{equation}
where
\begin{equation}
\xi^D\equiv \dfrac{Y_{d2}^\dagger}{c_\beta}-\dfrac{\sqrt{2} t_\beta M_D}{v},\,\qquad \xi^U\equiv \dfrac{Y_{u2}^\dagger}{c_\beta}-\dfrac{\sqrt{2} t_\beta M_U}{v}\,.
\label{eq:xiDef}
\end{equation}
Rotating the quark fields into the mass eigenstate bases $u_a$ and $d_a$ (without ``0'' superscripts), $M_Q\mapsto D_{\rm q}$ and $\xi^Q\mapsto \hat\xi^Q$; without loss of generality we may work, as in section \ref{EFT}, in a basis where $M_U$ is diagonal with real and positive elements $m_u$, $m_c$, and $m_t$. Then, the Yukawa lagrangian reads
\begin{equation}
\begin{aligned}
\label{eq:lagrangianCoupling}
-\mathcal{L}_{Q}\, &=   \bar{u}_b \left(  \CKM{bc} \, \hat\xi^D_{ca} P_R  - \hat \xi^{U*}_{cb} \, \CKM{ca} P_L \right) d_a \, H^+ \\
&+\bar{d}_b \left(  \hat\xi^{D*}_{cb} \, \CKMc{ac} \, P_L - \CKMc{cb} \, \hat \xi^U_{ca} P_R \right) u_a \, H^-\\
&+ \bigg( \bar{d}_b \bigg[ \bigg\{ \dfrac{ D_{{\rm d},ba}}{v} \sba  + \dfrac{1}{\sqrt{2}} \hat \xi^D_{ba} \cba \bigg\} P_R 
+ \bigg\{ \dfrac{ D_{{\rm d},ba}}{v} \sba  + \dfrac{1}{\sqrt{2}} \hat \xi^{D*}_{ba} \cba \bigg\} P_L \bigg] d_a
\\
&+ \bar{u}_b \bigg[  \bigg\{ \dfrac{ D_{{\rm u},ba}}{v} \sba + \dfrac{1}{\sqrt{2}}\hat \xi^U_{ba} \cba \bigg\} P_R 
+ \bigg\{ \dfrac{ D_{{\rm u},ba}}{v} \sba + \dfrac{1}{\sqrt{2}}\hat \xi^{U*}_{ba} \cba \bigg\} P_L \bigg] u_a \bigg) h
 \\
&+ \bigg( \bar{d}_b \bigg[ \bigg\{ \dfrac{ D_{{\rm d},ba}}{v} \cba - \dfrac{1}{\sqrt{2}} \hat \xi^D_{ba}\sba \bigg\} P_R  
+  \bigg\{ \dfrac{ D_{{\rm d},ba}}{v} \cba - \dfrac{1}{\sqrt{2}} \hat \xi^{D*}_{ba} \sba \bigg\} P_L  \bigg] d_a
 \\
&+ \bar{u}_b \bigg[ \bigg\{ \dfrac{ D_{{\rm u},ba}}{v} \cba - \dfrac{1}{\sqrt{2}}\hat \xi^U_{ba} \sba \bigg\} P_R
+ \bigg\{ \dfrac{ D_{{\rm u},ba}}{v} \cba - \dfrac{1}{\sqrt{2}}\hat \xi^{U*}_{ba} \sba \bigg\} P_L \bigg] u_a \bigg) H 
\\
&+ \frac{i}{\sqrt{2}} \bigg( \bar{d}_b \bigg[ \hat \xi^D_{ba} P_R - \hat \xi^{D*}_{ba} P_L \bigg] d_a + \bar{u}_b \bigg[ - \hat \xi^U_{ba} P_R  + \hat \xi^{U*}_{ba} P_L  \bigg] u_a \bigg) A\,,
\end{aligned}
\end{equation}
where $a,b=1,2,3$. The correspondence with the notation in Ref.~\cite{Crivellin:2013wna}, for a generic Yukawa coupling,
\begin{equation}
\begin{aligned} \label{effec_Y}
\bar q_b\,g_{\overline{q}_b q_a \phi}\,q_a\,\phi & \equiv \bar q_b\left( \Gamma_{q_b q_a }^{LR\, \phi} P_R + \Gamma_{q_b q_a }^{RL\, \phi} P_L \right)q_a\,\phi\,,
\end{aligned}
\end{equation}
where $P_{R,L}=(1\pm\gamma_5)/2$, is provided in Tab.~\ref{table:couplings_notation}. Due to Hermiticity of the Lagrangian, $\Gamma^{LR\phi \;*}_{q_a q_b} = \Gamma^{RL\phi}_{q_b q_a}$.

\begin{table*}\centering
\ra{1.3}
\begin{tabular}{@{}ccccc@{}}\toprule \midrule
  & & $\Gamma_{q_b q_a }^{LR \phi}$ & &  $\Gamma_{q_b q_a }^{RL 
  \phi}$ \\ 
 \cmidrule{1-2} \cmidrule{3-5}
 $\bar{q}_b q_a h $ & & $\dfrac{ D_{{\rm q},ba}}{v} \sba  + \dfrac{1}{\sqrt{2}} \hat \xi^Q_{ba} \cba$ & &  $\dfrac{ D_{{\rm q},ba}}{v} \sba  + \dfrac{1}{\sqrt{2}} (\hat \xi^Q_{ba})^* \cba$ \\ 
 $\bar{q}_b q_a H $ & & $\dfrac{ D_{{\rm q},ba}}{v} \cba - \dfrac{1}{\sqrt{2}} \hat \xi^Q_{ba}\sba $ & &  $\dfrac{ D_{{\rm q},ba}}{v} \cba - \dfrac{1}{\sqrt{2}} (\hat \xi^Q_{ba})^*\sba $ \\ 
 $\bar{q}_b q_a A $ & & $\frac{i}{\sqrt{2}} \hat\xi^Q_{ba} $ & &  $-\frac{i}{\sqrt{2}} (\hat\xi^Q_{ba})^* $ \\ 
  $\bar{u}_b d_a H^+$ & &  $\CKM{bc} \, \hat\xi^D_{ca}$ & &  $ -(\hat \xi^U_{cb})^* \, \CKM{ca} $ \\ 
  \midrule \bottomrule
\end{tabular}.
\caption{Scalar-quark-quark couplings extracted from Eq.~\eqref{eq:lagrangianCoupling} using the convention of Eq.~\eqref{effec_Y}, where $Q=D,U$.}
\label{table:couplings_notation}
\end{table*}
In this work we are interested in HQFV involving a third family quark. Therefore we will only consider the flavor-violating (complex) couplings in $\hat\xi^{U,\,D}$ between the third and the second families, and in addition for simplicity we set the diagonal coupling of the second generation to zero, that is, 
\begin{equation}
\hat\xi^{U}= \left( \begin{array}{ccc}
0 & 0 & 0 \\
0 & 0 & \hat\xi^U_{23}\\
0 & \hat\xi^U_{32} & \hat\xi^U_{33}
\end{array} \right)\,,\qquad \qquad
\hat\xi^{D}= \left( \begin{array}{ccc}
0 & 0 & 0 \\
0 & 0 & \hat\xi^D_{23}\\
0 & \hat\xi^D_{32} & \hat\xi^D_{33}
\end{array} \right)\,.
\label{eq:yukawa_restriction}
\end{equation}
The only a priori requirement placed on the entries of $\hat\xi^U$ and $\hat\xi^D$ is that they respect perturbativity, i.e. they are smaller than $4\pi$.

\subsection{Constraints on Quartic Couplings}
\label{sec:stability}
Since the Hamiltonian has to be bounded from below, the quartic part of the scalar potential in \refEQ{eq:scalar_potential} is required to be positive for all values of the fields and all scales. Furthermore, the considered vacuum should be the global minimum of the potential \cite{Ivanov:2015nea} (one could weaken the requirement and include a sufficiently long-lived metastable local minimum). The quartic couplings are also required to be perturbative, i.e. smaller than $4\pi$. We also require that the scattering of the different scalars at high energies, controlled by the quartic part of the potential, respects perturbative unitarity: in particular, that the eigenvalues of the tree level $2\to 2$ scattering matrix do not yield probabilities larger than 1 (see e.g. \cite{Huffel:1980sk,Ginzburg:2005dt,Kanemura:2015ska}, one loop corrections in a restricted 2HDM have been addressed in Ref.~\cite{Grinstein:2015rtl}).

\subsection{Oblique Parameters}\label{sec:oblique}
We include the so-called ``oblique parameters'' S, T and U \cite{Altarelli:1990zd,Peskin:1991sw}, which parametrise radiative corrections to electroweak gauge boson propagators. For the theoretical expressions, see Refs.~\cite{Grimus:2008nb,Haber:2010bw}; we use the experimental values \cite{Baak:2014ora}

\begin{equation}
	\begin{aligned}
		S = 0.05 \pm 0.11, \,\,\, T = 0.09 \pm 0.13, \,\,\,  U = 0.01 \pm 0.11,
	\end{aligned}
\end{equation}
with a correlation matrix 
\begin{equation}
\Sigma = \left( \begin{array}{ccc}
1.0 & 0.9 & -0.59 \\
0.9 & 1.0 & -0.83\\
-0.59 & -0.83 & 1.0
\end{array} \right)\,.
\end{equation}

\subsection{Higgs Signal Strengths}\label{sec:higgsdecays}
A necessary ingredient in the scalar sector is, of course, a neutral scalar with properties in agreement with the 125 GeV SM-like Higgs discovered at the LHC \cite{Aad:2012tfa,Chatrchyan:2012xdj}. We identify it with $h$, and thus the first requirement is $m_h = (125.09 \pm 0.32)$ GeV \cite{Aad:2015zhl}. The width is also required to satisfy $\Gamma_h<17$ MeV following the result at $2\sigma$ presented in Ref.~\cite{Khachatryan:2016ctc}. The most relevant information for the phenomenological aspects of the 125 GeV scalar is the set of signal strengths $\mu_{XY}$ for combined production (Y) and decay (X) channels,
\begin{equation}
\mu_{XY}=\dfrac{\sigma([pp]_Y\to h)_{\rm 2HDM}\,\BR(h\to X)_{\rm 2HDM}}{\sigma([pp]_{Y}\to h)_{\rm SM}\,\BR(h\to X)_{\rm SM}}\,,
\end{equation}
which are factorized in ``production $\times$ decay'' model dependent factors
\begin{equation}
\mu_{XY}=\kappa_Y^{P}\kappa_X^{BR},\qquad \kappa_Y^P=\dfrac{\sigma([pp]_Y\to h)_{\rm 2HDM}}{\sigma([pp]_{Y}\to h)_{\rm SM}},\quad \kappa_X^{BR}=\dfrac{\BR(h\to X)_{\rm 2HDM}}{\BR(h\to X)_{\rm SM}}\,.
\end{equation}
The relevant production modes are gluon-gluon fusion (ggF), vector boson fusion (VBF), Higgs-strahlung ($Wh$, $Zh$) and associated production with top quarks ($tth$); the corresponding factors are
\begin{equation}\label{eq:Higgs:prod}
\begin{aligned}
	&\kappa^{P}_{ggF} = \frac{\Gamma(h\to gg)_{\rm 2HDM}}{\Gamma(h\to gg)_{\rm SM}}\, ,\\
	&\kappa^{P}_{tth} = \frac{v^2}{2m_t^2} \left(|\Gamma_{tt}^{LRh}|^2 + |\Gamma_{tt}^{RLh}|^2\right)\, , \\
	&\kappa^{P}_{VBF} = \kappa^{P}_{WH} = \kappa^{P}_{ZH} = \sba^2\, .
\end{aligned}
\end{equation}
The corresponding factors for the relevant decay channels are
\begin{equation}\label{eq:Higgs:decay}
\begin{aligned}
	&\kappa^{BR}_{\gamma\gamma} = \frac{\Gamma(h\to \gamma\gamma)_{\rm 2HDM}}{\Gamma(h\to \gamma\gamma)_{\rm SM}}\, , \\
	&\kappa^{BR}_{bb} = \frac{v^2}{2m_b^2} \left(|\Gamma_{bb}^{LRh}|^2 + |\Gamma_{bb}^{RLh}|^2\right)\, , \\
	&\kappa^{BR}_{\tau\tau} = \frac{v^2}{2m_\tau^2} \left(|\Gamma_{\tau\tau}^{LRh}|^2 + |\Gamma_{\tau\tau}^{RLh}|^2\right)\, , \\
	&\kappa^{BR}_{WW} = \kappa^{BR}_{ZZ} = \sba^2\, .
\end{aligned}
\end{equation}
Both $\kappa^{P}_{ggF}$ and $\kappa^{BR}_{\gamma\gamma}$ arise from one loop amplitudes: the expressions can be found, for example, in Ref.~\cite{Gunion:1989we}. For $h\to\bar\tau\tau$, since we assume for simplicity SM-like Yukawa couplings in the lepton sector, $\kappa^{BR}_{\tau\tau}=\sba^2$ (the experimental uncertainties in that decay channel are, in any case, large).

The experimental results (values and uncertainties) from the combined ATLAS and CMS analyses of LHC Run I data \cite{Khachatryan:2016vau} are given in the following matrix:
\begin{equation}\label{eq:Higgsmu:RunI}
\mu_{XY}=\left( \begin{array}{ccccc}
1.1^{+0.23}_{-0.22} & 1.3^{+0.5}_{-0.5} & 0.5^{+1.3}_{-1.2} & 0.5^{+3.0}_{-2.5} & 2.2^{+1.6}_{-1.3} \\
1.13^{+0.23}_{-0.22} & 0.1^{+0.5}_{-0.5} & \times & \times & \times \\
0.84^{+0.17}_{-0.17} & 1.2^{+0.4}_{-0.4} & 1.6^{+1.2}_{-1.0} & 5.9^{+2.6}_{-2.2} & 5.0^{+1.8}_{-1.7} \\
1.0^{+0.6}_{-0.6} & 1.3^{+0.4}_{-0.4} & -1.4^{+1.4}_{-1.4} & 2.2^{+2.2}_{-1.8} & -1.9^{+3.7}_{-3.3} \\
\times & \times & 1.0^{+0.5}_{-0.5} & 0.4^{+0.4}_{-0.4} & 1.1^{+1.0}_{-1.0} \\
\end{array} \right).
\end{equation}
The ordering for decay channels (rows) is $\{\gamma\gamma,ZZ,WW,\tau\tau,bb\}$ and for production mechanisms (columns) $\{$ggF, VBF, $Wh$, $Zh$, $tth\}$. For the missing entries ``$\times$'' there is no measurement available in Ref.~\cite{Khachatryan:2016vau}. In addition to \refEQ{eq:Higgsmu:RunI}, we also include CMS and ATLAS data from LHC Run II on $h\to\bar bb$ and $h\to\bar\tau\tau$ in the analysis of section \ref{sec:numerical_analysis}: for $h\to\bar bb$, we consider CMS \cite{Sirunyan:2017elk} and ATLAS \cite{Aaboud:2017xsd} results for VBF production while for $h\to\bar\tau\tau$ we combine ggF and VBF production following Ref.~\cite{Sirunyan:2017khh}.

We also include a likelihood from CMS/ATLAS Run II results on $h \rightarrow \bar{b}b$ and $h \rightarrow \tau \tau$.
For $\bar{b}b$ we sum likelihoods from CMS \cite{Sirunyan:2017elk} and ATLAS \cite{Aaboud:2017xsd} results for the VBF production channel. For $\bar\tau \tau$ we combine the VBF and ggF production pathways and use only the CMS data~\cite{Sirunyan:2017khh}. Notice that the analysis of Higgs signal strengths only requires the 2HDM vs. SM modifying factors in Eqs.~\eqref{eq:Higgs:prod}--\eqref{eq:Higgs:decay}.

\subsubsection{Flavor-Changing Higgs Processes}\label{HiggsFlavorChanging}
In \refEQ{eq:lagrangianCoupling}, $\mathcal{L}_{Q}$ includes flavor-changing couplings of $h$ to $\bar bs$, $\bar sb$, $\bar tc$ and $\bar ct$ controlled by the off-diagonal entries of $\hat\xi^U$ and $\hat\xi^D$ in \refEQ{eq:yukawa_restriction}. 
The $h\to bs$ decay width at tree-level, $\Gamma(h\to bs)\equiv$ $\Gamma(h\to \bar{b}s) + \Gamma(h\to \bar{s}b)$, is
\begin{equation}\label{eq:hbs}
	\begin{aligned}
		\Gamma(h\to bs) \simeq \frac{3 m_h \cba^2}{16\pi} \left( |\hat\xi^D_{23}|^2 + |\hat\xi^D_{32}|^2 \right)\,,
	\end{aligned}
\end{equation}
where we have neglected final state masses. The $t\to ch$ decay width at tree level reads
\begin{equation}
	\Gamma(t\to ch) \simeq \frac{m_t \cba^2}{32\pi}\, |\hat\xi^U_{32}|^2 \, \left(1 - \frac{m^2_h}{m^2_t}\right)^2\,,
\end{equation}
where we have neglected the charm mass. For the conjugate process $\bar t\to h\bar c$, $\hat\xi^U_{32}\mapsto \hat\xi^U_{23}$.\\
In the analysis of section \ref{sec:numerical_analysis}, scalar decays are carried out using the inbuilt routines offered by 2HDMC \cite{Eriksson:2009ws}. The 2HDMC code does not support flavor-changing processes officially but the program is designed thoughtfully to allow for these processes. Nevertheless, some slight modifications had to be made, including promoting the Yukawa entries from real to complex. Furthermore, beyond \refEQ{eq:hbs}, $h\to bs$ receives QCD corrections at NLO that may increase the rate by $10-20\%$ \cite{Crivellin:2017upt}. The 2HDMC includes QCD corrections for this process, and they are turned on in the analysis of section \ref{sec:numerical_analysis}.

\section{Flavor Constraints\label{sec:ConstraintsFlavor}}

In order to study HQFV in the general 2HDM, flavor constraints have to be included. We discuss the most relevant ones in the following.

In the down quark sector we focus on the process $h\to bs$: in this case, the most stringent constraints come from the $|\Delta B|=2$ process of $B_s^0$--$\bar B_s^0$ mixing and from the $|\Delta B|=1$ radiative decay process $B\to X_s\gamma$. Since in the SM all flavor-changing processes are induced by W boson exchange, both processes occur at the one loop level. Their GIM and loop suppressions make them highly sensitive to the presence of new physics contributions: in the general 2HDM these new contributions appear at tree level in $B_s^0$--$\bar B_s^0$ and at one loop in $B\to X_s\gamma$. They are discussed in the following subsections.
We do not consider other processes involving final state leptons like, e.g,  $B_s\to\mu^+\mu^-$: since we focus on the quark sector, assuming SM-like tree level couplings of scalars to leptons highly suppresses new contributions to these processes.\\
Concerning HQFV in the up quark sector, as already mentioned, we focus on $t\to ch$: we incorporate existing bounds at the $10^{-3}$ level on $\BR(t\to ch)$ (see \refEQ{eq:topFlChDecays:Exp1}). One could also consider constraints arising from $D^0$--$\bar D^0$ mixing. However, with the Yukawa couplings considered in \refEQ{eq:yukawa_restriction}, the contribution to $D^0$--$\bar D^0$ mixing involving $\hat\xi^U$ vanishes.\footnote{There are one loop contributions mediated by the charged scalar which depend on $\hat\xi^D$, but they are irrelevant once the constraints from the down quark sector discussed above are considered. Notice, in any case, that while $B_s^0$--$\bar B_s^0$ transitions are dominated by short-distance physics (e.g. the contributions mediated by the top quark), long-distance effects (i.e. intermediate hadronic states) are quite likely dominating in $D^0$--$\bar D^0$ and only a rough constraint on the size of the short-distance scalar mediated contributions could have been considered.} We do not consider constraints from $t\to cg,c\gamma$ processes since  in this scenario they only arise at one loop while existing bounds on the corresponding branching ratios are similar to the ones for $t\to ch$, which arise instead at tree level.

\begin{table*}\centering
\ra{1.3}
\begin{tabular}{@{}llll@{}}\toprule\midrule
Observable & & Value   \\
\midrule
$\Delta M_{B_s, \, \rm obs}$  & & $(1.1688 \pm 0.0014) \times 10^{-11} \, \mbox{GeV}$ \cite{Patrignani:2016xqp}\\
$\Delta M_{B_s, \, \rm SM}$ & & $(1.32 \pm 0.08_{\rm th.}) \times 10^{-11} \, \mbox{GeV}$\\ 
$\beta_{s,\, \rm obs}$  & & $ (1.5 \pm 1.6) \times 10^{-2} \, \mbox{rad}$ \cite{Patrignani:2016xqp}\\
$\beta_{s,\, \rm SM}$ & & $(1.82 \pm 0.11_{\rm th.}) \times 10^{-2} \, \mbox{rad}$\\ 
$\mbox{BR}(B \rightarrow X_s \gamma)_{\rm obs}$  & & $(3.32 \pm 0.16) \times 10^{-4}$ \cite{Amhis:2016xyh}\\
$\mbox{BR}(B \rightarrow X_s \gamma)_{\rm SM}$ & & $(3.34 \pm 0.33_{\rm th.}) \times 10^{-4}$\\ 
 \midrule
  \bottomrule
\end{tabular} 
\caption{Experimental and SM predictions for $B_s$-meson mixing observables (mass splitting and the CP-violating angle) and radiative B-meson decays. The subscript \emph{th.} in some errors refers to theoretical.}
\label{table:FlavPhysicsValues}
\end{table*}

\subsection{Effective Operators}

We use an Effective Field Theory (EFT) approach to compute flavor constraints. An effective Hamiltonian is defined as
\begin{equation}
\Heff = (\text{PF}) \sum_i C_i(\mu) \; O_i(\mu),
\label{eq:H_eff}
\end{equation} 
where $\mu$ is the energy scale at which the matrix elements of the Hamiltonian are evaluated, $C_i(\mu)$ are the Wilson coefficients which encode the information of the underlying theory and $O_i(\mu)$ are the operators which mediate the process. For simplicity it is common to include powers of the weak coupling or CKM factors in the prefactor $(\text{PF})$: for example, for $b\to s$ transitions, it is common to set $(\text{PF})=-\frac{4G_F}{\sqrt 2}\CKMc{ts}\CKM{tb}$. 

The underlying theory and the EFT are typically matched at an energy scale $\mu\sim m_W$: the evolution (``running'') of the Wilson coefficients from the matching scale down to the $B$ meson scale $\mu_B\sim 4.2$ GeV is given by
\begin{equation}
\label{eq:RGE_Wilson}
	\frac{d}{d\ln \mu} C_i(\mu) = \gamma_{ji} C_j(\mu),
\end{equation}
where $\gamma_{ij}$ is the anomalous dimension matrix (ADM). The solution of this Renormalization Group Evolution equation, in vector notation, is given by
\begin{equation}
	\vec{C}(\mu) = \hat{U}(\mu,\mu_0) \vec{C}(\mu_0),
\end{equation}
where the evolution operator matrix $\hat{U}(\mu,\mu_0)$ is computed in terms of $\gamma_{ji}$ \cite{Buchalla:1995jx} and can be found using the publicly available \rm{Mathematica} code DSixTools \cite{Celis:2017cm}, see also Ref.~\cite{Aebischer:2017tk}.

\subsection{$B^0_s$--$\bar B^0_s$ Meson Mixing}
In neutral meson systems $M^0$ -- $\bar M^0$, $M^0\leftrightarrows\bar M^0$ transitions (or ``oscillations'') show that $M^0$ and $\bar M^0$ are not evolution eigenstates; the evolution eigenstates have slightly different mass and width. In the $B^0_s$--$\bar B^0_s$ system, the physical mass splitting $\DMBs$ is dominated by short-distance physics and can be computed perturbatively in terms of the appropriate effective hamiltonian 
\begin{equation}
M_{12}^{B_s}=\frac{\bra{B_s^0} \HEFF{|\Delta B|=2}\ket{\bar B_s^0}}{2\MBs},\qquad \DMBs=2|M_{12}^{B_s}|\,.
\end{equation}
The CP violating mixing phase is\footnote{Notice that the mixing phase is not rephasing invariant and thus it is only its combination with decay amplitudes which has a rephasing invariant physical meaning; as is usual, we refer nevertheless to the ``mixing  phase'' $2\beta_s$ since, in the adopted CKM phase convention, one has real decay amplitudes in transitions like the ``golden'' mode $B_s\to J/\Psi\,\phi$.}
\begin{equation}
2\beta_s=-\arg\left(\bra{B_s^0} \HEFF{|\Delta B|=2}\ket{\bar B_s^0}\right)\,.
\end{equation}
In the EFT description of $B_s^0$--$\bar B_s^0$ mixing, we adopt the usual operator basis:
\begin{equation}
\begin{aligned}
\label{eq:MM_basis}
	O_1 &= (\bar{s}_{\alpha} \gamma^\mu P_L b_{\alpha} )(\bar{s}_{\beta} \gamma^\mu P_L b_{\beta} ),  \\
	O_2 &= (\bar{s}_{\alpha}  P_L b_{\alpha} )(\bar{s}_{\beta} P_L b_{\beta} ),  \\
	O_3 &= (\bar{s}_{\alpha}  P_L b_{\beta} )(\bar{s}_{\beta} P_L b_{\alpha} ),  \\ 
	O_4 &= (\bar{s}_{\alpha}  P_L b_{\alpha} )(\bar{s}_{\beta} P_R b_{\beta} ),  \\ 
	O_5 &= (\bar{s}_{\alpha}  P_L b_{\beta} )(\bar{s}_{\beta} P_R b_{\alpha} ),  \\ 
\end{aligned}
\end{equation}
where we have explicitly denoted the color indices $\alpha$ and $\beta$. Exchanging $P_L \leftrightarrows P_R$ in $O_{1,2,3}$ one obtains the (additional) primed operators $O_{1,2,3}^{'}$ ($O_{4,5}$ do not give new operators under $P_L \leftrightarrows P_R$). The full Hamiltonian describing $B_s^0$--$\bar B_s^0$ is 
\begin{equation}
\label{eq:MM_effH}
\HEFF{|\Delta B|=2} =  \sum^5_{i=1} C_i(\mu) O_i(\mu) + \sum^3_{i=1} C_i'(\mu) O_i'(\mu).
\end{equation} 
Since, as discussed below, $W$ mediated contributions only affect $C_1$, while the new scalar contributions affect $C_2$, $C_2^{'}$ and $C_4$, we do not factor out the usual $G_F$ and $(\CKMc{ts}\CKM{tb})^2$ in \refEQ{eq:MM_effH}. For the $B_s$ system,
\begin{equation}
\bra {B_s^0}\HEFF{|\Delta B|=2} \ket{\bar B_s^0} =  \sum^5_{i=1} C_i(\mu) \; \bra {B_s^0}O_i(\mu) \ket{\bar B_s^0}+ \sum^3_{i=1} C_i'(\mu) \; \bra {B_s^0}O_i'(\mu) \ket{\bar B_s^0},
\end{equation}
the matrix elements of the operators in \refEQ{eq:MM_basis} are
\begin{equation}
\begin{aligned}
 \label{params}
\langle  B_s^0 \vert O_1(\mu) \vert   \bar B_s^0 \rangle  &= b_1 \, M_{B_s}^2  f_{B_s}^2 B^{B_s}_1(\mu),  \\
\langle B_s^0 \vert O_i(\mu) \vert  \bar B_s^0 \rangle  &= b_i \, \chi_{B_s}\,M_{B_s}^2  f_{B_s}^2 B^{B_s}_i(\mu),\quad i=2,3,4,5,\\
\end{aligned}
\end{equation}
\begin{equation}
	\vec b = \{8/3,\;-5/3,\;1/3,\;2,\;2/3\}\,, \qquad 
\chi_{B_s} (\mu) = \frac{M_{B_s}^2}{ (m_{b}(\mu) + m_{s}(\mu) )^2}\,.
\end{equation}
Non-perturbative QCD effects \cite{Becirevic:2001yv} are encoded in the bag factors $B_i^{B_s}$ (the vacuum insertion approximation corresponds to $B_i^{B_s}\to 1$); they are given in Tab.~\ref{table:meson_mass_decay_B_values} in appendix \ref{app:scan_parameters}, together with the decay constant $f_{B_s}$ and the meson mass $M_{B_s}$. The primed operators of Appendix~\ref{ap:evolution} have the same matrix elements as the unprimed ones (from parity invariance of QCD).

\subsubsection{Standard Model Contribution}
\begin{figure}[htb]
\begin{center}
\subfigure[SM box.\label{FIG:BsMixing:SMbox}]{\includegraphics[width=0.41\textwidth]{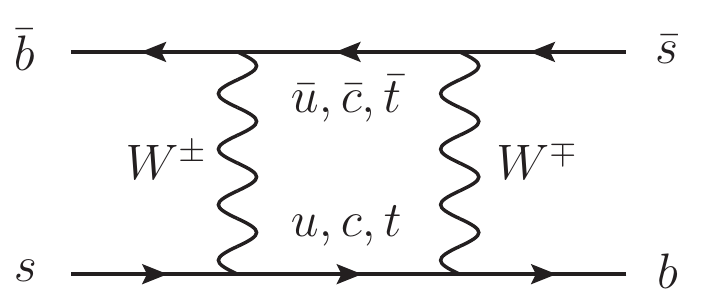}}\qquad\qquad
\subfigure[2HDM tree level.\label{FIG:BsMixing:2HDMtree}]{\includegraphics[width=0.35\textwidth]{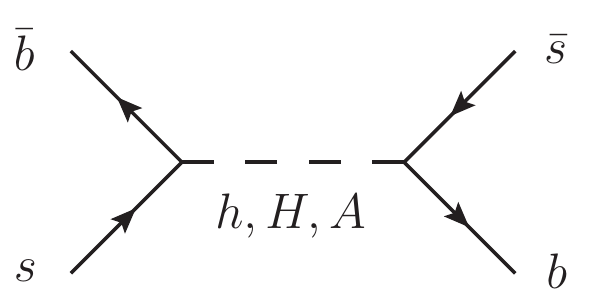}}\hfill\\
\subfigure[Mixed box.\label{FIG:BsMixing:2HDMmixedbox}]{\includegraphics[width=0.4\textwidth]{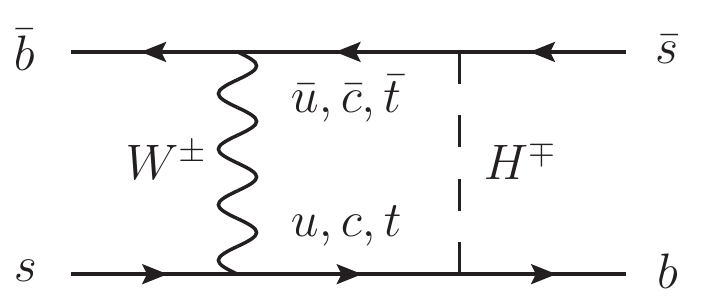}}\qquad
\subfigure[2HDM box.\label{FIG:BsMixing:2HDMbox}]{\includegraphics[width=0.4\textwidth]{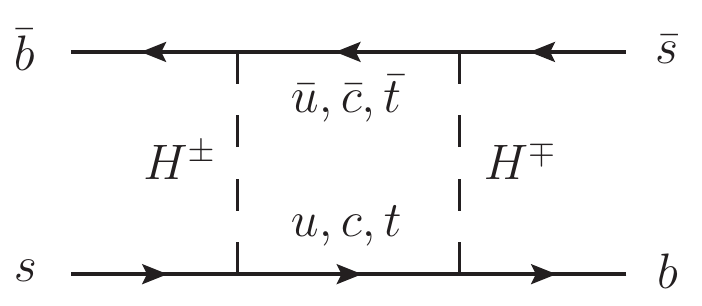}}
\end{center}
\caption{Contributions to $B_s^0$--$\bar B_s^0$ mixing.\label{FIG:BsMixing}}
\end{figure}

As anticipated, in the SM there are only contributions to the $O_1$ operator. The dominant contribution to $C_1$ (see Fig.~\ref{FIG:BsMixing:SMbox}) is
\begin{equation}
	C^{\rm SM}_1 (\mu_B) = \frac{G^2_F}{4\pi^2} \bigl(\CKMc{ts}\CKM{tb}\bigr)^2 m^2_W \hat\eta_{B}(\mu_B) S(x_t)\,,\quad x_t\equiv m_t^2/m_W^2\,,
\label{eq:meson_C_SM}
\end{equation}
with $S(x)$ the well-known Inami-Lim function \cite{doi:10.1143/PTP.65.297}. The RGE for $\Delta F=2$ is given in App.~\ref{ap:evolution}: one can read the evolution of $C_1$ from the matching scale $\mu_W\sim m_W$ down to $\mu_B\sim m_B$, given by $\hat\eta_B(\mu_B)=0.862$. Then,
\begin{equation}
 \bra {B_s^0}\HEFF{|\Delta B|=2} \ket{\bar B_s^0} = \frac{G^2_Fm^2_W}{6\pi^2}M_{B_s}^2 f^2_{B_s}\hat\eta_B B_1^{B_s} \bigl( \CKMc{ts}\CKM{tb}\bigr)^2 S(x_t)  \,.
\end{equation}
Individually, the evolution factor $\hat\eta_B$ and bag-factor $B^{B_s}_1$ are both scale dependent, but the combination of $\hat\eta_B B^{B_s}_1$ is a scale as well as scheme-independent quantity.
Numerically
\begin{equation}
\bra {B^0_s}\HEFF{|\Delta B|=2} \ket{\bar B_s^0}_{\rm SM} = (7.28 - 0.26i) \times 10^{-11} \; \mbox{GeV},
\end{equation}
which gives
\begin{equation}
	\Delta M_{B_s, \, \rm SM} = (1.36\pm 0.08) \times 10^{-11}  \; \mbox{GeV} = (20.64 \pm 1.28)\, \mbox{ps}^{-1}\,.
	\label{eq:mass_splitting_a}
\end{equation}
The theoretical error we choose is based upon the combination of QCD errors as laid out in Tab. II of Ref.~\cite{DiLuzio:2017fdq}, where a theoretical error of $6.2\%$ is stated. Using $\hat \eta_B = 0.839$ (see Refs.~\cite{Buras:1990fn,Lenz:2006hd}), $\Delta M_{B_s, \, \rm SM}$ agrees with Ref.~\cite{DiLuzio:2017fdq}. We have updated our final scan and predictions with this improved quantity, which gives $\Delta M_{B_s, \, \rm SM} = 1.32 \times 10^{-11}\, \rm GeV$. The SM final value in Eq.~\eqref{eq:mass_splitting_a} is larger than the observed one, specifically, its error translates into a $1.8 \sigma$ discrepancy with the SM, as alluded to in Ref.~\cite{DiLuzio:2017fdq}. The $B^0_s$--$\bar B^0_s$ mixing phase reads
\begin{equation}
	\beta_{s,\, \rm SM} = (1.82 \pm 0.11) \times 10^{-2} \; \mbox{rad}\,.
\end{equation}
In Tab.~\ref{table:FlavPhysicsValues} we summarize the values observed and computed for the SM. 

\subsubsection{Two-Higgs-Double Model Contributions}

At tree level, the $B^0_s$--$\bar B^0_s$ mixing process is mediated by neutral scalars $h$, $H$ and $A$, as shown in Fig.~\ref{FIG:BsMixing:2HDMtree}. The contributions to the Wilson Coefficients are \cite{Crivellin:2013wna}
\begin{equation}
\begin{aligned}
	C_2 (\mu_W) &= \sum^3_{k = 1} -\frac{1}{2m^2_{\phi^0_k}} \biggl( \Gamma^{LR,{\phi^0_k}\star}_{3 2} \biggr)^2, \\
	C_2' (\mu_W) &= \sum^3_{k = 1} -\frac{1}{2m^2_{\phi^0_k}} \biggl( \Gamma^{LR,{\phi^0_k}}_{2 3} \biggr)^2, \\
	C_4 (\mu_W) &= \sum^3_{k = 1} -\frac{1}{m^2_{\phi^0_k}} \Gamma^{LR,{\phi^0_k}}_{2 3} \Gamma^{LR,{\phi^0_k}\star}_{3 2}.  \\
\end{aligned}
\label{eq:mm_tree_level}
\end{equation}
where $\phi^0_k = (h,H,A)$. 
Beyond tree level, there are contributions from neutral and charged scalar particles from box diagrams as shown in Figs.~\ref{FIG:BsMixing:2HDMmixedbox} and~\ref{FIG:BsMixing:2HDMbox}; for the corresponding expressions we refer to Ref.~\cite{Crivellin:2013wna}.

\subsubsection{Explaining the Discrepancy within the Two-Higgs-Doublet Model} \label{sec:disc}
Before addressing the full numerical analysis of section \ref{sec:numerical_analysis}, it is interesting to study the parameter space in the 2HDM model that can explain the $1.8\sigma$ deviation between the observed value of $\DMBs$ and the SM prediction, see Tab.~\ref{table:FlavPhysicsValues}. Notice that the 2HDM contribution can partially cancel the SM contribution, and therefore yield a better agreement with the lower observed value. For degenerate $H$ and $A$ as expected from EWPT, the tree level contributions to the Wilson coefficients in Eq.~\eqref{eq:mm_tree_level} give
\begin{align} \label{2DHM_meson}
	\Delta M_{B_s, \, \rm 2HDM}\simeq & -A_B \biggl[\cba^2\,\biggl( \frac{1}{m^2_h} -\frac{1}{m^2_H}\biggr)+\frac{2}{m^2_H}\biggr] \nonumber\\
	&\times \biggl\{(U_{22}B_2^{B_s}\,b_2 + U_{32}B_3^{B_s}\,b_3)\,\biggl[ (\hat \xi^{D*}_{32})^2 + (\hat \xi^{D}_{23})^2 \biggr]
	+ 2\, (U_{44}B_4^{B_s}b_4) \, \hat \xi^{D*}_{32}\hat \xi^{D}_{23}\biggr\} \,,
\end{align}
where we have defined $A_B \equiv f_{B_s}^2\,M_{B_s}^3/(4\,(m_{b} + m_{s})^2) \simeq 0.105\, {\rm GeV}^3$, and $U_{ij}$ are elements of the evolution matrix in Appendix~\ref{ap:evolution}.

We plot in Fig.~\ref{FIG:2HDM:MesonMixNum} $\DMBs$, including the 2HDM contribution both at tree and loop level, versus $|\hat\xi^{32}_D|$ for different values of $\hat\xi^{23}_D$. In the top plots, we fix $m_H=m_A=200$ GeV and $\hat\xi^D_{23} = (\pm 1 \pm i) \times 10^{-4}$. Under this setup we can fit the experimental observation for the intervals $|\hat\xi^D_{32}| \sim [2\times 10^{-4}, 5\times 10^{-3}]$ for both $\sba = 0.9$ (left plot) and $\sba = 0.99$ (right plot). The total allowed interval is discontinuous and a second region as large as $|\hat\xi^D_{32}| \sim 3.5\times 10^{-2}$ is allowed for $\sba = 0.99$. We can see that $|\hat\xi^D_{32}| \sim 3.6\times 10^{-2}$ is the largest Yukawa we expect for $\sba \leq 0.99$. The bottom plots show $m_H=m_A=2000$ GeV at a constant Yukawa of $\hat\xi^D_{23} = (\pm 1 \pm i) \times 10^{-3}$. For $\sba=0.99$ (right plot) we can attain a Yukawa as large as $|\hat\xi^D_{32}| \sim 1.6\times 10^{-2}$. We have also checked that in these regions the 2HDM is able to satisfy the observed value of the $B^0_s$--$\bar B^0_s$ mixing phase.

\begin{figure}[htb!]
\begin{center}
\subfigure[$\sba=0.9$, $m_H=m_A=200$ GeV,\qquad $\hat\xi^D_{23}=(\pm 1\pm i)\times 10^{-4}$.\label{FIG:2HDM:MesonMixNum:1a1}]{\includegraphics[width=0.45\textwidth]{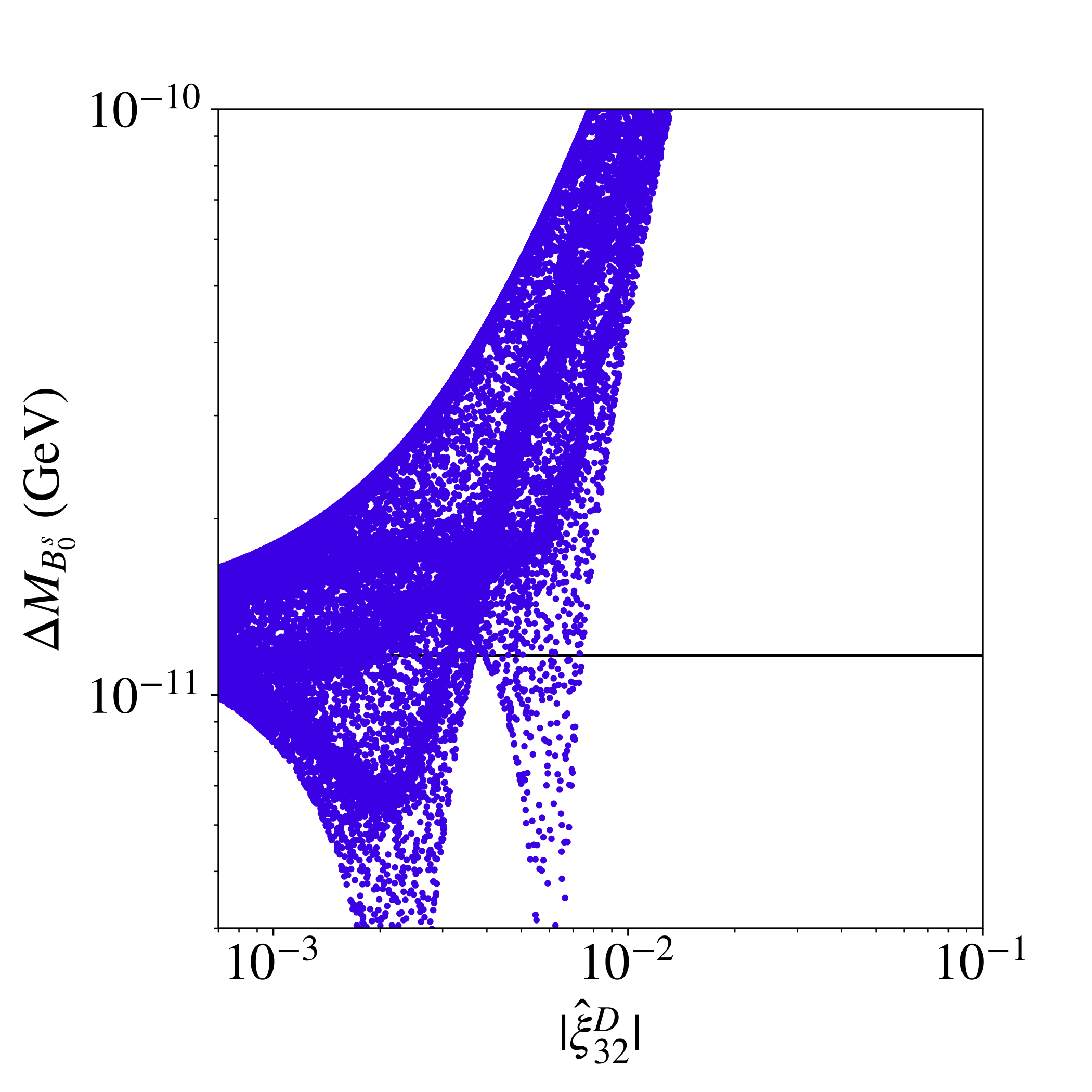}}\qquad
\subfigure[$\sba=0.99$, $m_H=m_A=200$ GeV,\qquad $\hat\xi^D_{23}=(\pm 1\pm i)\times 10^{-4}$.\label{FIG:2HDM:MesonMixNum:1b1}]{\includegraphics[width=0.45\textwidth]{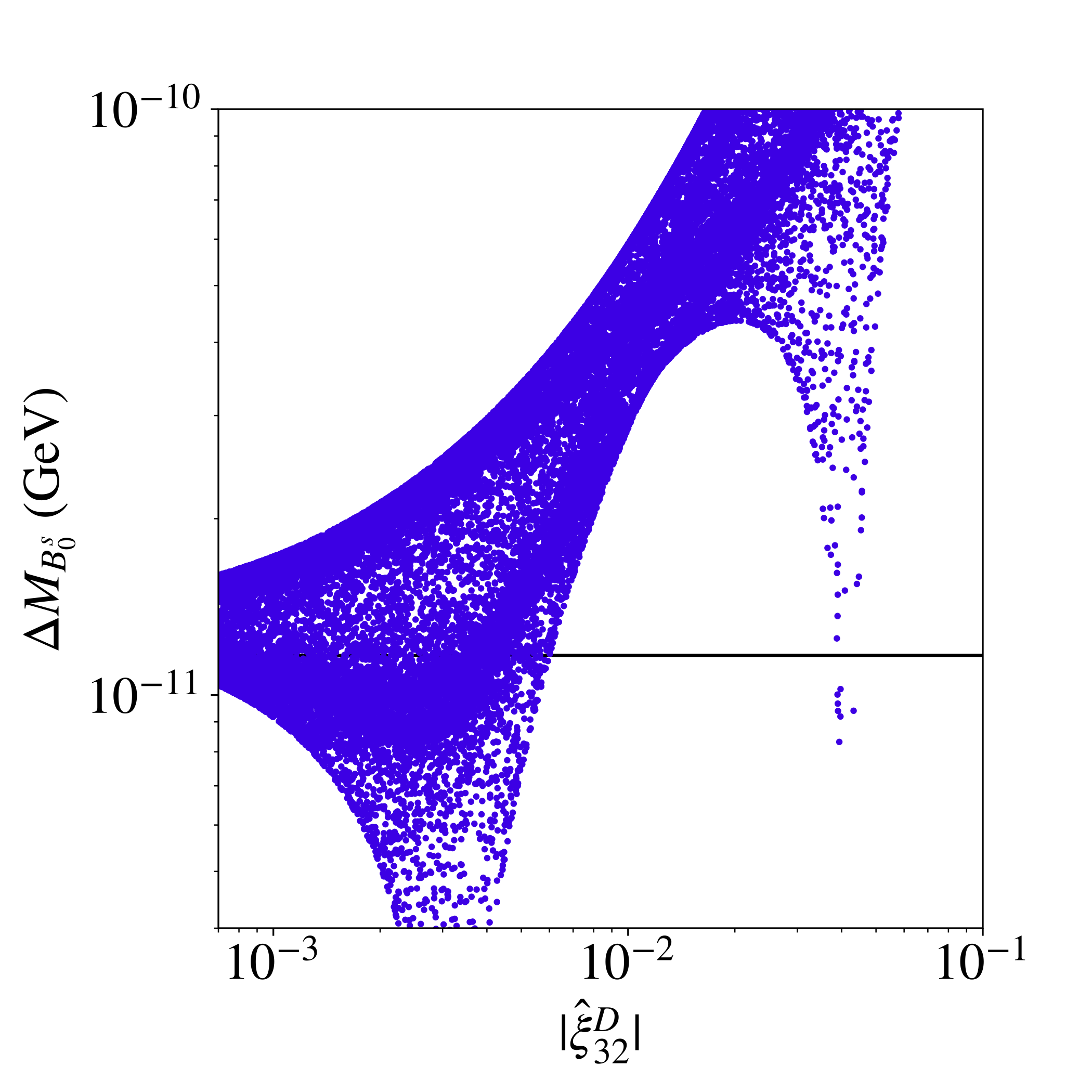}}\\
\subfigure[$\sba=0.9$, $m_H=m_A=2000$ GeV,\qquad $\hat\xi^D_{23}=(\pm 1\pm i)\times 10^{-3}$.\label{FIG:2HDM:MesonMixNum:2a1}]{\includegraphics[width=0.45\textwidth]{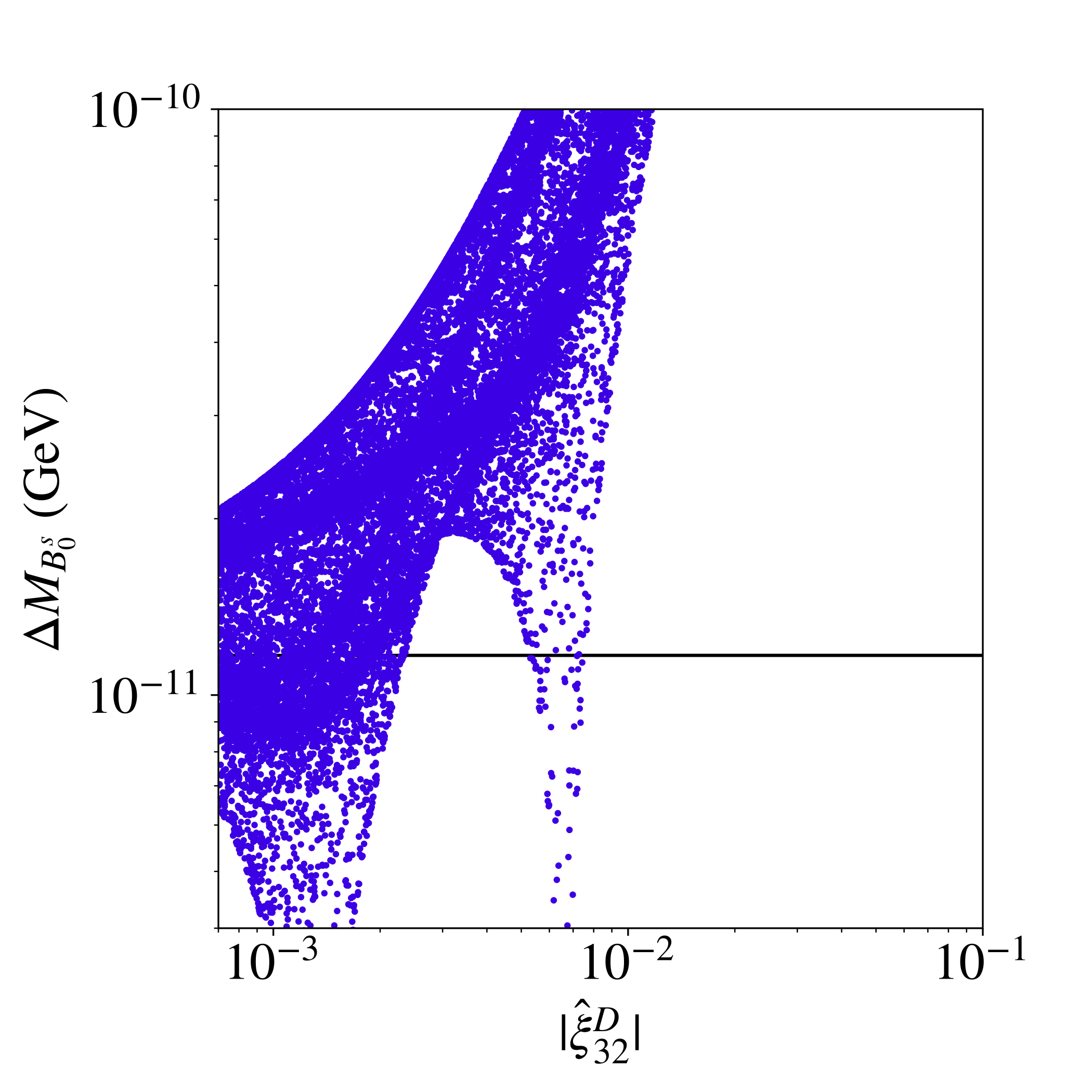}}\qquad
\subfigure[$\sba=0.99$, $m_H=m_A=2000$ GeV,\qquad $\hat\xi^D_{23}=(\pm 1\pm i)\times 10^{-3}$.\label{FIG:2HDM:MesonMixNum:2b1}]{\includegraphics[width=0.45\textwidth]{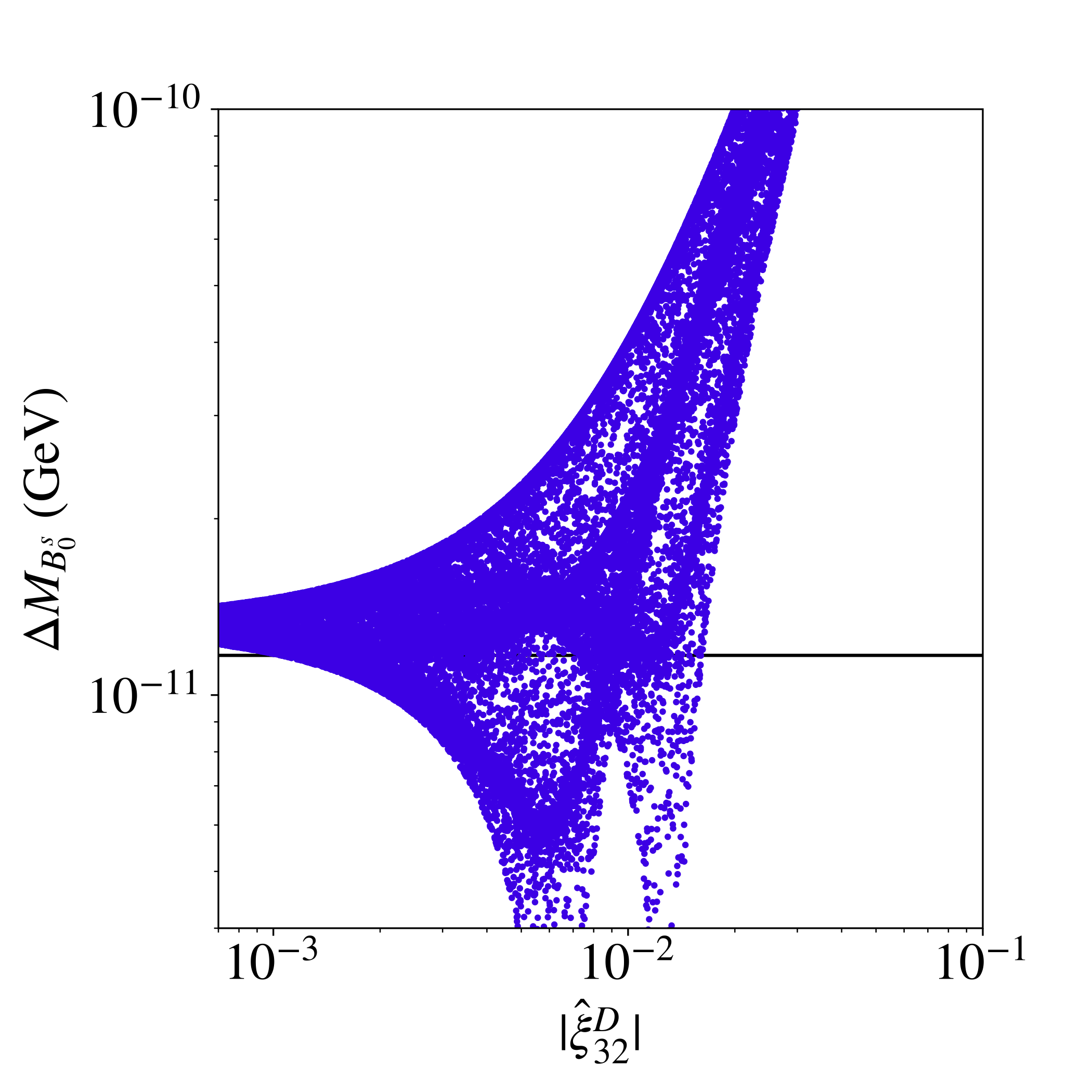}}\\
\end{center}
\caption{$\DMBs$ in the 2HDM versus $|\hat\xi^D_{32}|$. The horizontal line shows the observed value (the corresponding error is smaller than the width of the line itself). The different values of $\sba$, $m_H$, $m_A$ and $\hat\xi^D_{23}$ used in the analyses are shown in each case.
\label{FIG:2HDM:MesonMixNum}}
\end{figure}

\begin{figure}[htb!]
\begin{center}
\subfigure[$\hat\xi^D_{23} \gg \hat\xi^D_{32}$, or $\hat\xi^D_{23} \ll \hat\xi^D_{32}$.\label{FIG:2HDM:MesonMixHBS:1a}]
	{\includegraphics[width=0.45\textwidth]{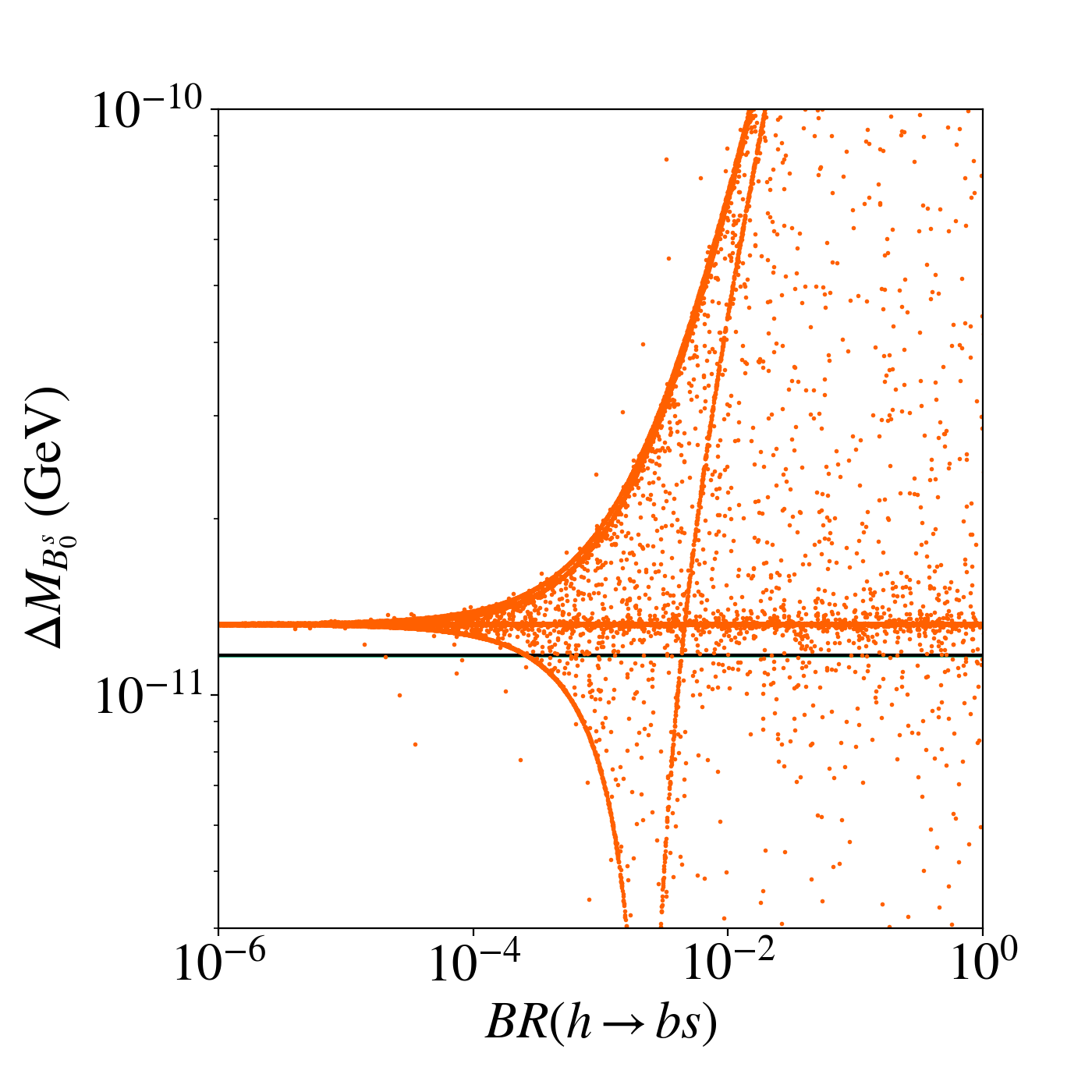}}\qquad
\subfigure[$\hat\xi^D_{23} = \hat\xi^D_{32}$.\label{FIG:2HDM:MesonMixHBS:1b}]{\includegraphics[width=0.45\textwidth]{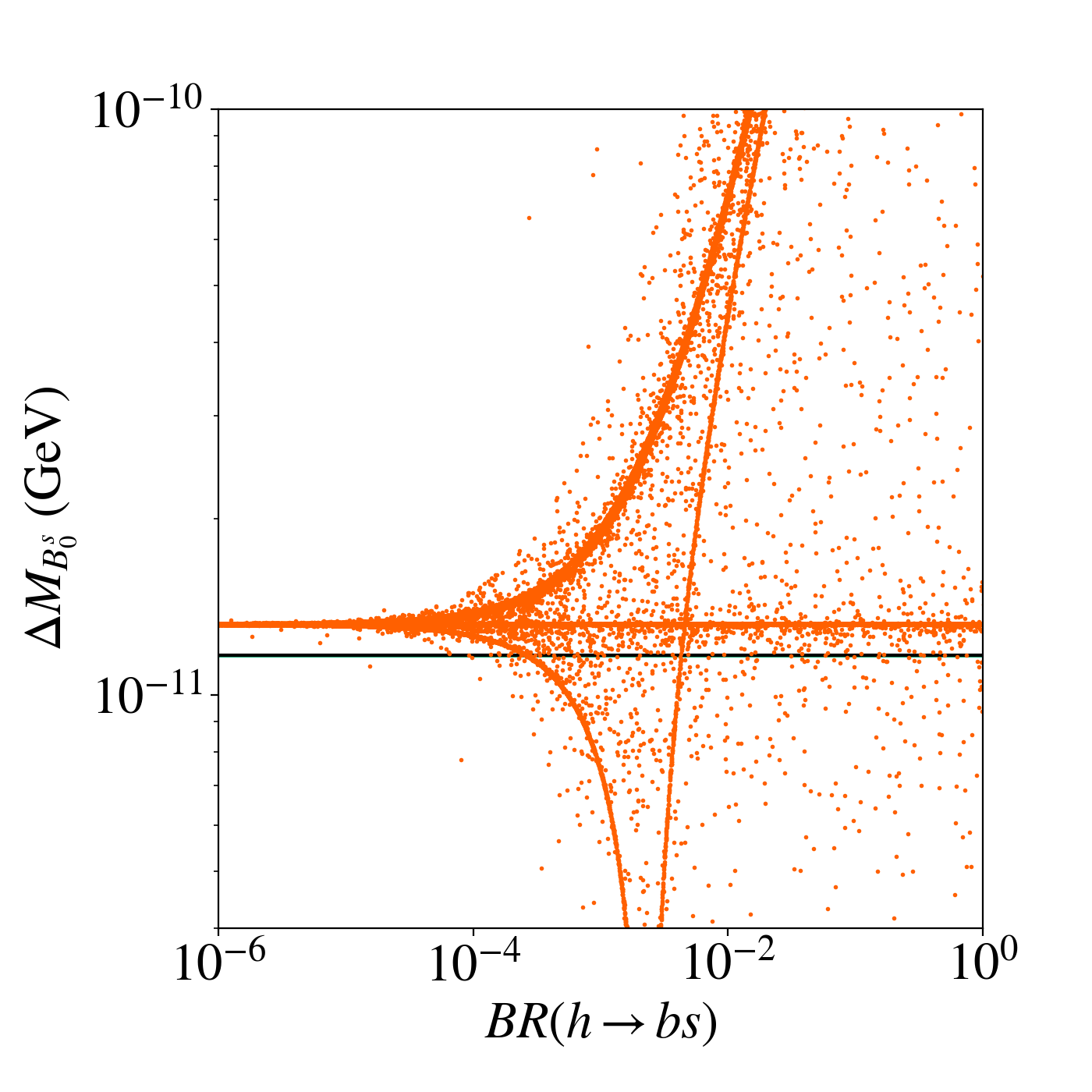}}\\
\end{center}
\caption{Mass splitting $\DMBs$, stemming from the SM and the 2HDM at tree level, versus $\rm{BR}(h\to bs)$. We set $m_H=m_A=2000$ GeV and $\sba = 0.9$. The horizontal line shows the observed value (the corresponding error is smaller than the width of the line itself). 
\label{FIG:2HDM:MesonMixHBS}}
\end{figure}

The 2HDM explanation of the discrepancy, if explained by the tree-level contribution, also implies a prediction of ${\rm BR} (h \rightarrow bs)$. For degenerate $H,\,A$, and much heavier than the light Higgs, the latter contribution to meson mixing dominates in Eq.~\eqref{2DHM_meson}. This is true unless $\cba\simeq0$, for which in any case there is no contribution to ${\rm BR} (h \rightarrow bs)$. Assuming a hierarchy in the off-diagonal Yukawas (taken real), for example $\hat \xi^D_{32} \gg \hat \xi^D_{23}$, so that the $C_2$ contribution to $\Delta M_{B_s, \, \rm 2HDM}$ dominates (and the mixed $C_4$ contribution can be neglected) we get from Eqs.~\eqref{eq:hbs} and \eqref{2DHM_meson}
\begin{equation}\label{eq:hbspred}
	\begin{aligned}
		{\rm BR}(h\to bs) \simeq &\,\frac{3 m^3_h}{16\pi \Gamma_h}\, \frac{|\Delta M_{B_s, \, \rm 2HDM}|}{A_B\,|U_{22}B_2^{B_s}\,b_2 + U_{32}B_3^{B_s}\,b_3|} \simeq \, 2.2 \times 10^{-4}\,,		
	\end{aligned}
\end{equation}
where we used $\Delta M_{B_s, \, \rm 2HDM}=\Delta M_{B_s, \, \rm obs}-\Delta M_{B_s, \, \rm SM}$, and $\Gamma_h \simeq 4.07\cdot 10^{-3}$ GeV. The prediction is identical if the other Yukawa dominates, $\hat \xi^D_{32} \ll \hat \xi^D_{23}$, so that $C'_2$ dominates. On the other hand, for equal Yukawas $\hat \xi^D_{32} = \hat \xi^D_{23}$, the mixed $C_4$ contribution cannot be neglected, and there is an extra term proportional to $U_{44}B_4^{B_s} b_4$ inside the denominator of Eq.~\eqref{eq:hbspred}, so that ${\rm BR}(h\to bs) \simeq 1.8\times{10^{-4}}$. As the angle $\beta-\alpha$ approaches one this lower limit grows. We confirm these predictions in Fig.~\ref{FIG:2HDM:MesonMixHBS}, where we only have the SM plus the 2HDM tree-level contributions. We therefore conclude, that, if the observed discrepancy is confirmed, if accommodated in a 2HDM with negligible contributions at loop level, it implies a prediction of ${\rm BR}(h\to bs)\simeq 10^{-4}$. In our numerical scan, we indeed can accommodate somewhat lower values, when new contributions from the heavy Higgses, and/or those beyond-tree-level containing the other Yukawas, are significant. Similar studies have been done in the context of SU(5) with two Higgs doublets~\cite{DiIura:2016wbx}.

\subsection{Radiative Decays: $\mbox{BR}(B\rightarrow X_s\gamma)$}

In addition to $B_s^0$--$\bar B_s^0$ mixing, we are also interested in the constraints imposed by the radiative decay $B\to X_s\gamma$, that is the transition $b\to s\gamma$ at the quark level. NNLO predictions (i.e. next-to-next-to-leading order in QCD) can be found in Refs.~\cite{Misiak:2006zs,Misiak:2015xwa}. In the context of the 2HDM, NNLO results can be found in Ref.~\cite{Hermann:2012fc}; earlier NLO predictions \cite{Ciuchini:1997xe,Borzumati:1998he} are sufficient for the scope of the present work.

The basis of operators that describes this $|\Delta B|=1$ process includes four quark current-current ($O_{1,2}$) and penguin ($O_{3-6}$) operators, together with photonic ($O_7$) and gluonic ($O_8$) dipole operators (see, e.g., Ref.~\cite{Misiak:2006ab}). Effective Wilson coefficients $C_{7,8[\rm eff]}$ are usually defined such that the perturbative contribution to $\text{Br}(B\to X_s\gamma)$ is proportional to $|C_{7[\rm eff]}|^2$ at leading order. Expressions for the LO and NLO contributions to the Wilson coefficients (at the matching scale $\mu_W\sim m_W$) can be found in Eqs. (16) and (17) of Ref.~\cite{Borzumati:1998he} . Leading order contributions involving neutral scalars can be found in Ref.~\cite{Crivellin:2017upt}.\footnote{For comparison with the notation of Ref.~\cite{Borzumati:1998he}, 		$(XY^*)^{\phi}_{u_i} = -1/(m_{u_i} m_{b})\, \Gamma^{LR \, \phi *}_{u_i \, d_3} \Gamma^{RL \, \phi}_{u_i \, d_2}$ and
		$(YY^*)^{\phi}_{u_i} = 1/(m^2_{u_i})\, \Gamma^{RL \, \phi *}_{u_i \, d_3} \Gamma^{RL \, \phi}_{u_i \, d_2}$.} The perturbative $b\to s\gamma$ decay rate is given by
\begin{equation}\label{eq:bsgamma}
	\Gamma(b \rightarrow s \gamma) = \frac{G_F^2}{32\pi^4}  
 \vert \CKMc{ts}\CKM{tb}\vert ^2 \alpha_{\rm em} \, m_b^5 \, \left(\vert C_{7{\rm eff}}\, (\mu_b)\vert ^2 + \vert C_{7{\rm eff}}^{\prime} (\mu_b) \vert ^2 \right).
\end{equation}
The inclusive $\bar B\to X_s\gamma$ decay rate is measured with photon energies $E_\gamma>1.6$ GeV, in which case the non-perturbative contributions relating the quark level and the meson decay rates are below the 5\% level \cite{Benzke:2010js}. Attending to the different sources of theoretical uncertainty, in order to place constraints on the 2HDM contributions, we use the perturbative quark level decay rate in \refEQ{eq:bsgamma} with a conservative theoretical error of $10\%$. The corresponding SM calculation is given in Tab.~\ref{table:FlavPhysicsValues} and is in very good agreement with the observed value.

\section{Numerical Analysis\label{sec:numerical_analysis}}

\subsection{Parameter Scan}
Given the large number of parameters of our general 2HDM (9 from the potential, 12 from the Yukawas in the 2-3 plane) we use a global fit using MultiNest \cite{Feroz:2008xx} to scan over the allowed parameter space. We also use 2HDMC (Two-Higgs-Doublet Model Calculator) \cite{Eriksson:2009ws} to perform some phenomenological calculations. We do not include the SM one loop contribution to $h \rightarrow bs$ or $t \rightarrow c h$, but we compute the new 2HDM contributions. We then plot our results using {\rm pippi}~\cite{Scott:2012qh}. The parameters and priors scanned over are given in Tab.~\ref{table:higgs_basis_prior}. We use the Higgs basis. To ensure that we carry out our scan over both quadrants in the physical angle we choose $-\pi/2 \leq (\beta-\alpha) \leq \pi/2$.

\begin{table*}\centering
\ra{1.3}
\begin{tabular}{@{}lll@{}}\toprule\midrule
  Parameter & Range & Prior \\
  \midrule
  $\Lambda_{1,2,3,4,5,7}$ & $\pm [10^{-7}, 4\pi]$&Log\\
  $\beta - \alpha$ & $[-\pi/2, \pi/2]$&Flat\\ 
  $M_{22}^2$ (GeV$^2$) & $[10^4,10^7]$ & Flat \\ 
  Re$(\hat\xi^{D,U}_{ij})$ & $\pm [10^{-7}, 4\pi]$ &Log \\
  Im$(\hat\xi^{D,U}_{ij})$ & $\pm [10^{-7}, 4\pi]$  &Log\\
  \midrule
  \bottomrule
  \end{tabular}
\caption{Parameters scanned over. We also indicate whether the priors are flat or log. In the Yukawa sector, $i,j = 2,3$, and all other couplings are zero.}
\label{table:higgs_basis_prior}
\end{table*}

We need to provide likelihood functions $\mathcal{L}$ (or $\chi^2 = -2\ln \mathcal{L}$) to scan the parameter space of the model. To ensure that the masses of the scalars are positive, as well as to impose stability of the scalar potential, we use a hard cut-off: for a calculated value $\mathcal{O}_{\rm calc}$ and lower bound $B_i$ 
\be
    \chi_{\rm bounds}^2= 
\begin{cases}
    0,& \text{if } \mathcal{O}_{\rm calc}> B_i\\
    \chi_{\rm max}      & \text{if } \mathcal{O}_{\rm calc} \leq B_i,
\end{cases}
\ee
where $\chi_{\rm max}$ is large enough that the scanner effectively invalidates the point. The reverse of this may be used for an upper bound.
Unitarity and perturbativity are imposed by a soft cut-off 
\be
    \chi_{\rm bounds}^2= 
\begin{cases}
    0,& \text{if } \mathcal{O}_{\rm calc}< B_i/0.64\\
    \biggl(\frac{0.64\mathcal{O}_{\rm calc}}{B_i} - 1\biggr)^2,              & \text{if } \mathcal{O}_{\rm calc} \geq B_i/0.64\,,
\end{cases}
\ee
where $B_i$ is the upper bound at 68\% confidence (improving the guidance provided to the scanner).
For observables that have been measured we use a centered distribution with the observed value at $\mathcal{O}_{\rm obs}$ and error $\sigma$\begin{equation}
    \chi_{\rm observations}^2= \biggl(\frac{\mathcal{O}_{\rm calc}-\mathcal{O}_{\rm obs}}{\sigma}\biggr)^2.
\end{equation}
The final $\chi^2$-like function is built from all $M$ bounds and $N$ observations,
\begin{equation}
	\chi^2 = \sum^M_i \chi_{{\rm bounds,i}}^2 + \sum^N_i \chi_{\rm observations,i}^2\,.
\end{equation}
For $B_s^0$--$\bar B_s^0$ mixing and $B\to X_s\gamma$, we sum the errors of experimental and calculated values in quadrature. 

\subsection{Results}

\begin{figure}[!htb]
\centering
{%
\setlength{\fboxsep}{20pt}%
\includegraphics[scale=1.0]{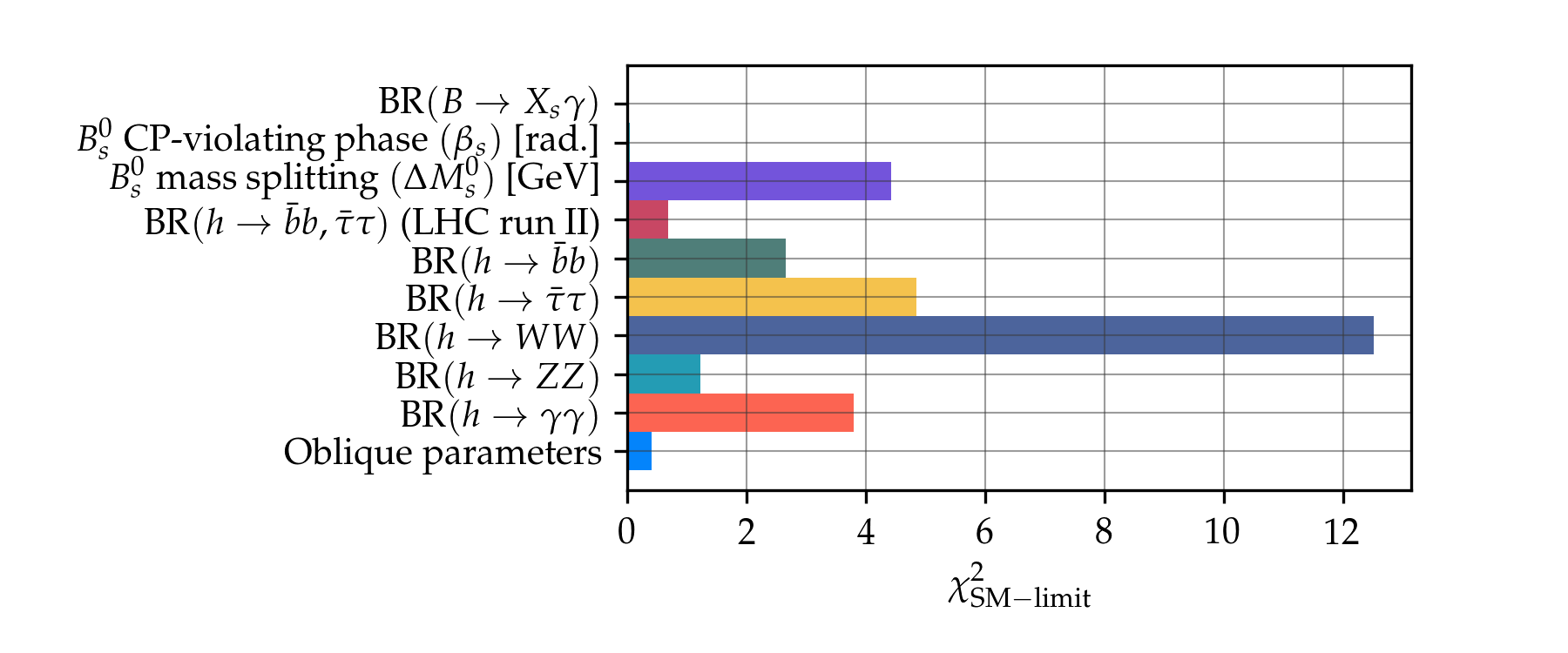}
}
\caption{$\chi^2$ contributions due to experimental constraints in the limit of the Standard Model, $s_{\beta-\alpha} \rightarrow 1$ and $\hat \xi^U_{ij}=\hat \xi^D_{ij}=0$.}
\label{fig:sm_chi2}
\end{figure}

To start with, we show in Fig.~\ref{fig:sm_chi2} the experimental contributions to the total $\chi^2$ value that we calculate in the SM limit, that is $\sba \to 1$ and $\hat \xi^U_{ij}=\hat \xi^D_{ij}=0$. The largest pulls here come from SM Higgs decays, as expected predominantly from $h\rightarrow WW$, due to the fact that the experimental values of some of the production channels are slightly off from the SM, see Table~\ref{eq:Higgsmu:RunI}. Using LHC Run II data~\cite{Aad:2019lpq, Aaboud:2018jqu, Aaboud:2017jvq} the $h\rightarrow WW$ signal strengths by production channel (as in Table~\ref{eq:Higgsmu:RunI}) are $(1.10^{+0.21}_{-0.21}, 0.62^{+0.36}_{-0.35}, 2.3^{+1.2}_{-1.0}, 2.9^{+1.9}_{-1.3}, 1.5^{+0.6}_{-0.6})$. This almost halves the $h\rightarrow WW$ channel $\chi^2_{\rm SM-limit}$ contribution to $7.15$. In any case, the SM is consistent with this data at the $\sim 2\sigma $ level.

\begin{figure}[!htb]
\centering
{%
\setlength{\fboxsep}{20pt}%
\includegraphics[scale=1.0]{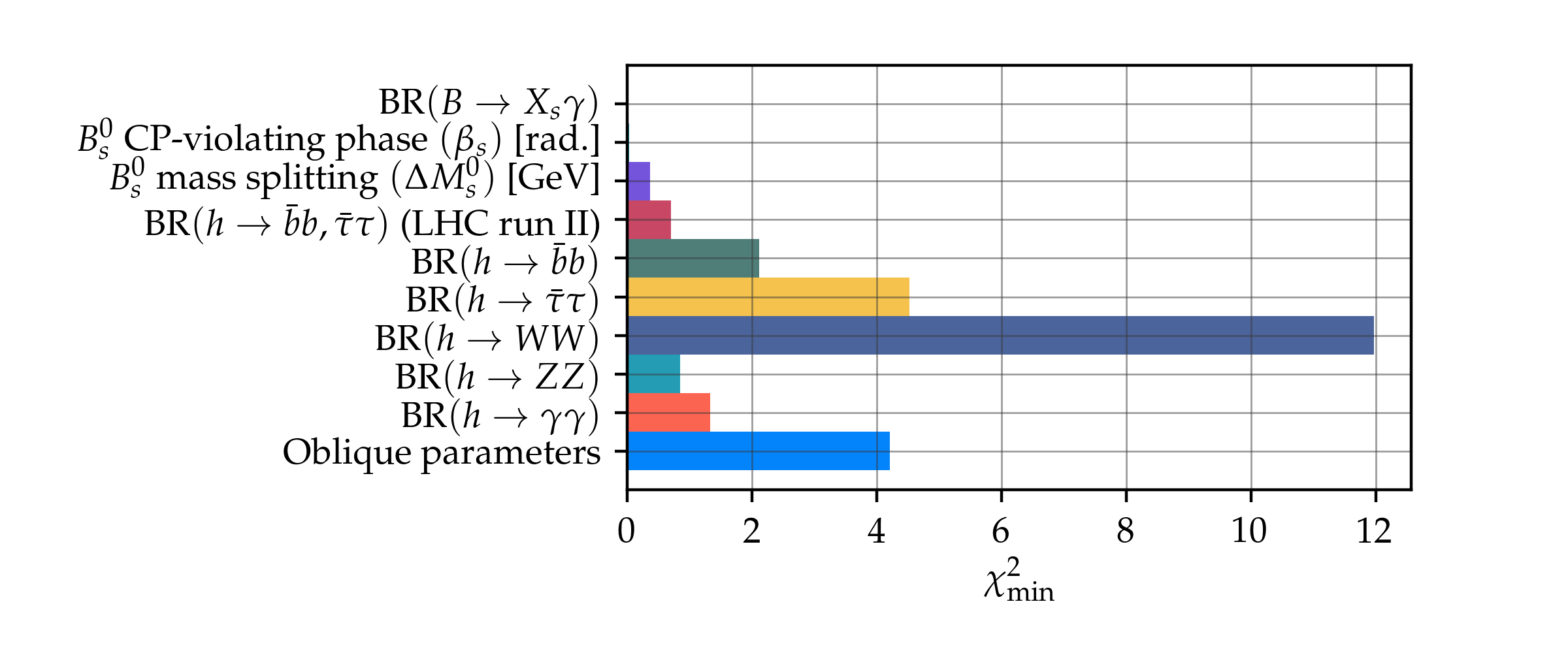}
}
\caption{The contributions from each of the constraints to the best-fit $\chi^2$ in our global scan of the 2HDM.}
\label{fig:best_chi2}
\end{figure}

In Fig.~\ref{fig:best_chi2} we show the pull from each constraint at our best fit point for the 2HDM. Relative to the $\chi^2_{\text{SM-limit}}$ showed in Fig.~\ref{fig:sm_chi2}, we see that the Higgs decay channels are very similar, except for the decrease in the $h \rightarrow \gamma\gamma$ channel. There is an increased contribution in the pull from oblique parameters, but this is minor as it is a contribution from all three parameters. Notably, flavor observables are well minimised at the best-fit point. Especially the $B^0_s$ meson mixing discrepancy present in the SM (as discussed in Sec.~\ref{sec:disc}) is reduced in the 2HDM.

\begin{figure}[h]
\centering
{%
\includegraphics[scale=0.6]{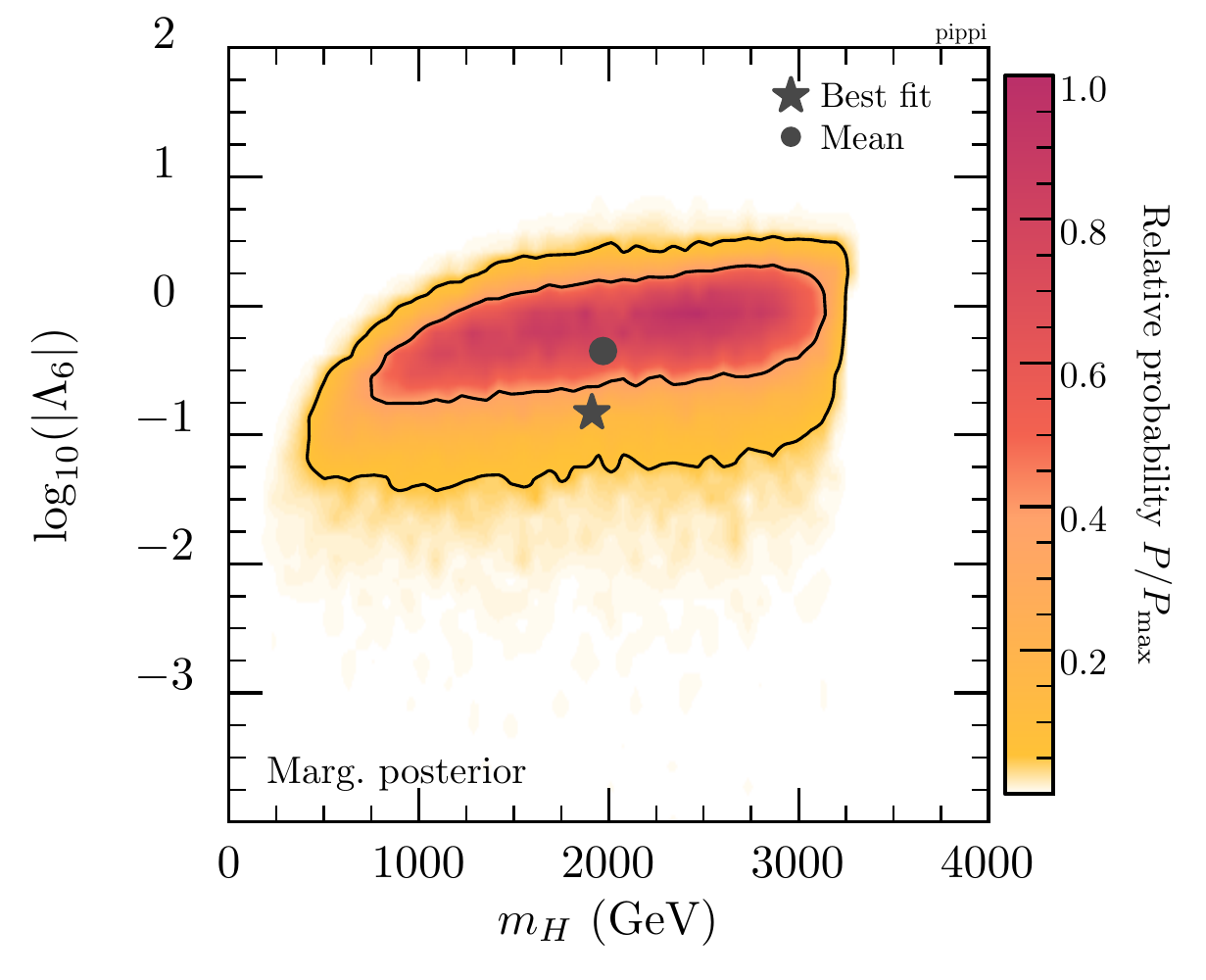}~~~\includegraphics[scale=0.6]{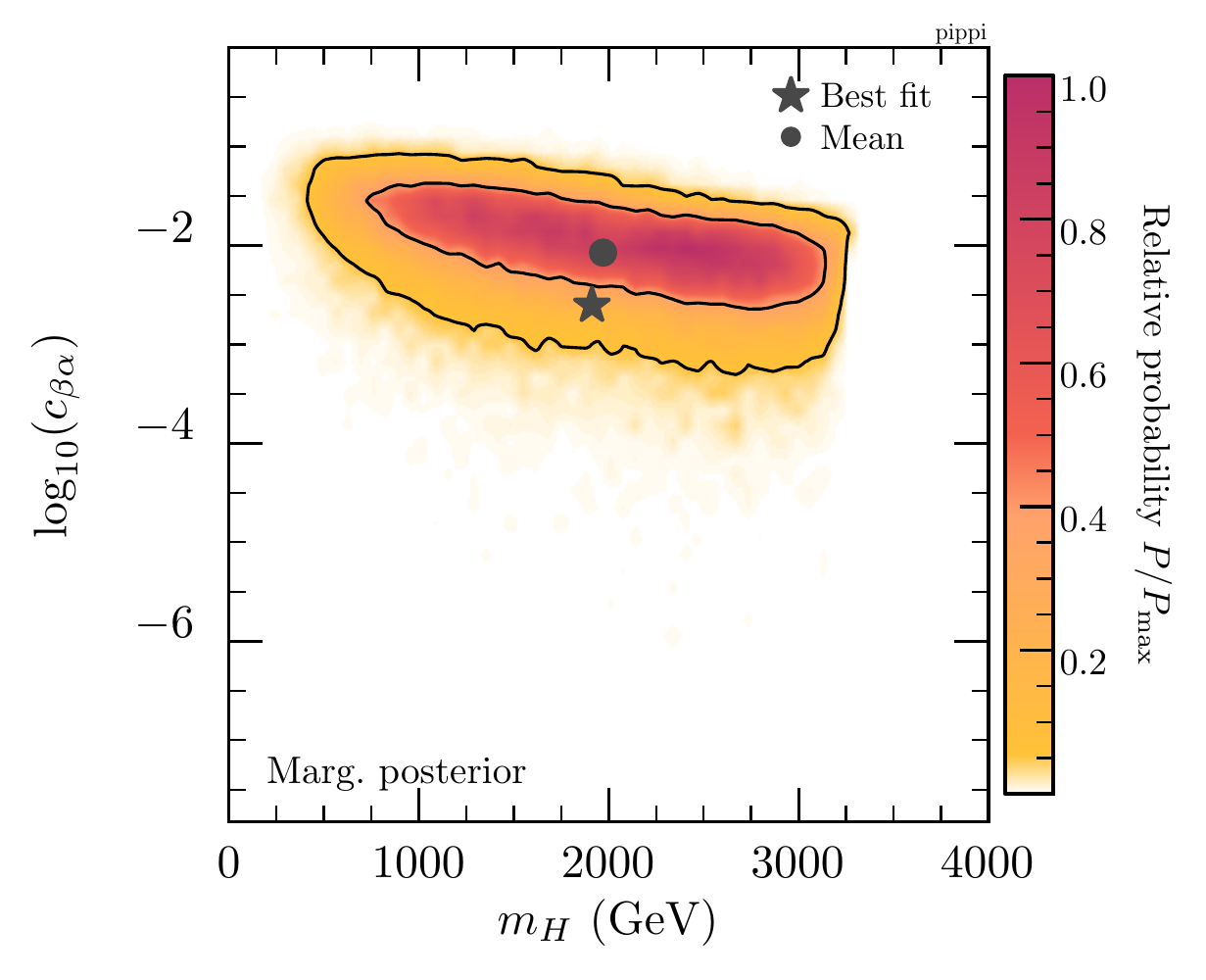}\\
\includegraphics[scale=0.6]{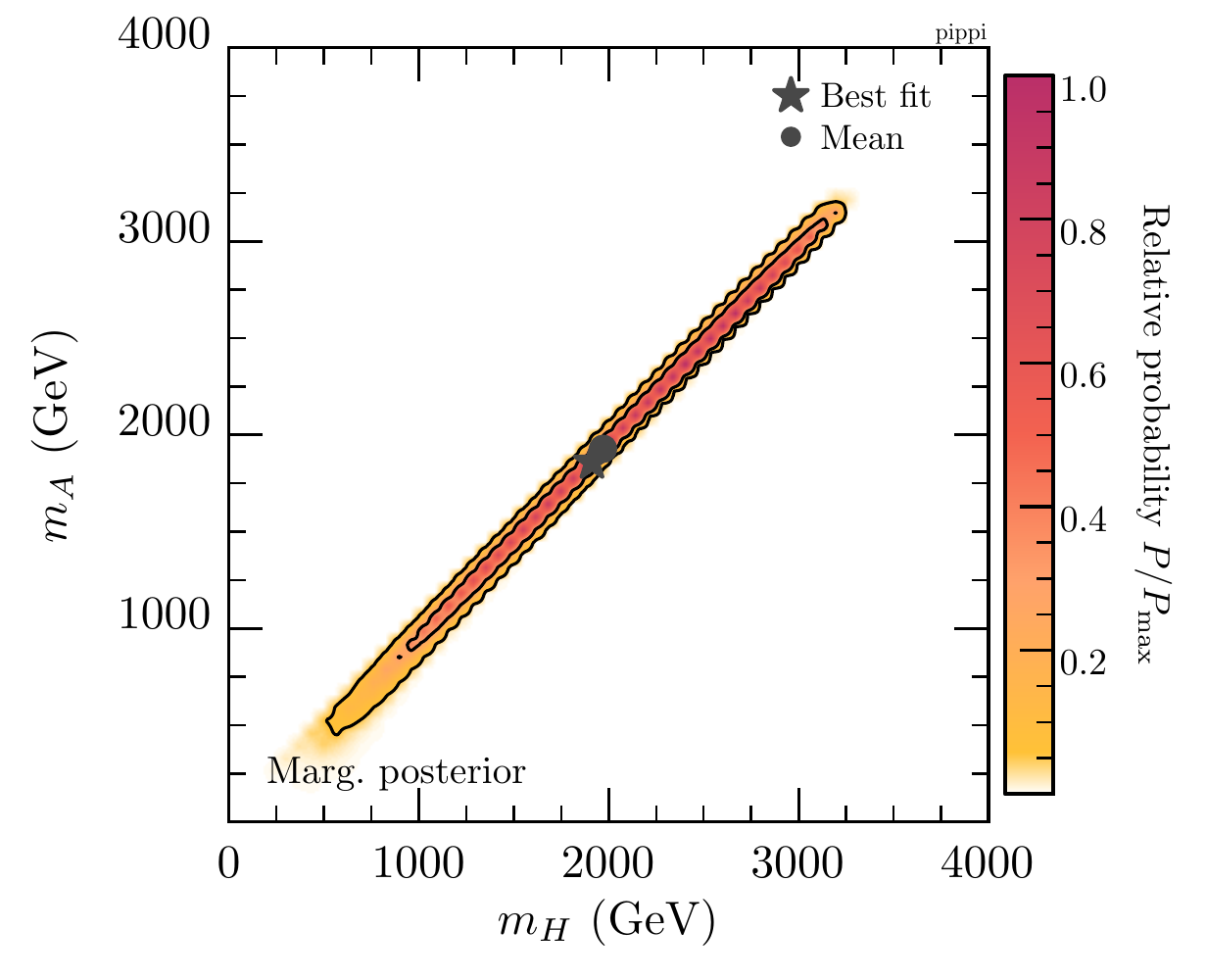}~~\includegraphics[scale=0.6]{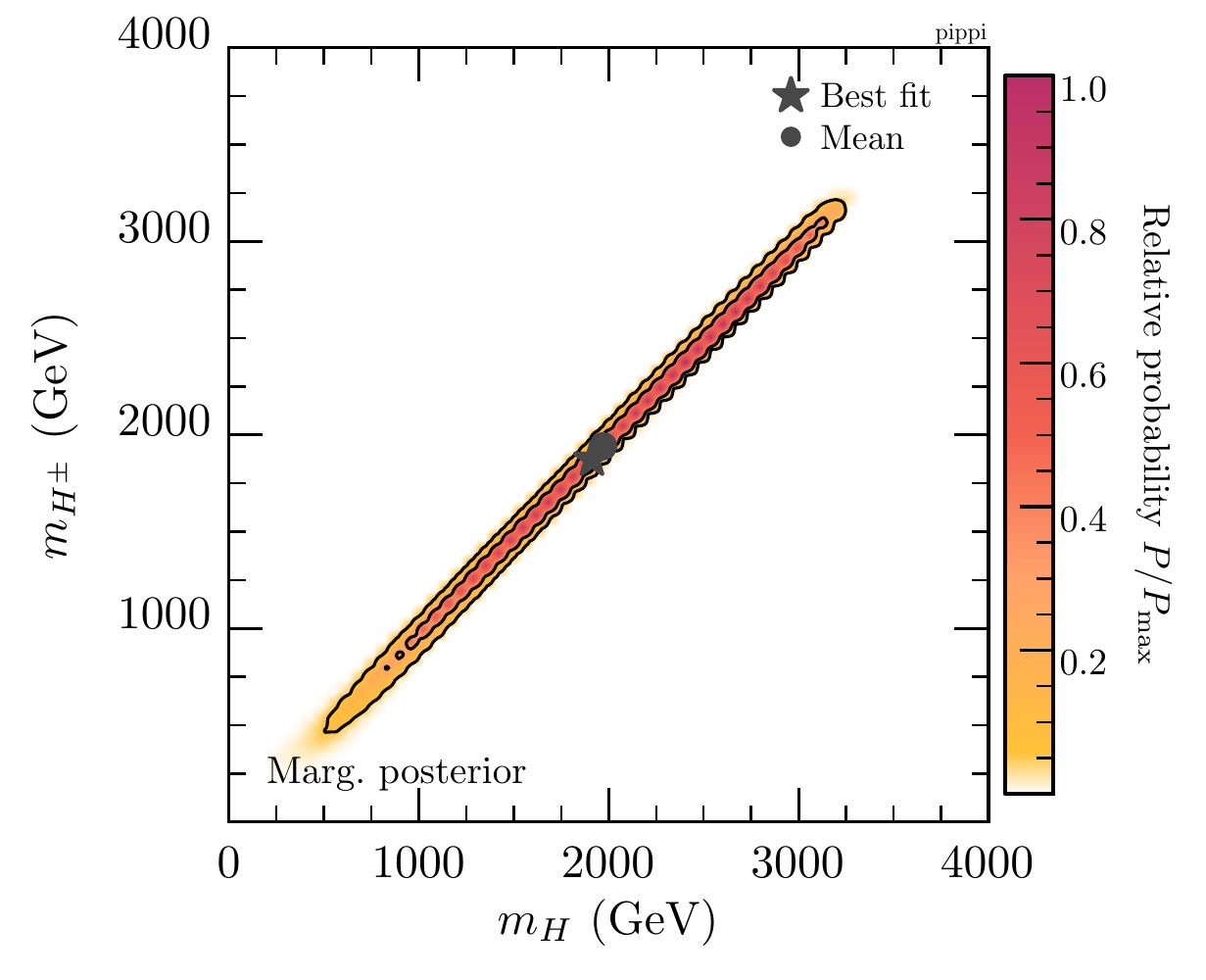}
}%
\caption{\emph{Top left [right]:} $\log_{10}(|\Lambda_6|)$ [$\log_{10}(c_{\beta-\alpha})$] versus $m_H$. \emph{Bottom left [right]:} Relationships between the extra scalar particles of the 2HDM, $m_A$ [$m_H^{\pm}$] versus $m_H$. The $1\sigma$ and $2 \sigma$ probability regions are represented by the solid lines.}
\label{fig:res1}
\end{figure}

In the top panels of Fig.~\ref{fig:res1}, we plot $\log_{10}(|\Lambda_6|)$ (left) and $\log_{10}(\cba)$ (right) versus $m_H$. On the top-left there is a correlation between $\Lambda_6$ and $m_H$ (as expected from Eq.~\eqref{eq:sinbminua} for a sufficiently SM-like Higgs boson, i.e., in the alignment limit $\sba \to 1$). The bottom panels display correlations between the extra scalars. They each obey a linear relationship imposed by the oblique parameter constraints. The size of our masses extends up to $\sim 3200$ GeV due to the priors on $M_{22}$ and the perturbativity limits used on the quartic couplings.

\begin{figure}[h]
\centering
{%
\includegraphics[scale=0.6]{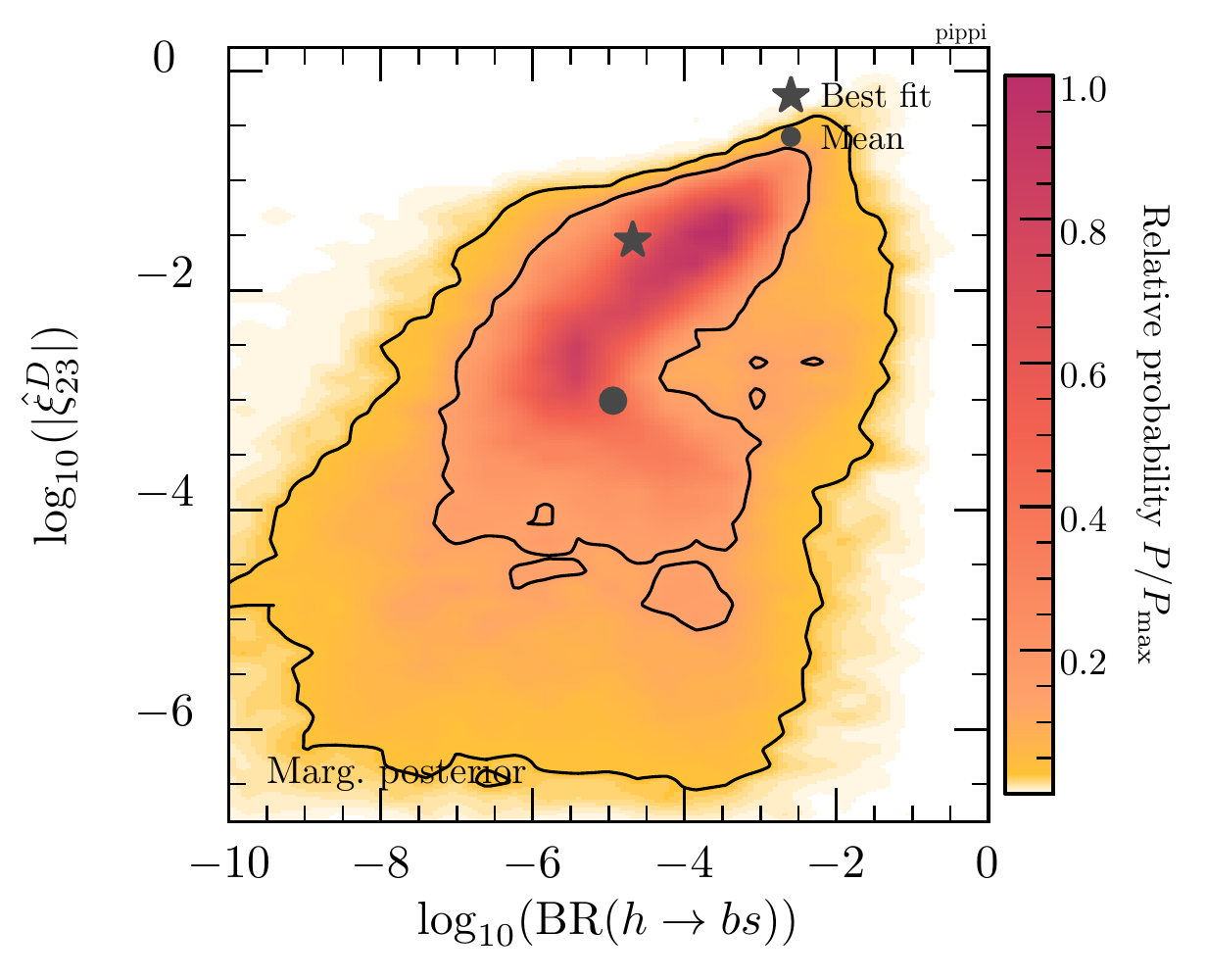}~~\includegraphics[scale=0.6]{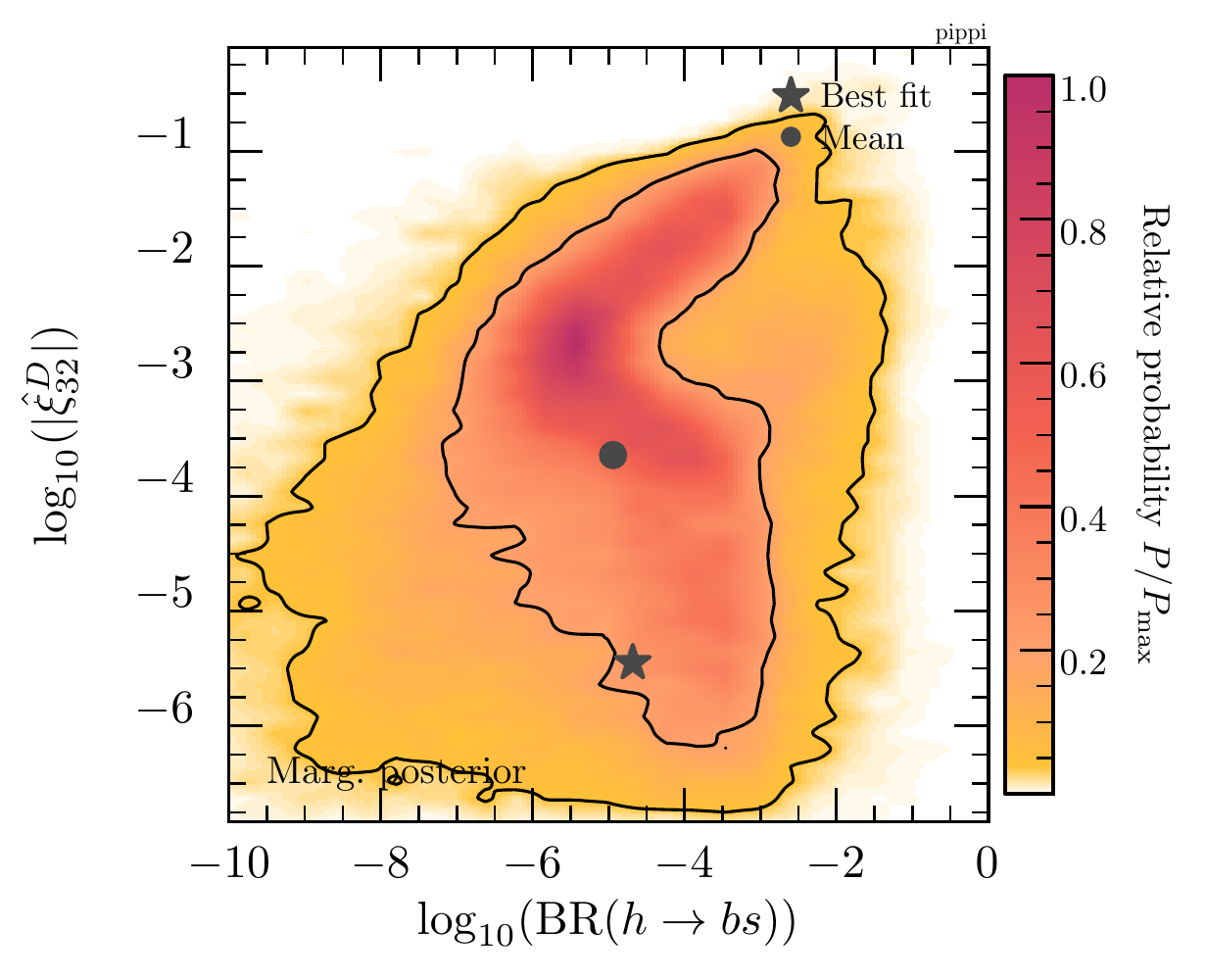}\\
\includegraphics[scale=0.6]{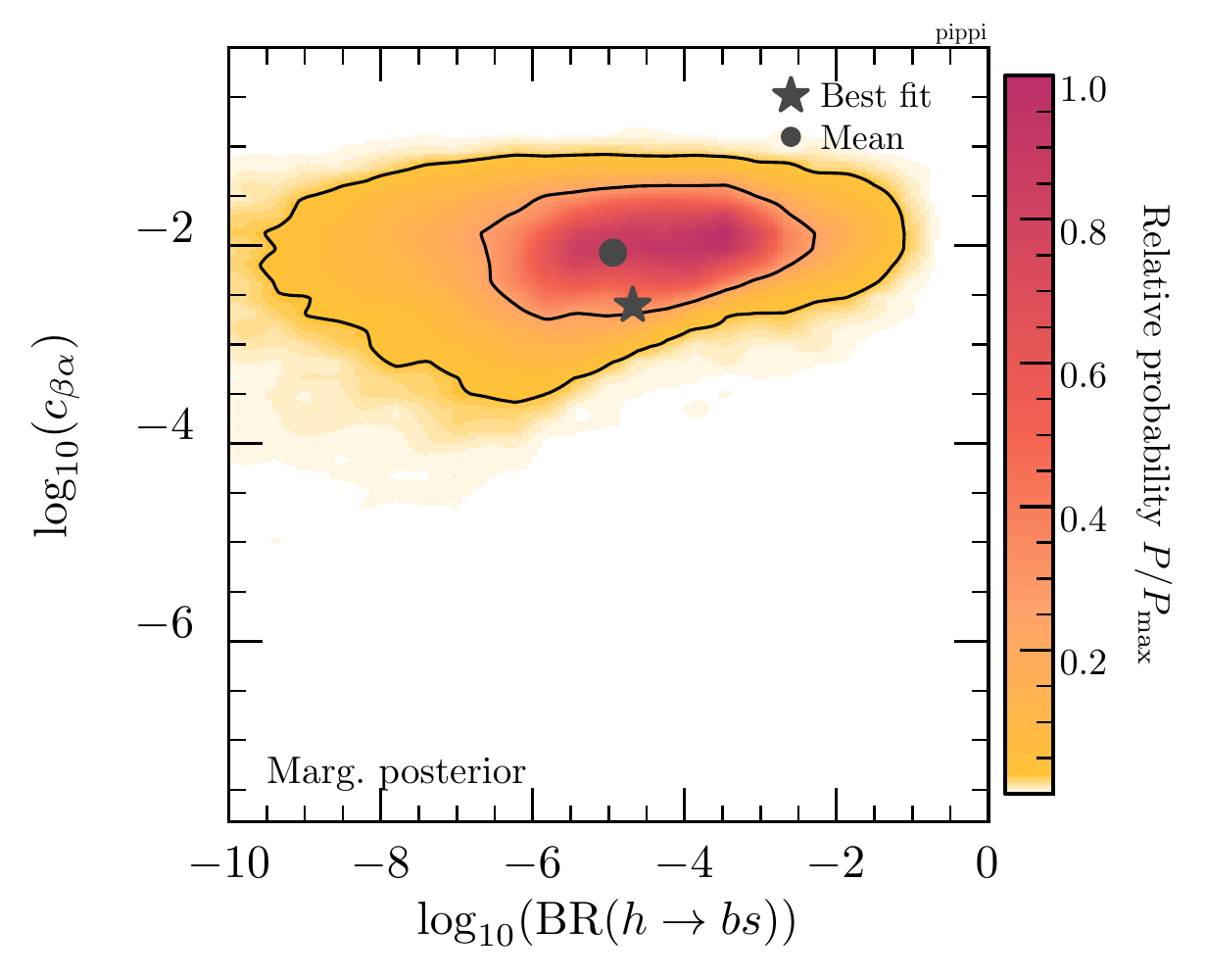}~~\includegraphics[scale=0.6]{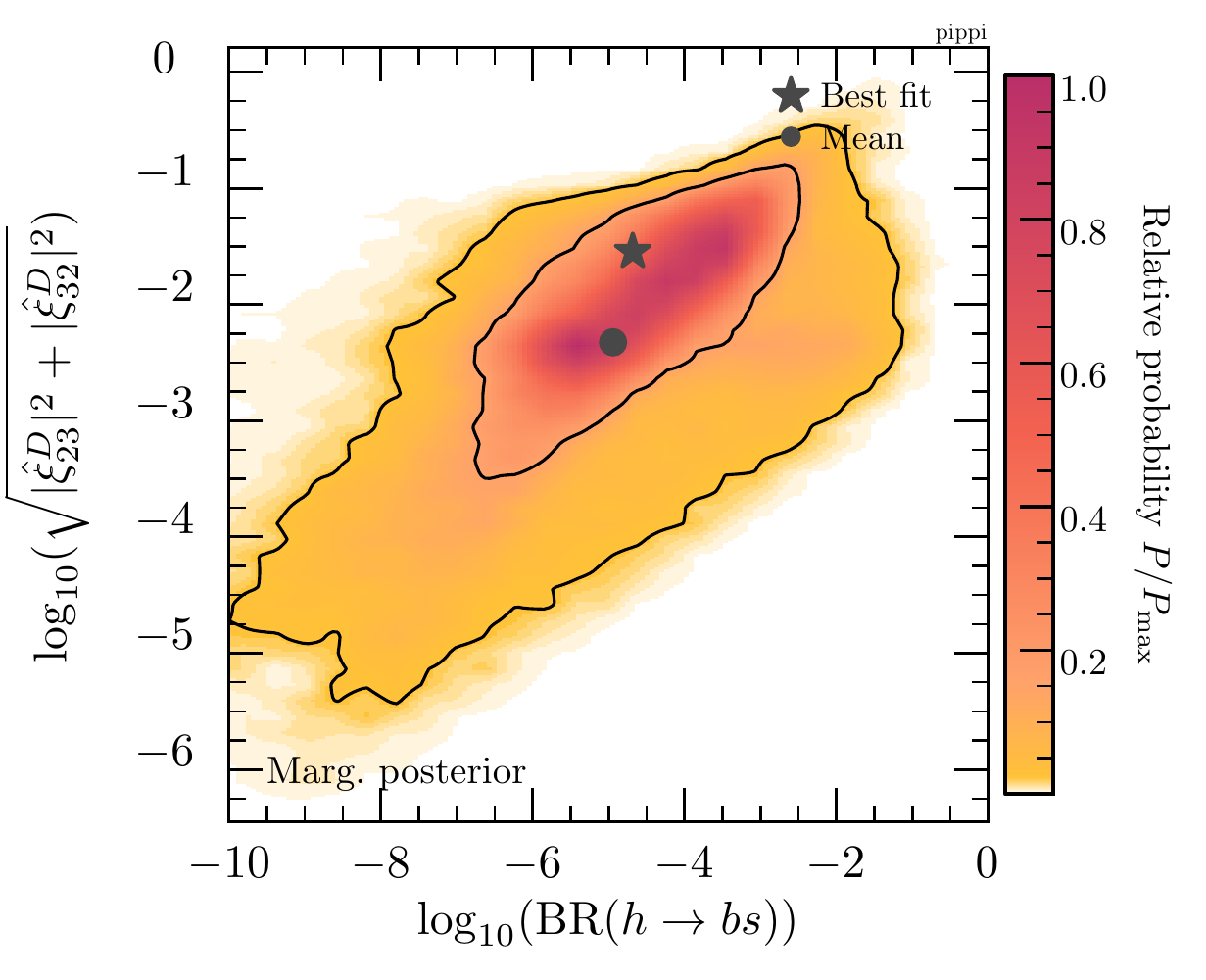}
}%
\caption{ Off-diagonal down-quark Yukawa couplings $\hat\xi^D$ versus $\log_{10}[{\rm BR}(h\rightarrow bs)]$. \emph{Top left [right]:} $\log_{10}(|\hat\xi_{23}^D|)$ [$\log_{10}(|\hat\xi_{32}^D|)$] \emph{Bottom left:} The logarithm of the physical angle $c_{\beta-\alpha}$ versus $\log_{10}[{\rm BR}(h\rightarrow bs)]$. \emph{Bottom right:} Logarithm of the modulus of the off-diagonal contributions to $\hat\xi^D$ versus $\log_{10}[{\rm BR}(h\rightarrow bs)]$.}
\label{fig:res2b}
\end{figure}

In Fig.~\ref{fig:res2b} we plot the logarithm of the absolute value of the off-diagonal Yukawa combinations ($\log_{10}(|\hat\xi_{23}^D|)$ and $\log_{10}(|\hat\xi_{32}^D|)$) versus $\log_{10}[{\rm BR}(h\rightarrow bs)]$. We attain an upper (lower) limit on $h\rightarrow bs$ of $\sim 10^{-3}$ ($\sim 10^{-7}$) at $1 \sigma$. The lower value is in the same range as the SM prediction at one loop (which we do not include in the scan).

\begin{figure}[h]
\centering
{%
\includegraphics[scale=0.6]{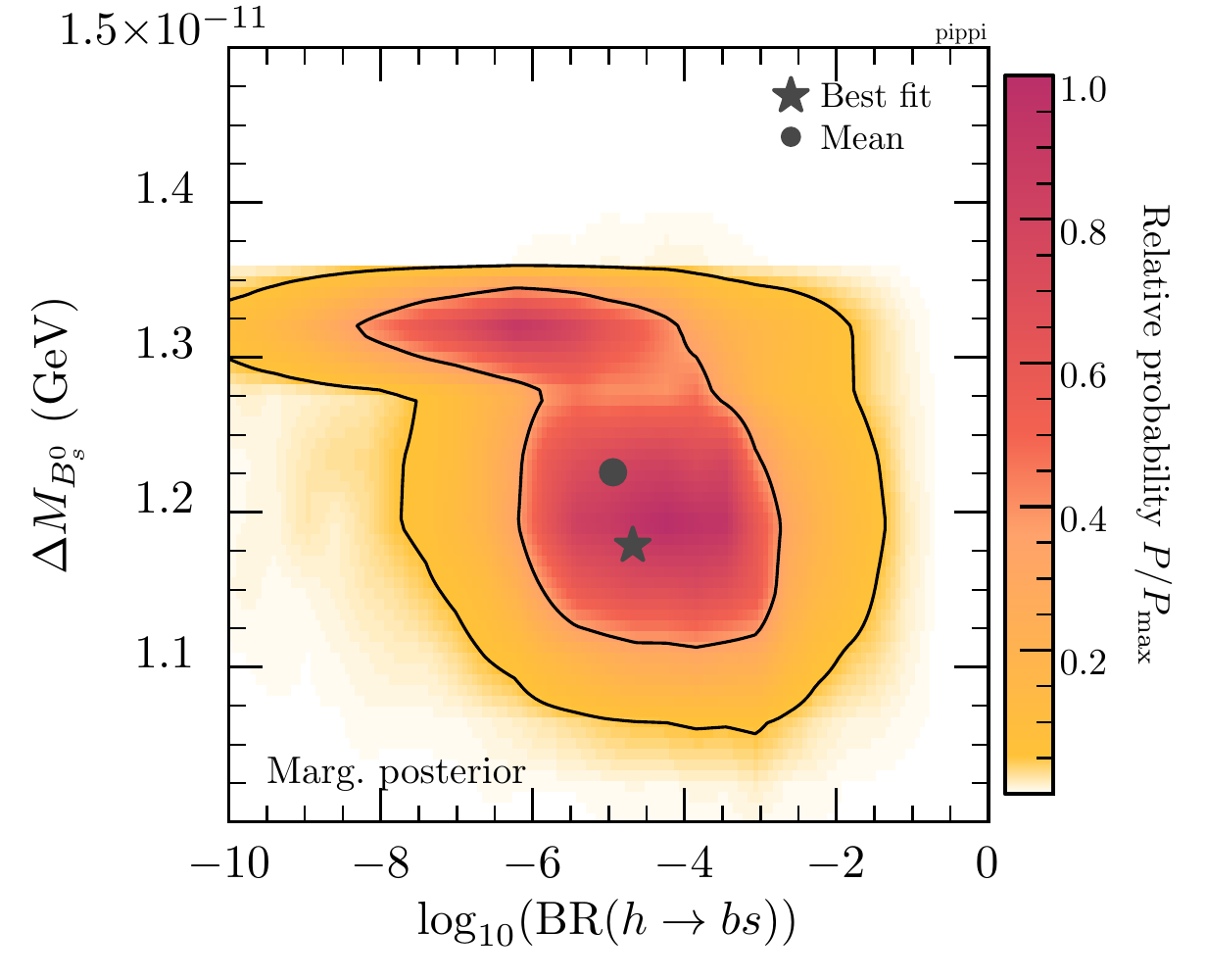}~~\includegraphics[scale=0.6]{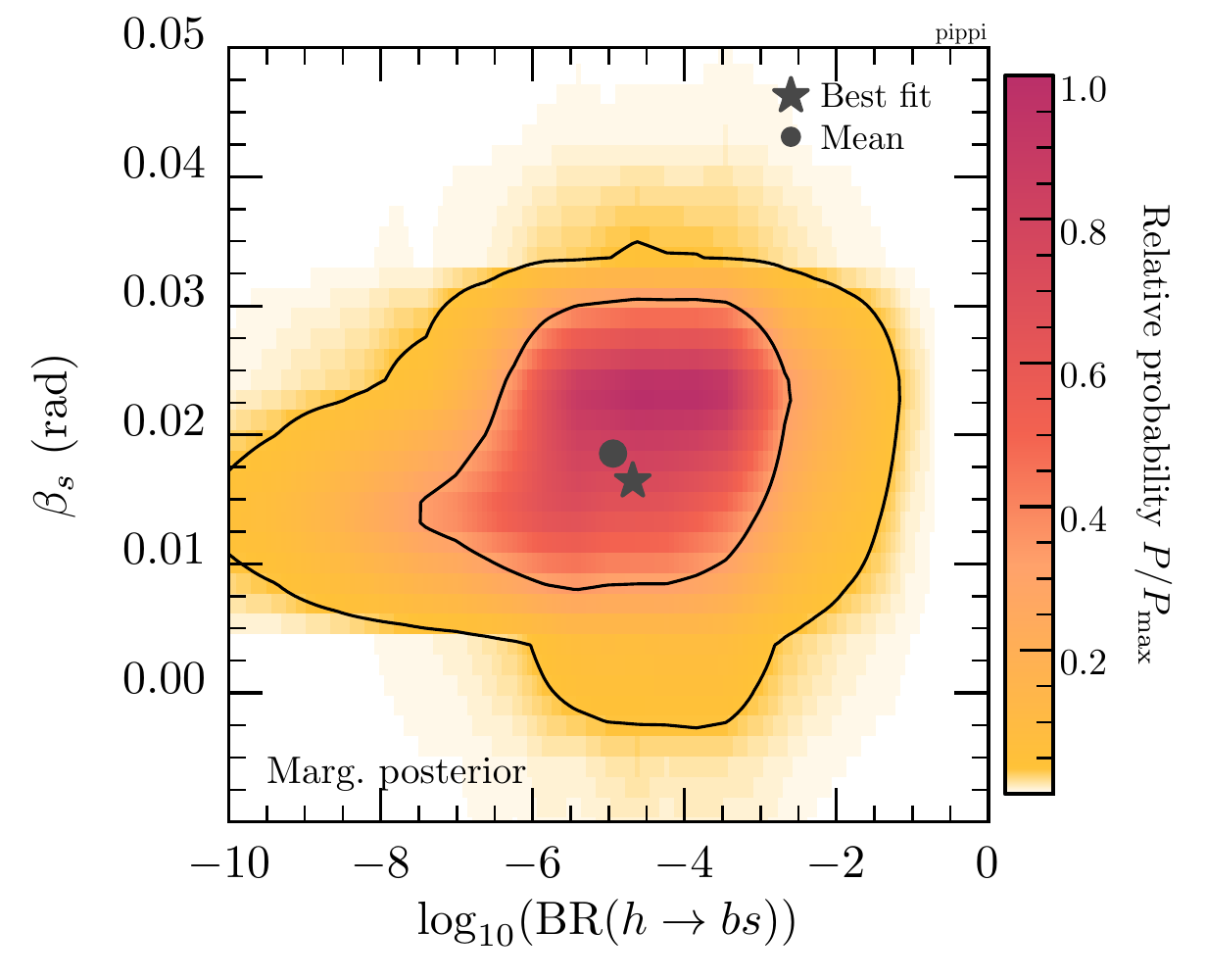}\\
\includegraphics[scale=0.6]{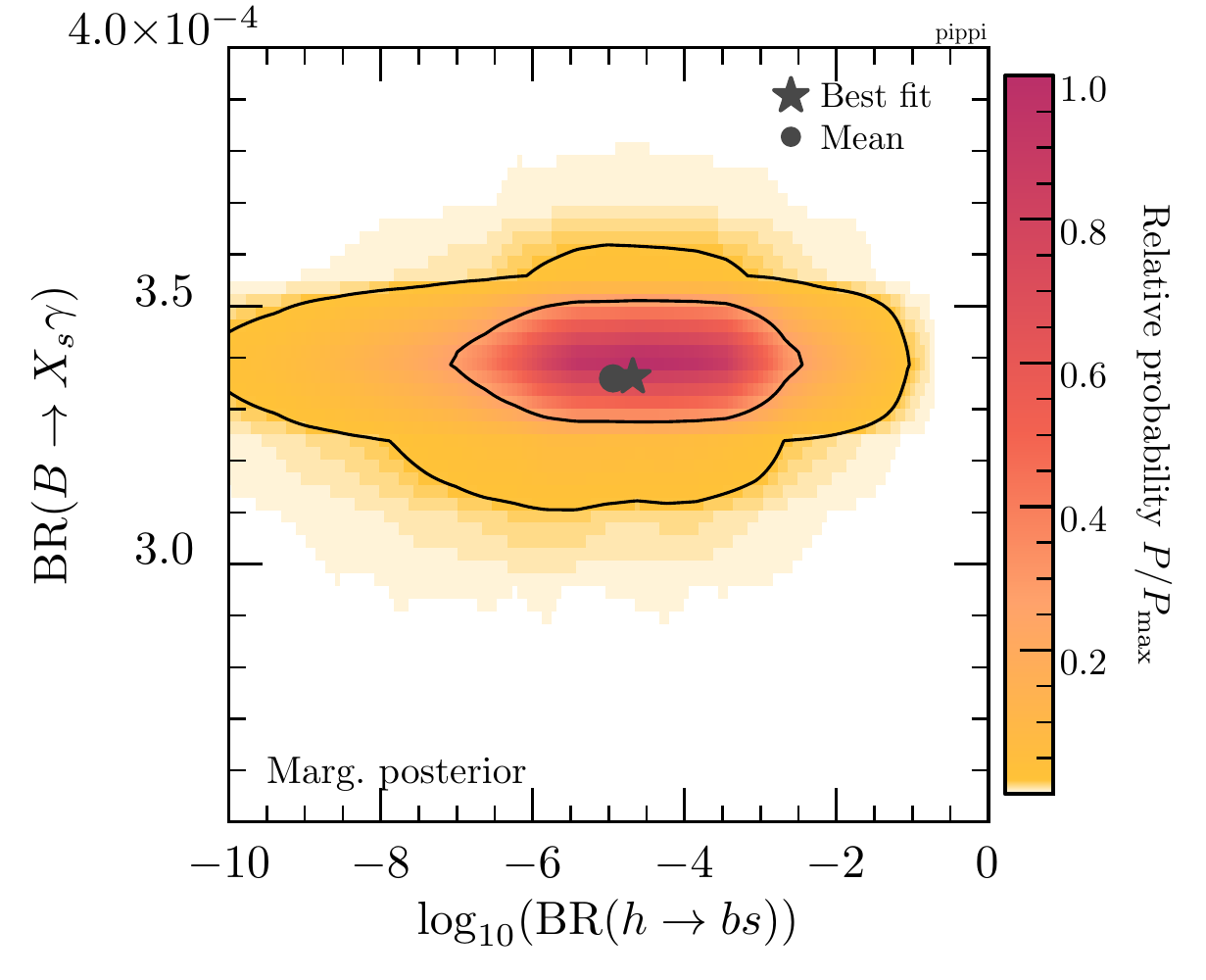}
}
\caption{Different observables and parameter versus $\log_{10}[{\rm BR}(h\rightarrow bs)]$. \emph{Top left [right]:} $B^0_s$ meson mixing mass splitting [CP phase]. \emph{Bottom} Radiative B decay BR($B\rightarrow X_s\gamma$).}
\label{fig:res2a}
\end{figure}

Exploring the constraints that caused these limits, we show in Fig.~\ref{fig:res2a} the posterior distributions of relevant flavor physics observables (the mass splitting $\DMBs$, the CP-violating phase $\beta_s$ and the radiative B-decay, $B \rightarrow X_s \gamma$) with respect to the $h\rightarrow bs$ decay. 
For $\DMBs$ we observe two almost disconnected solution regions, as we expect from Fig.~\ref{FIG:2HDM:MesonMixNum}. In the upper region, the predicted $\DMBs$ mass splitting coincides with the SM value, which is $1.8\sigma$ off the observed one. In the lower region, the 2HDM can accommodate the observed value, and what is more interesting, this yields a lower bound ${\rm BR}(h\rightarrow bs)$, at the level of $10^{-6}$ at $1\sigma$. 

\begin{figure}[h]
\centering
{%
\includegraphics[scale=0.6]{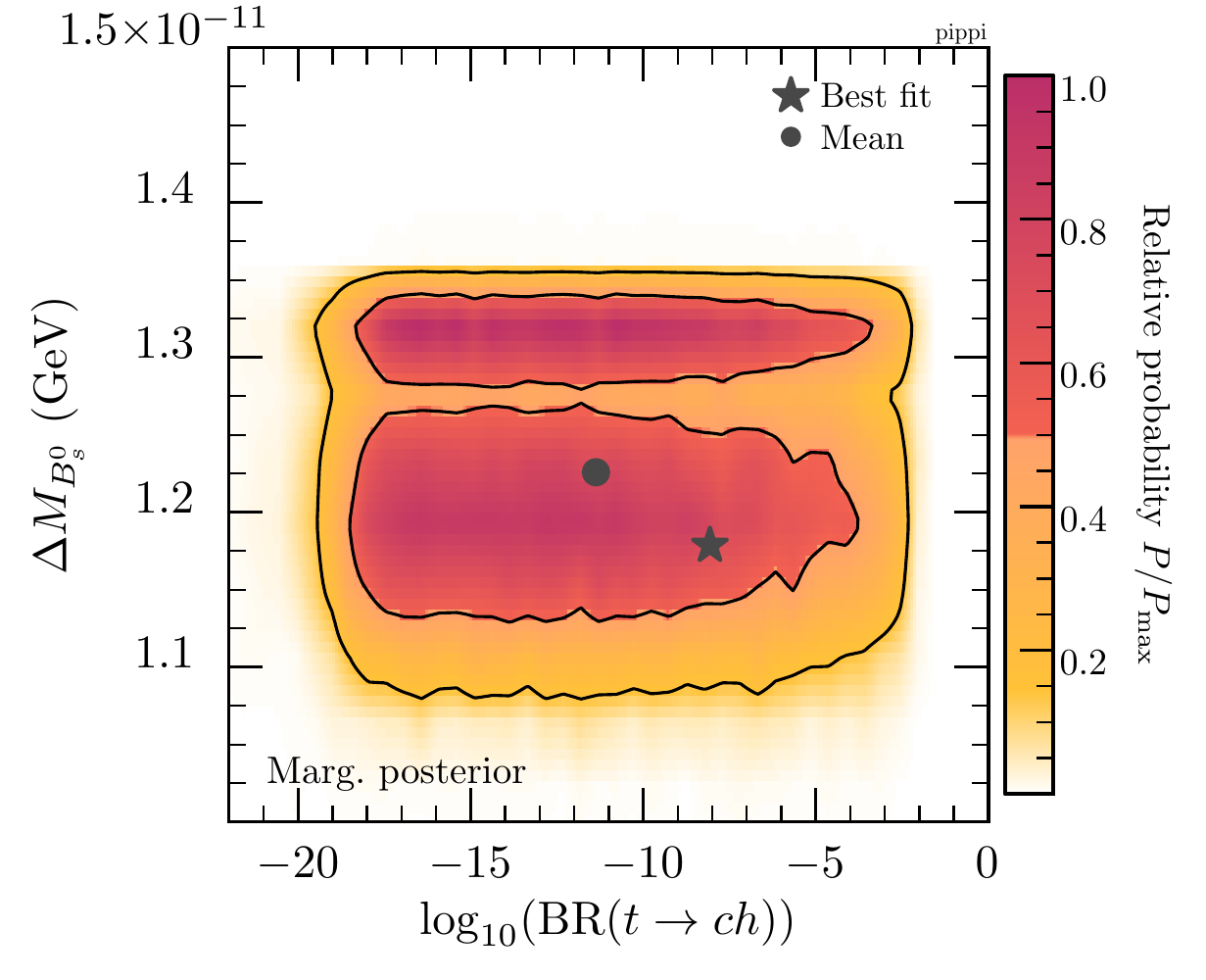}~~\includegraphics[scale=0.6]{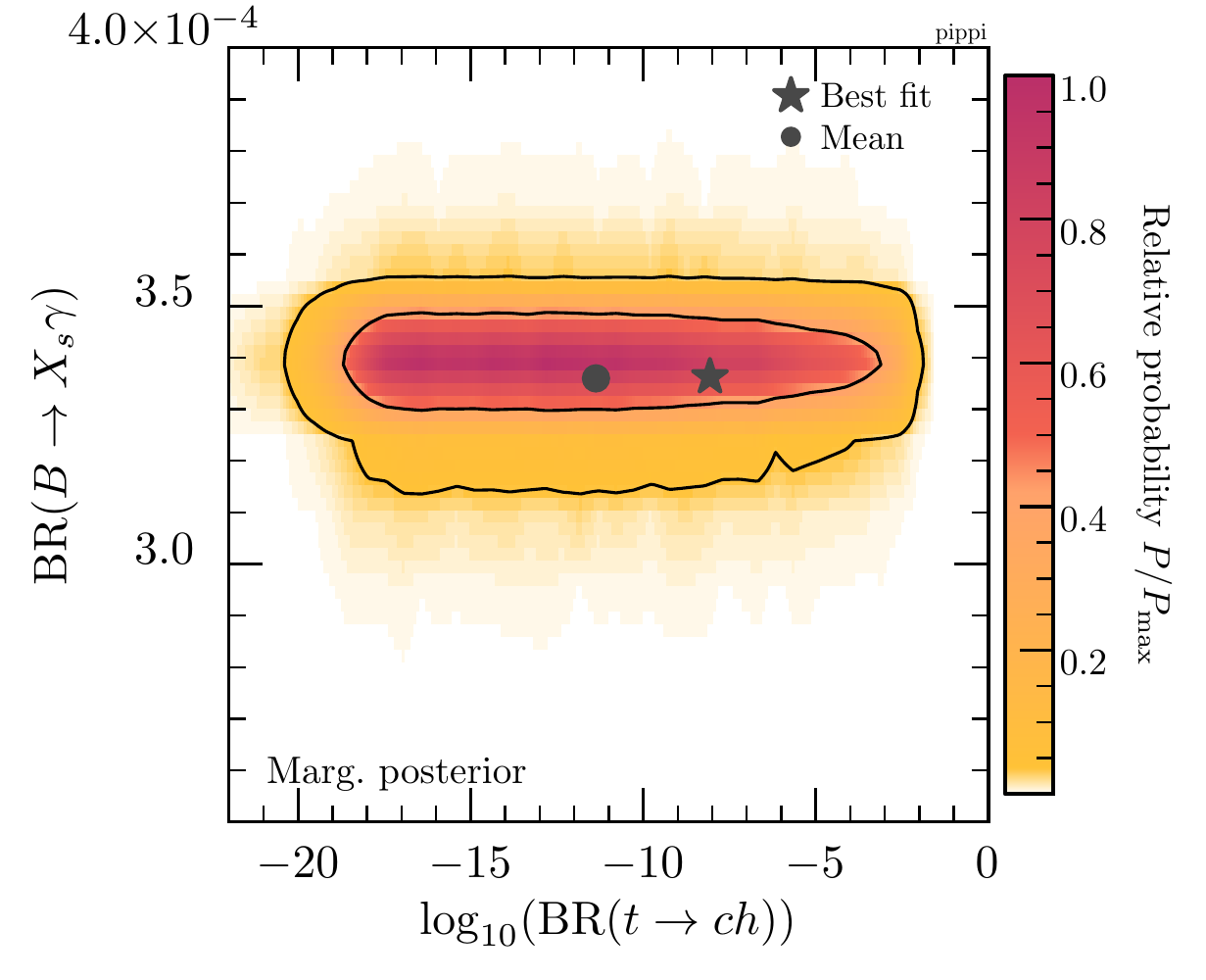}
\caption{Different observables and parameters versus $\log_{10}[{\rm BR}(t\rightarrow ch)]$. \emph{ Left:} $B^0_s$ meson mixing mass splitting. \emph{Right:} Radiative B decays.}
\label{fig:res3b}}
\end{figure}

In Fig.~\ref{fig:res3b} we plot the $B^0_s$ meson mixing mass splitting and $B \rightarrow X_s\gamma$ versus ${\rm BR}(t\rightarrow ch)$. In radiative B-decays, the combinations $\xi^U_{23}\,  \xi^U_{33}\, m_t$ with tops and $\xi^U_{23}\, \xi^D_{33}\, m_b$ with bottoms in the loop, enter. On the other hand, Higgs data favours somewhat large diagonal Yukawa contributions. This in turn implies some (weak) upper bounds on $\xi^U_{23}$. The upper limit on the branching BR($t\rightarrow ch$) comes from the LHC observed upper limit, $2.2\times 10^{-3}$ (see Eq.~\eqref{eq:topFlChDecays:Exp1}), hence, indirect constraints are weaker. As such, there is still almost an order of magnitude of precision before we may begin exploring the allowed 2HDM region at colliders. In this case, no lower bounds have been found from our scans, these are again just from the priors. 

\begin{figure}[h]
\centering
{%
\includegraphics[scale=0.6]{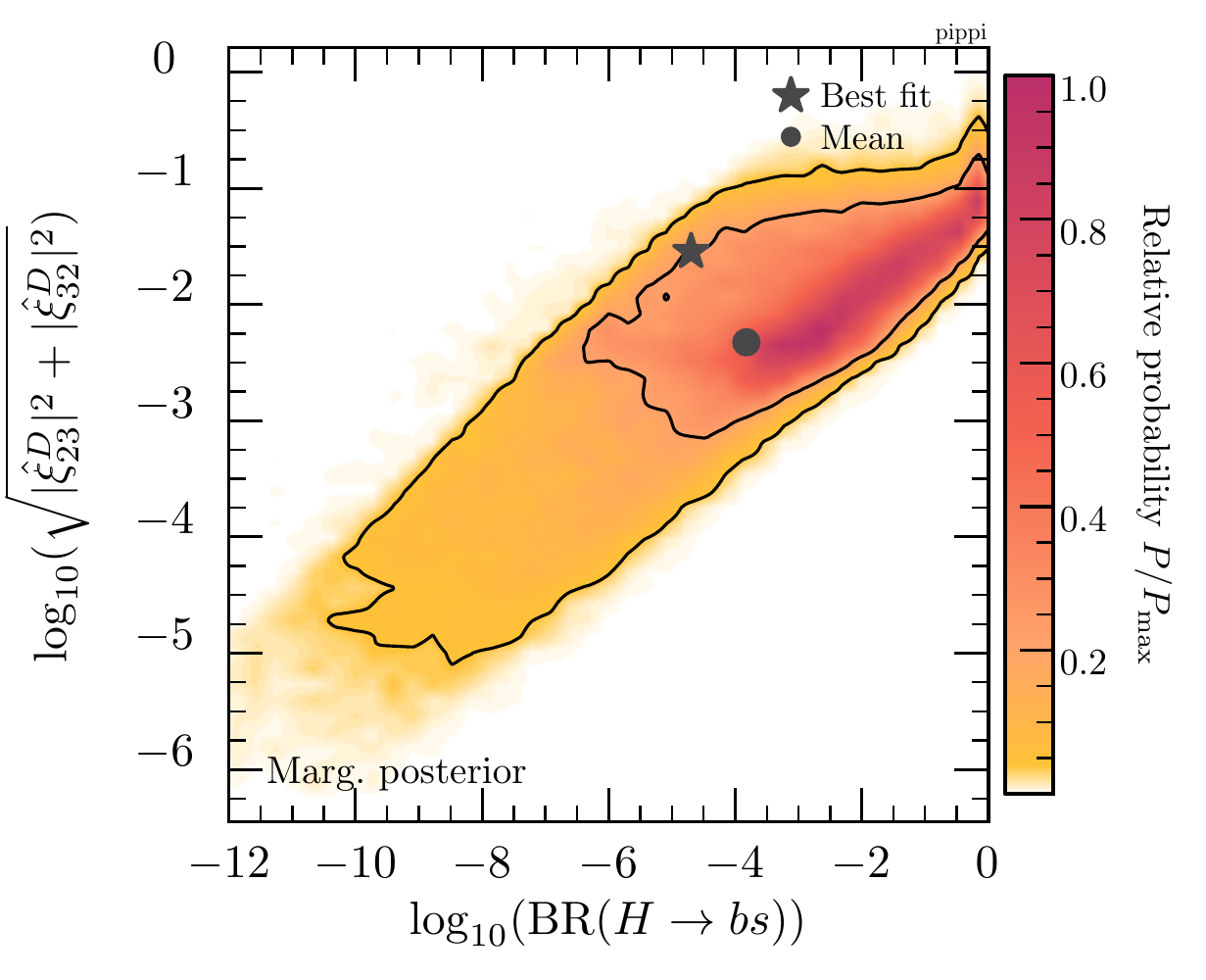}~~\includegraphics[scale=0.6]{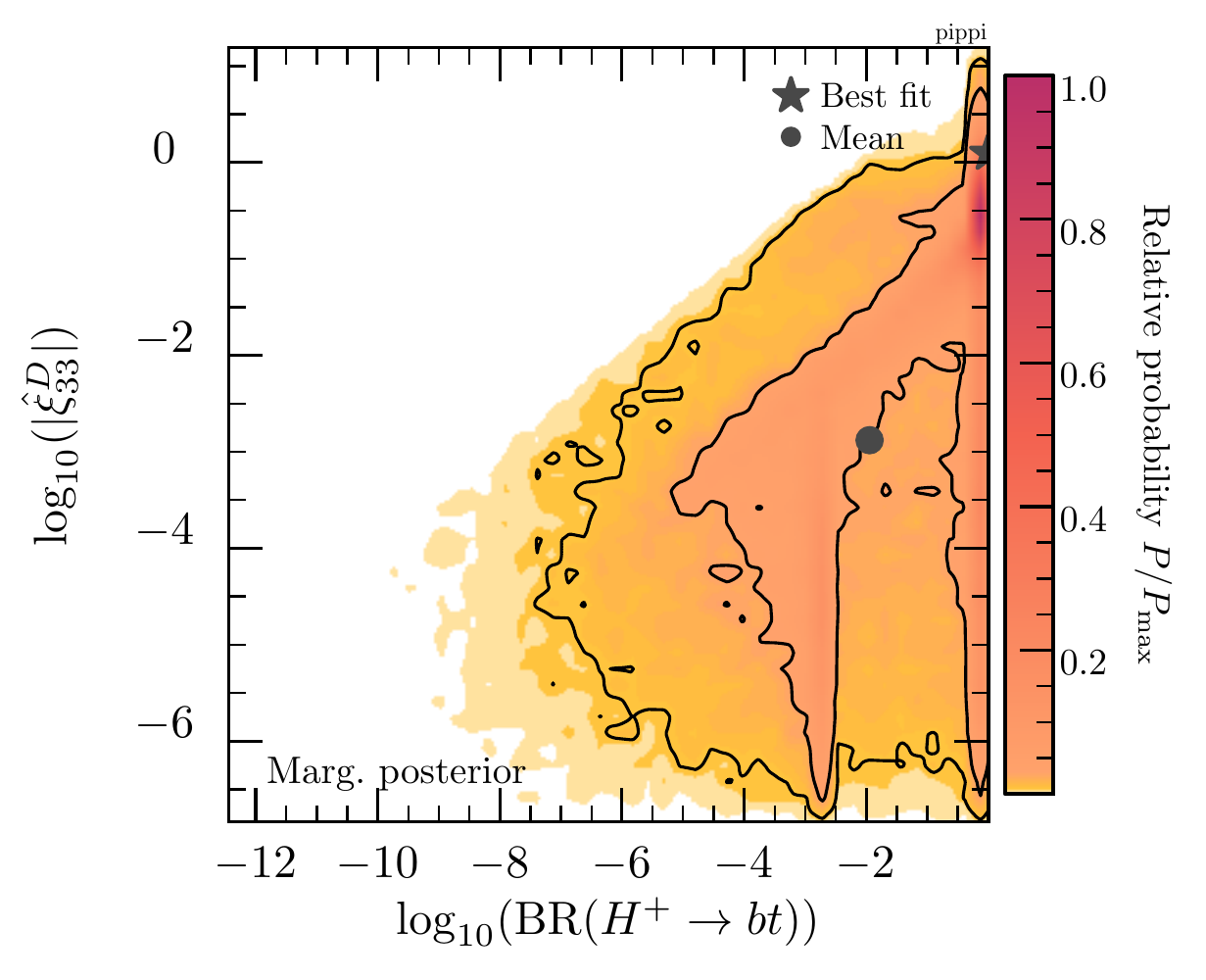}
}%
\caption{\emph{Left:}  $\log_{10}(\hat\xi^D_{23}|^2+|\hat\xi^D_{32}|^2)^{1/2}$ versus $\log($BR($H\rightarrow bs$)). \emph{Right:}  $\log_{10}(|\hat\xi^D_{33}|)$ versus $\log$(BR($H^+\rightarrow tb$)).}
\label{fig:res4}
\end{figure}

It is also interesting to investigate flavor violation in the new scalar sector, that is decays involving $H, A$ and $H^\pm$. Fig.~\ref{fig:res4} displays the modulus of the relevant off-diagonal Yukawas versus BR($H\rightarrow bs$). Similar plots are obtained for $A\rightarrow bs,\, tc$, and $H\rightarrow tc$. It is remarkable that these flavor-changing decays can saturate the decay widths of the heavy scalars. This may be relevant for direct searches. We also note that $H^+ \rightarrow bt$ has the largest lower bound.

\section{Conclusions}\label{conc}

In this work, we have investigated quark flavor violation involving the second and third families from an effective field theory point of view. We concentrated on the interesting processes $h \rightarrow bs$ and $t \rightarrow ch$. After outlining the possible tree-level simplified models, which involve new scalars and/or vector-like quarks, and estimating their contributions to HQFV processes, we have focused on the most promising scenario to produce large signals: a general (or Type-III) 2HDM model.

We carried out a comprehensive global scan of the 2HDM model imposing theoretical and experimental constraints. We focused primarily on $B$-physics constraints coming from $B^0_s$ meson mixing (mass splitting and CP-violating phase) and the radiative decay $B \rightarrow X_s \gamma$, which impose the most significant restrictions on the non-diagonal Yukawa elements $\hat\xi_{23,32}^D$ and $\hat\xi_{23,32}^U$. We have also obtained that the $\sim 2 \sigma$ mass-splitting discrepancy with respect to the SM in the $B_s$ meson system can be accommodated in the 2HDM at tree level, yielding a prediction of ${\rm BR}(h\to bs)\simeq 10^{-4}$ if loop level and heavy Higgs contributions are not significant.

The final values obtained in out full parameter scan are BR$(h \rightarrow bs)<10^{-3}\, (3\times 10^{-2})$ and BR$(t \rightarrow ch)<6 \times 10^{-4}\, (10^{-2})$ at  $1$ and $2 \sigma$ (lower bounds, if present, are at the level of the one-loop SM prediction). This parameter space is already accessible and can be further examined at future colliders. Beyond the two hallmark decays, possibly the easiest HQFV process to observe is $H^+ \rightarrow bt$ due to its large production cross section and the possible large branching fraction.

Any observed (therefore sizeable) signal of quark flavor violation involving the Higgs boson would clearly point to physics beyond the SM. As we have studied in this work, the stringent limits from low energy observables imply that it would most possibly stem from a 2HDM. We have demonstrated that the allowed parameter space in the up and the down sectors allowed by current upper limits are well within reach.


\begin{acknowledgments}
We are grateful to A.~Azatov for useful discussions. This work has been supported in part by the Australian Research Council. MN acknowledges support from Funda\c{c}\~ao para a Ci\^encia e a Tecnologia (FCT, Portugal) through the projects UID/FIS/00777/2019, CERN/FIS-PAR/0004/2017, and PTDC/FIS-PAR/29436/2017. JHG acknowledges financial support from the H2020-MSCA-RISE project “InvisiblesPlus”, and he thanks the Theoretical Physics Department of Fermilab, where this project was completed, for the kind hospitality.
\end{acknowledgments}

\clearpage
\appendix
\section{Derivative Operators for Vector-like Quarks\label{app:VLQ_details}} 
In Tab.~\ref{tab_der} we list the quantum numbers ($2T+1,T_{3},Y$) of the different SM (EFT) quark objects on which the covariant derivative 
acts in order to derive the $Z, W$ couplings. Details on the procedure are given in Ref.~\cite{Herrero-Garcia:2016uab}. In \emph{Diff.} we just take the difference
of the pair $(T_{3},Y)_{\mathrm{EFT}}-(T_{3},Y)_{\mathrm{SM}}$ which
will give us the ``left-over'' combination of $W_{3}$ and $B$
fields, and therefore the $Z$ and $W$ interactions. 

\begin{table} 
\centering
\begin{tabular}{|c|c|c|c|c|c|c|c|c|c|} 
\hline 
& Part & SM & EFT & Diff. & $d_{R}Z$ & $d_{L}Z$ & $u_{R}Z$ & $u_{L}Z$ & $W$\tabularnewline
\hline 
\hline 
$d_{R}\text{\ensuremath{\Phi}}$ & $d_{R}$ & $(1,0,-1/3)$ & $(2,-\frac{1}{2},\frac{1}{6})$ & $-(\frac{1}{2},-\frac{1}{2})$ &  $-1$& & & & \tabularnewline
\hline 
$d_{R}\text{\ensuremath{\tilde{\Phi}}}$ & $d_{R}$ & $(1,0,-1/3)$ & $(2,\frac{1}{2},-\frac{5}{6})$ & $(\frac{1}{2},-\frac{1}{2})$ &  $+1$& &&  & \tabularnewline
\hline 
$u_{R}\text{\ensuremath{\Phi}}$ & $u_{R}$ & $(1,0,2/3)$ & $(2,-\frac{1}{2},\frac{7}{6})$ & $-(\frac{1}{2},-\frac{1}{2})$ &  && $-1$ & &  \tabularnewline
\hline 
$u_{R}\text{\ensuremath{\tilde{\Phi}}}$ & $u_{R}$ & $(1,0,2/3)$ & $(2,\frac{1}{2},\frac{1}{6})$ & $(\frac{1}{2},-\frac{1}{2})$ &  && $+1$ &  & \tabularnewline
\hline 
$\Phi^{\dagger}Q$ & $d_{L}$ & $(2,-\frac{1}{2},\frac{1}{6})$ & $(1,0,-1/3)$ & $(\frac{1}{2},-\frac{1}{2})$ &  &$+1$  & & & -1\tabularnewline
\hline 
$\tilde{\Phi}^{\dagger}Q$ & $u_{L}$ & $(2,\frac{1}{2},\frac{1}{6})$ & $(1,0,2/3)$ & $-(\frac{1}{2},-\frac{1}{2})$ &  &  & & -1 & -1\tabularnewline
\hline 
\multirow{2}{*}
{$\Phi^{\dagger}\vec{\tau}Q$} & $-d_{L}$ & $(2,-\frac{1}{2},\frac{1}{6})$ & $(3,0,-1/3)$ & $(\frac{1}{2},-\frac{1}{2})$ &  & $+1$ &  & & \multirow{2}{*}{}\tabularnewline
\cline{2-8} 
& $\sqrt{2}u_{L}$ & $(2,\frac{1}{2},\frac{1}{6})$ & $(3,1,-1/3)$ & $(\frac{1}{2},-\frac{1}{2})$ &  &  &  &$+2$& \tabularnewline
\hline 
\multirow{2}{*}
{$\tilde{\Phi}^{\dagger}\vec{\tau}Q$} & $\sqrt{2}d_{L}$ & $(2,-\frac{1}{2},\frac{1}{6})$ & $(3,-1,2/3)$ & $-(\frac{1}{2},-\frac{1}{2})$ &  &  $-2$&  & & \multirow{2}{*}{}\tabularnewline
\cline{2-8} 
& $u_{L}$ & $(2,\frac{1}{2},\frac{1}{6})$ & $(3,0,2/3)$ & $-(\frac{1}{2},-\frac{1}{2})$ &  &  &&  $-1$& \tabularnewline
\hline 
\end{tabular}
\caption{Derivation of FCNC interactions generated by \emph{Derivative operators}. $Z$ couplings are in units of $y_q v/m_q\,\times e/(2c_{W}s_{W})$, while $W$ ones are in units of $V y_q v/m_q\,\times e/(2\sqrt{2}s_{W})$.}  \label{tab_der}
\end{table}


\section{Parameter Values} \label{app:scan_parameters}

The SM values used for the calculation are presented in Tab.~\ref{table:SM_values} and relevant parameters for meson mixing are given in Tab.~\ref{table:meson_mass_decay_B_values}.
The complex CKM matrix we use in our calculation is attained from UTFit 2016 SM Fits~\cite{Bona:2006ah}, and reads
\begin{equation}
	\CKMmat = 
\left(
\begin{array}{ccc}
0.97431 & 0.22512 & 0.00365e^{-65.88i} \\
-0.22497e^{0.0352i} & 0.97344e^{-0.001877i} & 0.04255 \\
0.00869e^{-22.0i} & -0.04156^{1.040i} & 0.999097 \\
\end{array}
\right)\,.
\end{equation}

\begin{table*}\centering
\ra{1.3}
\begin{tabular}{@{}lllll@{}}\toprule
\midrule
Parameter & Value & & Parameter & Value  \\ 
\cmidrule{1-2}\cmidrule{4-5}
  $m_u$ & $2.2 \times 10^{-3}$ GeV & & $\alpha_{\rm em}(m_Z)$ & $1/127.934$   \\
  $m_c$  & $1.67$ GeV & & $\alpha^{0}$ & $1/137.036$ \\
  $\overline{m_c}(m_c)$  \cite{Bazavov:2018omf} & $1.273$ GeV & & $G_{\rm F}$ & $1.16638\times 10^{-5}\,\text{GeV}^{-2}$\\
  $m_t$  & $173.5$ GeV & & $\alpha_s(m_Z)$ & $0.1182$   \\
  $\overline{m_t}(m_t)$ \cite{Jegerlehner:2012kn} & $(173.5 - 10.38)$ GeV & &$m_W$ & $80.385$ GeV \\
  $m_d$ & $4.7 \times 10^{-3}$ GeV & & $m_Z$ & $91.1876$ GeV \\
  $m_s$ & $0.096$ GeV & \\ 
  $m_b$ & $4.78$ GeV & \\ 
  $\overline{m_b}(m_b)$\cite{Bazavov:2018omf} & $4.197$ GeV \\
  \midrule
  \bottomrule
\end{tabular}
\caption{Standard Model values used for global scan, attained from Ref.~\cite{Patrignani:2016xqp} where not explicitly stated otherwise. Parameters dependant on scale are normalised in the $\overline{\rm MS}$ scheme while non scheme dependant masses are assumed to be given as pole masses. \note{where are the ckm values from? FLAG?}}
\label{table:SM_values}
\end{table*}

\begin{table}\centering
\ra{1.3}
\begin{tabular}{@{}llllllll@{}}
\toprule
\midrule
\multicolumn{2}{c}{Meson Mass}\cite{Patrignani:2016xqp} & & \multicolumn{2}{c}{Decay Constant }\cite{Patrignani:2016xqp} & & \multicolumn{2}{c}{Bag factors} \cite{Lubicz:2008am} \\
	\cmidrule{1-2} \cmidrule{4-5} \cmidrule{7-8}
	$M_{B_s}$ & 5.36689 GeV & & $f_{B_s}$ & 0.224 GeV & & $B_1^{B_s}(\mu_b)$ & $0.87$\\
	& & & & & & $B_2^{B_s}(\mu_b)$ &  $0.80$ \\
	& & & & & & $B_3^{B_s}(\mu_b)$ &  $0.93$\\ 
 	& & & & & & $B_4^{B_s}(\mu_b)$ &  $1.16$ \\ 
 	& & & & & & $B_5^{B_s}(\mu_b)$ &  $1.75$ \\
\midrule
 \bottomrule
\end{tabular}
\caption{Values of $B_i(\mu)$ are renormalised in the $\overline{\rm MS}$ scheme. The B-meson decay constant and mass are also provided.}
\label{table:meson_mass_decay_B_values}
\end{table}

\section{Evolution Matrix for Meson Mixing} \label{ap:evolution}

We extract the RGE matrix for the Meson Mixing basis introduced in Eq.~\eqref{eq:MM_basis} using DSixTools \cite{Aebischer:2017tk}. This matrix represents the running of the operators from $\mu_i = m_W$ to $\mu_f = m_B$: 
\begin{equation}
\begin{aligned}
U&(m_B, m_W) = \\ 
  & \left(
\begin{array}{cccccccc}
 0.862096 & 0 & 0 & 0 & 0 & 0 & 0 & 0 \\
    0 & 1.41304 & -0.197994 & 0 & 0 & 0 & 0 & 0 \\
    0 & -0.0516513 & 0.682309 & 0 & 0 & 0 & 0 & 0 \\
    0 & 0 & 0 & 1.79804 & 0.288788 & 0 & 0 & 0 \\
    0 & 0 & 0 & 0 & 0.931673 & 0 & 0 & 0 \\
    0 & 0 & 0 & 0 & 0 & 0.862096 & 0 & 0 \\
    0 & 0 & 0 & 0 & 0 & 0 & 1.41304 & -0.197994 \\
    0 & 0 & 0 & 0 & 0 & 0 & -0.0516513 & 0.682309 \\
\end{array} \right)
\end{aligned}
\end{equation}

\bibliographystyle{JHEP}

\bibliography{HQFV}

\end{document}